\documentclass[a4paper,11pt]{article}
\pdfoutput=1 

\usepackage{aas_macros}
\usepackage{jcappub} 
\usepackage[T1]{fontenc} 

\usepackage[dvipsnames]{xcolor}
\usepackage[normalem]{ulem} 
\usepackage{soul}  
\usepackage{tikz}
\usepackage{wrapfig}

\newcommand{\cwr}[1]{%
  \ifmmode
    \tikz[baseline=(X.base)]{
      \node[inner sep=0pt, text=red] (X) {$#1$};
      \draw[Red, line width=0.5pt] (X.west|-X.center) -- (X.east|-X.center);
      \draw[Red, line width=0.5pt] (X.west|-X.south) -- (X.east|-X.south);
    }
  \else
    \textcolor{Red}{\uline{\sout{#1}}}
  \fi
}
\newcommand{\remove}[1]{%
  \ifmmode
    \tikz[baseline=(X.base)]{
      \node[inner sep=0pt, text=red] (X) {$#1$};
      \draw[Red, line width=0.5pt] (X.west|-X.center) -- (X.east|-X.center);
      \draw[Red, line width=0.5pt] (X.west|-X.south) -- (X.east|-X.south);
    }
  \else
    \textcolor{Red}{\uline{\sout{#1}}}
  \fi
}

\newcommand{\JCAP}[1]{\textcolor{ForestGreen}{#1}}

\newcommand{\hMpc}{\mathrm{h}^{-1}\mathrm{Mpc}}
\newcommand{\bo}{\bar{\omega}}
\newcommand{\dz}{\Delta z}

\usepackage{orcidlink}
\usepackage{multirow}
\usepackage{amsmath}
\usepackage{bbm}
\usepackage{stmaryrd}
\usepackage{bm}
\usepackage{graphicx}  
\usepackage{tikz}      
\usetikzlibrary{positioning, decorations.pathreplacing} 
\usepackage{subcaption}
\usepackage{adjustbox}
\usepackage{comment}

\usepackage{color,hyperref}

\title{\boldmath Full calibration of the tomographic redshift distribution from the HSC PDR3 Shape Catalog with DESI}

\author[1,2]{J.~Choppin de Janvry\,\orcidlink{0009-0008-8066-446X}}
\author[1,3]{S.~Gontcho A Gontcho\,\orcidlink{0000-0003-3142-233X}}
\author[2,4]{U.~Seljak}
\author[1,2]{A.~Baleato Lizancos\,\orcidlink{0000-0002-0232-6480}}
\author[1]{E.~Chaussidon\,\orcidlink{0000-0001-8996-4874}}
\author[5]{W.~d'Assignies\,\orcidlink{0000-0002-9719-1717}}
\author[6]{J.~DeRose\,\orcidlink{0000-0002-0728-0960}}
\author[7]{S.~Heydenreich\,\orcidlink{0000-0002-7273-4076}}
\author[8,9]{E.~Paillas\,\orcidlink{0000-0002-4637-2868}}
\author[4]{D.~Valcin\,\orcidlink{0000-0003-0129-0620}}
\author[10]{T.~Zhang}
\author[1]{J.~Aguilar}
\author[11]{S.~Ahlen\,\orcidlink{0000-0001-6098-7247}}
\author[12,13]{D.~Bianchi\,\orcidlink{0000-0001-9712-0006}}
\author[14]{D.~Brooks}
\author[15,16]{F.~J.~Castander\,\orcidlink{0000-0001-7316-4573}}
\author[1]{T.~Claybaugh}
\author[1]{A.~Cuceu\,\orcidlink{0000-0002-2169-0595}}
\author[17]{A.~de la Macorra\,\orcidlink{0000-0002-1769-1640}}
\author[14]{P.~Doel}
\author[1,4]{S.~Ferraro\,\orcidlink{0000-0003-4992-7854}}
\author[5]{A.~Font-Ribera\,\orcidlink{0000-0002-3033-7312}}
\author[18,19]{J.~E.~Forero-Romero\,\orcidlink{0000-0002-2890-3725}}
\author[15,16,20]{E.~Gaztañaga\,\orcidlink{0000-0001-9632-0815}}
\author[21]{G.~Gutierrez}
\author[22,23]{H.~K.~Herrera-Alcantar\,\orcidlink{0000-0002-9136-9609}}
\author[24,25,26]{K.~Honscheid\,\orcidlink{0000-0002-6550-2023}}
\author[27]{M.~Ishak\,\orcidlink{0000-0002-6024-466X}}
\author[28]{R.~Joyce\,\orcidlink{0000-0003-0201-5241}}
\author[28]{S.~Juneau\,\orcidlink{0000-0002-0000-2394}}
\author[29]{R.~Kehoe}
\author[30]{D.~Kirkby\,\orcidlink{0000-0002-8828-5463}}
\author[1]{T.~Kisner\,\orcidlink{0000-0003-3510-7134}}
\author[1]{A.~Kremin\,\orcidlink{0000-0001-6356-7424}}
\author[14]{O.~Lahav}
\author[26]{C.~Lamman\,\orcidlink{0000-0002-6731-9329}}
\author[1]{M.~Landriau\,\orcidlink{0000-0003-1838-8528}}
\author[31]{L.~Le~Guillou\,\orcidlink{0000-0001-7178-8868}}
\author[5,32]{M.~Manera\,\orcidlink{0000-0003-4962-8934}}
\author[28]{A.~Meisner\,\orcidlink{0000-0002-1125-7384}}
\author[5,33]{R.~Miquel}
\author[20]{S.~Nadathur\,\orcidlink{0000-0001-9070-3102}}
\author[1,23]{N.~Palanque-Delabrouille\,\orcidlink{0000-0003-3188-784X}}
\author[34,35,36]{W.~J.~Percival\,\orcidlink{0000-0002-0644-5727}}
\author[1,4,37]{C.~Poppett}
\author[38]{F.~Prada\,\orcidlink{0000-0001-7145-8674}}
\author[39]{I.~P\'erez-R\`afols\,\orcidlink{0000-0001-6979-0125}}
\author[40]{G.~Rossi}
\author[41]{E.~Sanchez\,\orcidlink{0000-0002-9646-8198}}
\author[1]{D.~Schlegel}
\author[42,43]{M.~Schubnell}
\author[1]{J.~Silber\,\orcidlink{0000-0002-3461-0320}}
\author[28]{D.~Sprayberry}
\author[43]{G.~Tarl\'{e}\,\orcidlink{0000-0003-1704-0781}}
\author[28]{B.~A.~Weaver}
\author[1]{R.~Zhou\,\orcidlink{0000-0001-5381-4372}}

\affiliation[1]{Lawrence Berkeley National Laboratory, 1 Cyclotron Road, Berkeley, CA 94720, USA}
\affiliation[2]{Department of Physics, University of California, Berkeley, 366 LeConte Hall MC 7300, Berkeley, CA 94720-7300, USA}
\affiliation[3]{University of Virginia, Department of Astronomy, Charlottesville, VA 22904, USA}
\affiliation[4]{University of California, Berkeley, 110 Sproul Hall \#5800 Berkeley, CA 94720, USA}
\affiliation[5]{Institut de F\'{i}sica d’Altes Energies (IFAE), The Barcelona Institute of Science and Technology, Edifici Cn, Campus UAB, 08193, Bellaterra (Barcelona), Spain}
\affiliation[6]{Physics Department, Brookhaven National Laboratory, Upton, NY 11973, USA}
\affiliation[7]{Department of Astronomy and Astrophysics, UCO/Lick Observatory, University of California, 1156 High Street, Santa Cruz, CA 95064, USA}
\affiliation[8]{Instituto de Estudios Astrof\'isicos, Facultad de Ingenier\'ia y Ciencias, Universidad Diego Portales, Av. Ej\'ercito Libertador 441, Santiago, Chile}
\affiliation[9]{Steward Observatory, University of Arizona, 933 N. Cherry Avenue, Tucson, AZ 85721, USA}
\affiliation[10]{Department of Physics \& Astronomy and Pittsburgh Particle Physics, Astrophysics, and Cosmology Center (PITT PACC), University of Pittsburgh, 3941 O'Hara Street, Pittsburgh, PA 15260, USA}
\affiliation[11]{Department of Physics, Boston University, 590 Commonwealth Avenue, Boston, MA 02215 USA}
\affiliation[12]{Dipartimento di Fisica ``Aldo Pontremoli'', Universit\`a degli Studi di Milano, Via Celoria 16, I-20133 Milano, Italy}
\affiliation[13]{INAF-Osservatorio Astronomico di Brera, Via Brera 28, 20122 Milano, Italy}
\affiliation[14]{Department of Physics \& Astronomy, University College London, Gower Street, London, WC1E 6BT, UK}
\affiliation[15]{Institut d'Estudis Espacials de Catalunya (IEEC), c/ Esteve Terradas 1, Edifici RDIT, Campus PMT-UPC, 08860 Castelldefels, Spain}
\affiliation[16]{Institute of Space Sciences, ICE-CSIC, Campus UAB, Carrer de Can Magrans s/n, 08913 Bellaterra, Barcelona, Spain}
\affiliation[17]{Instituto de F\'{\i}sica, Universidad Nacional Aut\'{o}noma de M\'{e}xico,  Circuito de la Investigaci\'{o}n Cient\'{\i}fica, Ciudad Universitaria, Cd. de M\'{e}xico  C.~P.~04510,  M\'{e}xico}
\affiliation[18]{Departamento de F\'isica, Universidad de los Andes, Cra. 1 No. 18A-10, Edificio Ip, CP 111711, Bogot\'a, Colombia}
\affiliation[19]{Observatorio Astron\'omico, Universidad de los Andes, Cra. 1 No. 18A-10, Edificio H, CP 111711 Bogot\'a, Colombia}
\affiliation[20]{Institute of Cosmology and Gravitation, University of Portsmouth, Dennis Sciama Building, Portsmouth, PO1 3FX, UK}
\affiliation[21]{Fermi National Accelerator Laboratory, PO Box 500, Batavia, IL 60510, USA}
\affiliation[22]{Institut d'Astrophysique de Paris. 98 bis boulevard Arago. 75014 Paris, France}
\affiliation[23]{IRFU, CEA, Universit\'{e} Paris-Saclay, F-91191 Gif-sur-Yvette, France}
\affiliation[24]{Center for Cosmology and AstroParticle Physics, The Ohio State University, 191 West Woodruff Avenue, Columbus, OH 43210, USA}
\affiliation[25]{Department of Physics, The Ohio State University, 191 West Woodruff Avenue, Columbus, OH 43210, USA}
\affiliation[26]{The Ohio State University, Columbus, 43210 OH, USA}
\affiliation[27]{Department of Physics, The University of Texas at Dallas, 800 W. Campbell Rd., Richardson, TX 75080, USA}
\affiliation[28]{NSF NOIRLab, 950 N. Cherry Ave., Tucson, AZ 85719, USA}
\affiliation[29]{Department of Physics, Southern Methodist University, 3215 Daniel Avenue, Dallas, TX 75275, USA}
\affiliation[30]{Department of Physics and Astronomy, University of California, Irvine, 92697, USA}
\affiliation[31]{Sorbonne Universit\'{e}, CNRS/IN2P3, Laboratoire de Physique Nucl\'{e}aire et de Hautes Energies (LPNHE), FR-75005 Paris, France}
\affiliation[32]{Departament de F\'{i}sica, Serra H\'{u}nter, Universitat Aut\`{o}noma de Barcelona, 08193 Bellaterra (Barcelona), Spain}
\affiliation[33]{Instituci\'{o} Catalana de Recerca i Estudis Avan\c{c}ats, Passeig de Llu\'{\i}s Companys, 23, 08010 Barcelona, Spain}
\affiliation[34]{Department of Physics and Astronomy, University of Waterloo, 200 University Ave W, Waterloo, ON N2L 3G1, Canada}
\affiliation[35]{Perimeter Institute for Theoretical Physics, 31 Caroline St. North, Waterloo, ON N2L 2Y5, Canada}
\affiliation[36]{Waterloo Centre for Astrophysics, University of Waterloo, 200 University Ave W, Waterloo, ON N2L 3G1, Canada}
\affiliation[37]{Space Sciences Laboratory, University of California, Berkeley, 7 Gauss Way, Berkeley, CA  94720, USA}
\affiliation[38]{Instituto de Astrof\'{i}sica de Andaluc\'{i}a (CSIC), Glorieta de la Astronom\'{i}a, s/n, E-18008 Granada, Spain}
\affiliation[39]{Departament de F\'isica, EEBE, Universitat Polit\`ecnica de Catalunya, c/Eduard Maristany 10, 08930 Barcelona, Spain}
\affiliation[40]{Department of Physics and Astronomy, Sejong University, 209 Neungdong-ro, Gwangjin-gu, Seoul 05006, Republic of Korea}
\affiliation[41]{CIEMAT, Avenida Complutense 40, E-28040 Madrid, Spain}
\affiliation[42]{Department of Physics, University of Michigan, 450 Church Street, Ann Arbor, MI 48109, USA}
\affiliation[43]{University of Michigan, 500 S. State Street, Ann Arbor, MI 48109, USA}

\emailAdd{jean.choppindejanvrydev@gmail.com}
\emailAdd{satya@virginia.edu}
\emailAdd{useljak@berkeley.edu}

\abstract{The calibration of tomographic redshift distributions 
is essential for cosmological analysis of weak lensing data.  
In this work, we calibrate all four tomographic bins of the Hyper Suprime Camera (HSC) weak lensing catalog with the Dark Energy Spectroscopic Instrument (DESI) Data Release 1 and 2 using the clustering redshifts technique. We include $z>1.2$ redshift sources such as emission line galaxies (ELG) and quasars (QSO) sources in our calibration, which were not available in the previous HSC calibration (Rau et al. 2022, \cite{HSCClusteringRau2023}), allowing a complete calibration of all the redshift bins. We find the first tomographic bin exhibits a small shift towards low redshifts. The second bin is in good agreement with the photometric calibration, while third and fourth bin exhibit a shift towards higher redshifts. However, these shifts are considerably smaller than the shifts obtained in the HSC Year 3 cosmic shear analyses. We evaluate the impact of galaxy bias and magnification effects from all the samples on the measurements, finding them to be small, and we propose corrections to reduce them further. Specifically, we relax the assumption of linear bias and only assume no redshift evolution of the cross-correlation coefficient, allowing us to leverage smaller clustering scales. We model the redshift distributions with splines and compare our results to previous analyses as well as to other parameterizations found in literature. For the two high-redshift tomographic bins, we find the shifts to higher redshifts with respect to the measurements performed in Rau+2022 to be $\Delta z_3=-0.043^{+0.022}_{-0.020}$ and $\Delta z_4=-0.050^{+0.012}_{-0.012}$.}
\begin{document} 
\maketitle
\flushbottom


\section{Introduction}
\label{sec:intro}

With the current and upcoming generations of large-scale optical imaging surveys, such as the Hyper Suprime-Cam Subaru Strategic Program (HSC-SSP, \cite{HSC2022datarelease3}), the Dark Energy Survey (DES, \cite{DES2018datarelease1}), the Kilo-Degree Survey (KiDS, \cite{KiDS2025datarelease5}), the Nancy Grace Roman Space Telescope \cite{roman2019wfirst}, Euclid \cite{EuclidI2024overview} and the Vera C. Rubin Observatory's Legacy Survey of Space and Time (LSST, \cite{LSST2019survey}) probing hundreds of millions to billions of galaxies with accurate shape information, the strength of cosmological parameter inference from weak lensing analysis is becoming increasingly important. Some of the most precise single measurements of the lensing amplitude $S_8\equiv\sigma_8\sqrt{\Omega_m/0.3}$ currently come from cosmic shear analyses \cite{CosmoShearLi2023HSCY3, Dalal2023CosmoShearHSC, Amon2022CosmoShearDES, Secco2022CosmoShearDES, Asgari2021KidsCosmoShear, wright2025kidslegacyCosmoShear}, where one measures the correlation in the distortion of galaxy images due to weak gravitational lensing. 

These optical surveys rely on characterization of their normalized sample redshift distributions projected along the line of sight $n(z)$ when measuring two-point statistics of galaxy clustering and gravitational shear fields for cosmological parameter inference, most often performed through the two-point correlation functions \cite{CosmoShearLi2023HSCY3} or the angular power spectrum \cite{Dalal2023CosmoShearHSC}. Since optical surveys rely on broadband photometry covering optical and near-infrared wavelengths, narrow spectral galaxy features such as emission lines are difficult to identify due to the limited coverage per observed object and the broadband nature of the optical filters. Therefore, imaging surveys employ photometric redshift (photo-$z$) algorithms to infer individual redshifts. Typically, photometric redshifts are used to split the galaxies into tomographic redshift bins, and the true redshift distribution along the line of sight needs to be estimated for each bin. 
One can reconstruct the overall redshift distribution $n(z)$ from observed data for each tomographic bin. In the case of the fiducial HSC Year 3 analysis \cite{HSCClusteringRau2023, CosmoShearLi2023HSCY3, Dalal2023CosmoShearHSC}, four tomographic bins are used, spanning linear bins from $z_p\sim0.3$ to $z_p\sim1.5$, where $z_p$ here is a photometric redshift from a given photo-$z$ method. Henceforth, Bin 1 refers to $0.3\leq z_p<0.6$, Bin 2 to $0.6\leq z_p<0.9$, Bin 3 to $0.9\leq z_p<1.2$ and Bin 4 to $1.2\leq z_p<1.5$ respectively. 
Many different algorithms have been developed for photometric redshifts, ranging from inferring redshifts based on prior knowledge of galaxy spectral energy distributions (SEDs) (\cite{LePhare1, LePhare2, Brammer2008EAZY, MizukiHSCTanaka2015}) to mapping redshifts to photometry information with machine learning (\cite{CarrascoKind2014PhotozMLz, Hsieh2014DEMp, Collister2004ANNz}). Often, since the performance of the algorithms does not allow for perfect point estimates of the redshift, photo-$z$ methods will instead provide a probability density function (PDF) for each source, which allows for more information on how likely the point estimate is for a given photo-$z$ algorithm. If photometric redshifts were perfectly calibrated, the convolution of these PDFs would accurately describe the PDF $n(z)$ for the tomographic bins. This is not possible in practice due to the performance limitations of photo-$z$ algorithms with limited spectroscopic training sets. 

Thus, challenges arise when debiasing and calibrating the redshift distributions obtained by the chosen photo-$z$ algorithm(s), and as a result, weak lensing cosmological analyses can suffer from this systematic. Differences in tomographic redshift calibration can lead to significant cosmological parameter shifts, as demonstrated by Figure 1 of \cite{HSCClusteringRau2023} or by the differences in calibrations on the KiDS-Legacy dataset \cite{wright2025kidslegacyClusterZ, wright2025kidslegacyCosmoShear}, compared to the analysis performed on KiDS-1000 \cite{Asgari2021KidsCosmoShear}. 
 Another approach taken by the HSC team was to parametrize photometric redshift uncertainty using a free shift parameter of the redshift first moment for the higher redshift tomographic bins lacking good redshift calibration. While this can encompass systematics, it in turn weakened the obtained cosmological information on $S_8$ and $\Omega_m$ \cite{CosmoShearLi2023HSCY3, Dalal2023CosmoShearHSC}. Moreover, this self-calibration approach requires that the parameterization of the error in the redshift distribution is accurate.

To mitigate the weaknesses of photo-$z$ algorithms and to debias the photo-$z$ distributions, a number of methods have been developed to calibrate the sample redshift distribution of each bin. 
A popular approach is to rely on the optical photometry information given by the survey and known spectroscopic redshifts from external datasets to infer redshift distributions from the galaxy population color-space, often using data reduction algorithms and machine learning methods. This technique is commonly executed with Self-Organizing Maps (SOM) \cite{WrightSOMz, SOMPZDES2024, Hildebrandt2021ClusteringRedshiftsSOM, Masters2017C3R2SOM, Buchs2019SOMRedshifts}, or through direct weighted calibration \cite{LimaDIRMethod2008}. 
The limiting factor of this approach is often the source depth and color-space coverage of spectroscopic surveys, as spectroscopic redshift targets are more prone to failures and need longer integration times on faint targets with few easily discernible features for spectroscopy. 

Another method is to use the clustering of galaxies: by leveraging angular cross-correlations between a spectroscopic sample in small redshift slices and the galaxies in each of the tomographic bins of the weak lensing survey dataset, the technique known as clustering redshifts allows to debias \cite{Newman2008ClusterZ,ménard2014clusteringbasedredshiftestimationmethod, Gatti2018DESY1ClusteringRedshifts, Gatti2018DESY1systematics,Gatti2021ClusteringDESWLsrcDistrib, Cawthon2022DESY3Boss,Davis2017DESyear1ClusterZWLsrcdistrib,EuclidClusterRedshifts2025,KidsCrossCorrvdB2020,wright2025kidslegacyClusterZ,HSCClusteringRau2023,Morrison2017TheWiZZClusterZ, Schmidt2013ClusteringRedshifts, McQuinnClusterZ2013} the $n(z)$ distributions. 
This is the method employed in this work. 
However, clustering redshifts is also subject to systematics, since the method benefits from information of small angular scales (typically $0.1-5\hMpc$) yielding higher signal to noise, though galaxy bias modeling at these scales remains difficult. 
In this work, common possible systematics that can affect and bias the clustering redshifts calibration of the photometric sample are investigated, such as the impacts of galaxy bias and magnification. 
This methodology is applied to the S19A Shape Catalog \cite{HSCShapeCataloguePDR3Li2022} from the Hyper Suprime-Cam Subaru Strategic Survey Public Data Release 3 \cite{HSC2022datarelease3} (S19A HSC-SSP PDR3) as our photometric catalog to calibrate. 
Hereafter, we refer to this dataset simply as HSC. 
The Dark Energy Spectroscopic Instrument Data Release 1 \cite{DESI.DR1.I.Presentation} and Data Release 2 \cite{DESI.DR2.Presentation.InPrep} (DESI DR1, DR2) will serve as our spectroscopic calibration samples.

The Dark Energy Spectroscopic Instrument (DESI) is a Stage IV spectroscopic survey \cite{DESI2016b.Instr, DESI2022.KP1.Instr, Corrector.Miller.2023} carrying out a eight year program that started collecting data in 2021. DESI uses a focal plane hosting 5000 robotic positioners containing optical fibers \cite{FiberSystem.Poppett.2024} to obtain spectra, and therefore accurate spectroscopic redshifts \cite{Spectro.Pipeline.Guy.2023}, from galaxies. DESI plans to probe $17,000$ deg$^2$ of the sky in its main phase of operations \cite{SurveyOps.Schlafly.2023}, covering spectroscopic redshift ranges up to $z\sim4.2$. The DESI experiment studies six different types of tracers: the Milky Way Survey, observing stars in our local galaxy, the Bright Galaxy Survey (BGS, $0.01\lesssim z\lesssim0.6$) \cite{DESI.Selection.BGS.Hahn2023}, the Luminous Red Galaxies (LRG, $0.3\lesssim z\lesssim1.2$) \cite{DESI.Selection.LRG.Zhou2023}, the Emission Line Galaxies (ELG, $0.7\lesssim z\lesssim1.6$) \cite{DESI.Selection.ELG.Raichoor2023}, the Quasi-Stellar Objects (QSO, $0.7\lesssim z\lesssim2.1$) \cite{DESI.Selection.QSO.Chaussidon2023} and the Lyman-$\alpha$ forest (Ly$\alpha$, $z\gtrsim 2.1$). With its accurate redshift catalog, DESI has already allowed for significant breakthroughs on cosmological measurements \cite{DESI.DR2.II.BAO, DESI2024.VII.KP7B}.

The Hyper Suprime-Cam Subaru Strategic Program\footnote{\href{https://hsc-release.mtk.nao.ac.jp/doc/}{\textcolor{blue}{hsc-release.mtk.nao.ac.jp/doc/}}}\cite{HSCOverviewAihara2017} is a Stage 3 optical imaging survey using the Hyper Suprime-Cam wide-field camera installed on the Subaru $8.2$m telescope on the summit of Maunakea, Hawai'i, and covers three depth layers: the Wide field, the Deep field, and Ultra-Deep field. The S19A (from the the internal September 2019 data release) Shape Catalog \cite{HSCShapeCataloguePDR3Li2022} compiles galaxy shapes from $i_{\mathrm{band}}$ imaging data obtained from 2014 to 2019 in the Wide field of the HSC-SSP. It is crucial to obtain well calibrated redshift distributions of this dataset, as constraints from cosmic shear heavily depend on this larger area.

In the tomographic bin analysis performed by the HSC collaboration \cite{HSCClusteringRau2023}, there was no calibration from angular clustering of galaxies above $z\gtrsim1.2$ due to the redshift limitations of the LRG dataset used in that study, therefore not providing calibration information from clustering redshifts for most of Bin 3 and none for Bin 4. 
On the color-redshift calibration aspect, lack of high redshift spectroscopic sources and deep infrared photometry led to degeneracies that made for difficult calibration.
In turn, this makes the mean value from the fiducial calibration of a given redshift bin uncertain, meaning that major shifts are possible on Bins 3 and 4.
Indeed, analysis by \cite{Zhang2022PhotometricRedshiftShifts} showed a single shift parameter was sufficient to capture $n(z)$ uncertainty per bin in the HSC distributions.
The cosmic shear analyses on the two-point correlation function \cite{CosmoShearLi2023HSCY3} and on the angular power spectrum \cite{Dalal2023CosmoShearHSC} of HSC used a conservative uniform prior $\mathcal{U}([-1,1])$ for the tomographic bins redshift shift parameters.
Both analyses report hints of photo-$z$ miscalibration, with possible redshift expectation values being higher than initially obtained from the redshift bins only calibrated with photometry. Reported marginalized mode of shifts with asymmetric $34\%$ confidence interval and maximum a posteriori (MAP) values of shifts obtained by both HSC cosmic shear analyses have returned\footnote{Note the following convention: a negative shift implies the found expectation value (from cosmic shear analysis) for a bin is higher than the available measurements.}: 
$\dz_3=-0.075^{+0.056}_{
-0.059} (-0.046)$ and $\dz_4=-0.157^{+0.094}_{-0.111} (-0.144)$ 
for the angular power spectrum \cite{Dalal2023CosmoShearHSC} and 
$\dz_3=-0.115^{+0.052}
_{-0.058} (-0.120)$ and $
\dz_4=-0.192^{+0.088}
_{-0.088} (-0.190)$ 
for the angular two point correlation functions \cite{CosmoShearLi2023HSCY3} in Bin 3 and Bin 4. 
More recent works \cite{Rana2025HSCY3CosmicShearRatios} used Cosmic Shear Ratios (CSR) in order to provide calibration for the photometric redshift bins, while \cite{Zhang2025RedshiftGCPointMass} leverages galaxy clustering and weak lensing (GC-WL) to infer constraints on the source redshift bins. 
While the results found by that calibration are consistent with those obtained by the cosmic shear analyses, the error bars loosely constrain the redshift expectation: the obtained results were $\Delta z_3 = -0.002^{+0.085}_{-0.217},\;\Delta z_4 = -0.292^{+0.229}_{-0.324}$ for the CSR study and $\Delta z_3 =-0.112^{+0.046}_{-0.049},\;\Delta z_4 =-0.185^{+0.071}_{-0.081}$ for the GC-WL analysis. 
In this work, the focus will be on constraining these two shift parameters, comparing results to the HSC analyses as well as previous calibrations, and producing $n(z)$ distributions for each of the bins. 

The outline of this paper is as follows. 
Section \ref{sec:data} presents the datasets employed in this study: HSC’s Shape catalog and DESI DR1, DR2, as well as the quality cuts applied to both catalogs. 
Section \ref{sec:modeling} describes the modeling of the $n(z)$ distributions with two-point correlation functions and discusses systematic effects, including galaxy bias from both samples and magnification effects. 
Section \ref{sec:2pcf} describes our measurements of the angular correlation vectors. 
Section \ref{sec:parametrization} presents a spline based approach to model the $n(z)$ distributions and discusses other methods to regularize the clustering redshifts measurements. 
Finally, results are showcased in section \ref{sec:results}.

\section{Data}
\label{sec:data}

\begin{figure}[h]
    \centering
    \includegraphics[width=0.95\textwidth]{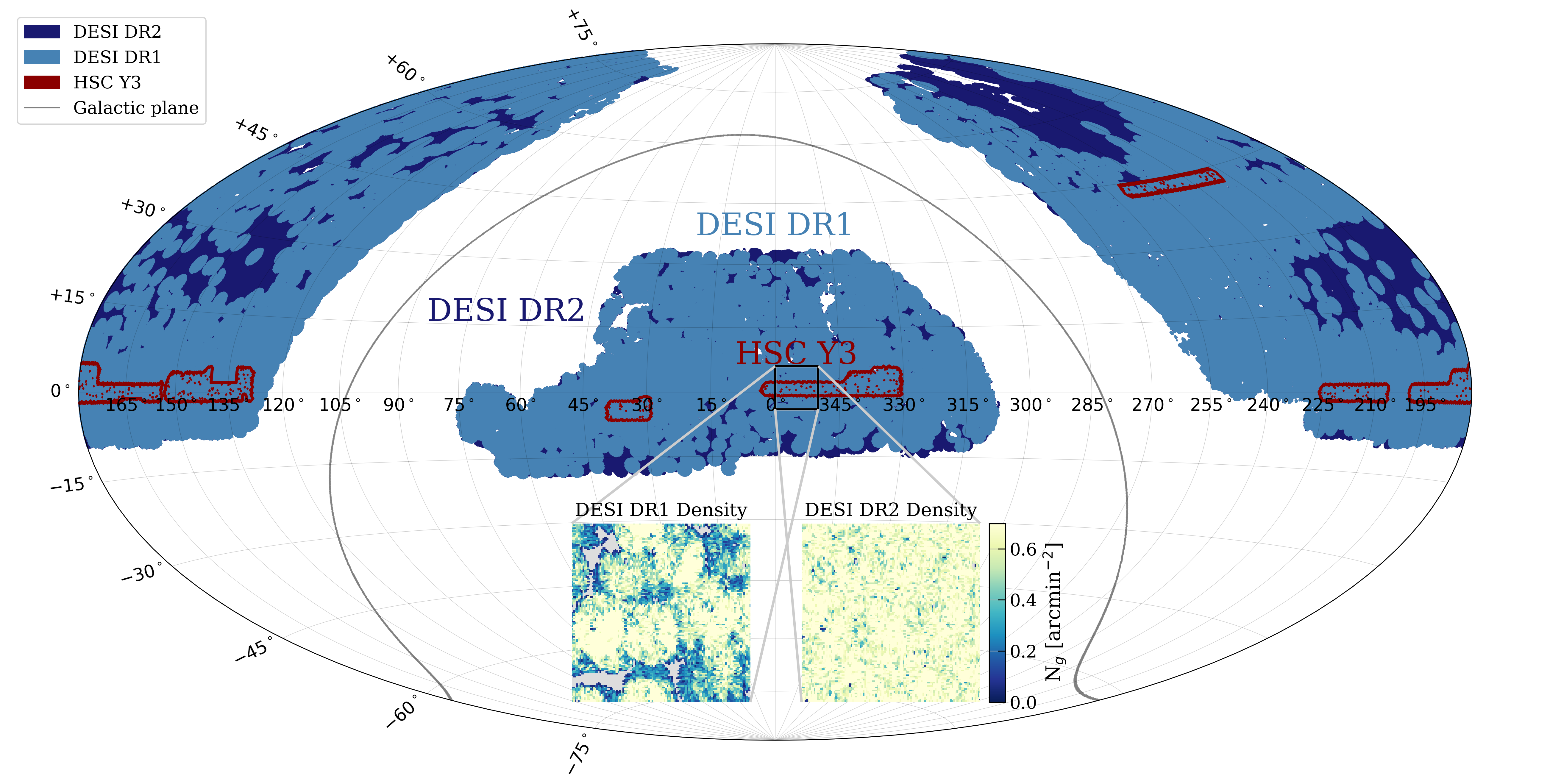}
    \caption{Sky coverage of DESI DR2 \cite{DESI.DR2.Presentation.InPrep} (dark blue), DR1 \cite{DESI.DR1.I.Presentation} (light blue, overlapping DR2) and HSC Year 3 Shape Catalog \cite{HSCShapeCataloguePDR3Li2022} (red outline), with the galactic plane (gray line). An area centered around $\mathrm{RA}\sim355^\circ$, $\mathrm{DEC}\sim0^\circ$ is highlighted with density maps to compare the coverage of DESI DR1 and DR2 of the same area covering a chunk of the footprint of HSC Y3's \texttt{VVDS} field. The HSC data lands in areas of high overall completeness for DESI, as these areas were targeted in priority. DESI DR2 improves over DR1 with better overall completeness, especially for the ELGs and in the northern \texttt{HECTOMAP} field (roughly $\mathrm{RA}\sim40^\circ$, $\mathrm{DEC}\sim240^\circ$).}
    \label{fig:footprint}
\end{figure}

This section describes the datasets and catalogs used in this analysis. 
DR1 refers to DESI Data Release 1\footnote{\href{https://data.desi.lbl.gov/doc/releases/dr1/}{\textcolor{blue}{data.desi.lbl.gov/doc/releases/dr1/}}}\cite{DESI.DR1.I.Presentation}, and DR2 refers to DESI Data Release 2. Data for DR1 was obtained with main survey observations spanning May 2021 to June 2022, while data for DR2 spans May 2021 to April 2024. To this, data releases also include survey validation observations from December 2020 to May 2021.
Compliance with DESI internal policies required the use of DESI DR2 to be limited to the calibration of the two highest HSC tomographic bins: $0.9 < z_{\mathrm{p}}\leq1.2$ and $1.2 < z_{\mathrm{p}}\leq1.5$. We use the public DESI DR1 dataset for the remaining tomographic bins: $0.3 < z_{\mathrm{p}}\leq0.6$ and $0.6 < z_{\mathrm{p}}\leq0.9$, where $z_p$ is the photometric redshift derived from HSC presented in section \ref{sec:data:hsc}. 
For measurements on tomographic bins different from the fiducial HSC \cite{HSCClusteringRau2023, CosmoShearLi2023HSCY3, Dalal2023CosmoShearHSC} ones (for example, in section \ref{sec:modeling:photoz_gal_bias}), DR1 is used for any tomographic bin such that $z_{\mathrm{p}}\lesssim0.9$, and DR2 elsewhere ($z_{\mathrm{p}}>0.9$), where $z_{\mathrm{p}}$ is the center of the tomographic bin in photometric redshift. 
The DESI dataset is further described hereafter, in section \ref{sec:data:desi}. 
The photometric sample we calibrate is the S19A HSC-SSP PDR3 Shape Catalog (HSC) \cite{HSCShapeCataloguePDR3Li2022}, described in \ref{sec:data:hsc}. The three catalogs' footprints are displayed in Figure \ref{fig:footprint}.

\subsection{DESI Large Scale Structure Catalog}
\label{sec:data:desi}

For both DESI DR1 and DR2, calibrations are performed with the Large Scale Structure (LSS) catalogs of the DESI survey \cite{DESIConstructionLSS2024}. The LSS catalogs provide weighted clustering and random catalogs tailored to the specifics of the DESI survey for each of the target classes used. 
The LSS catalogs take into account imaging systematics, selection and footprint effects, possible redshift failures, tiling overlap, as well as other effects -- the full description of which can be found in the DESI data model\footnote{\href{https://desidatamodel.readthedocs.io/en/latest/DESI_ROOT/survey/catalogs/RELEASE/LSS/SPECPROD/LSScats/VERSION/data_clustering.html}{\textcolor{blue}{desidatamodel.readthedocs.io/LSSCats}}} and in in refs \cite{DESIConstructionLSS2024, DESI.2024.II.SampleDef}. 

\begin{figure}
    \centering
    \includegraphics[width=0.95\textwidth]{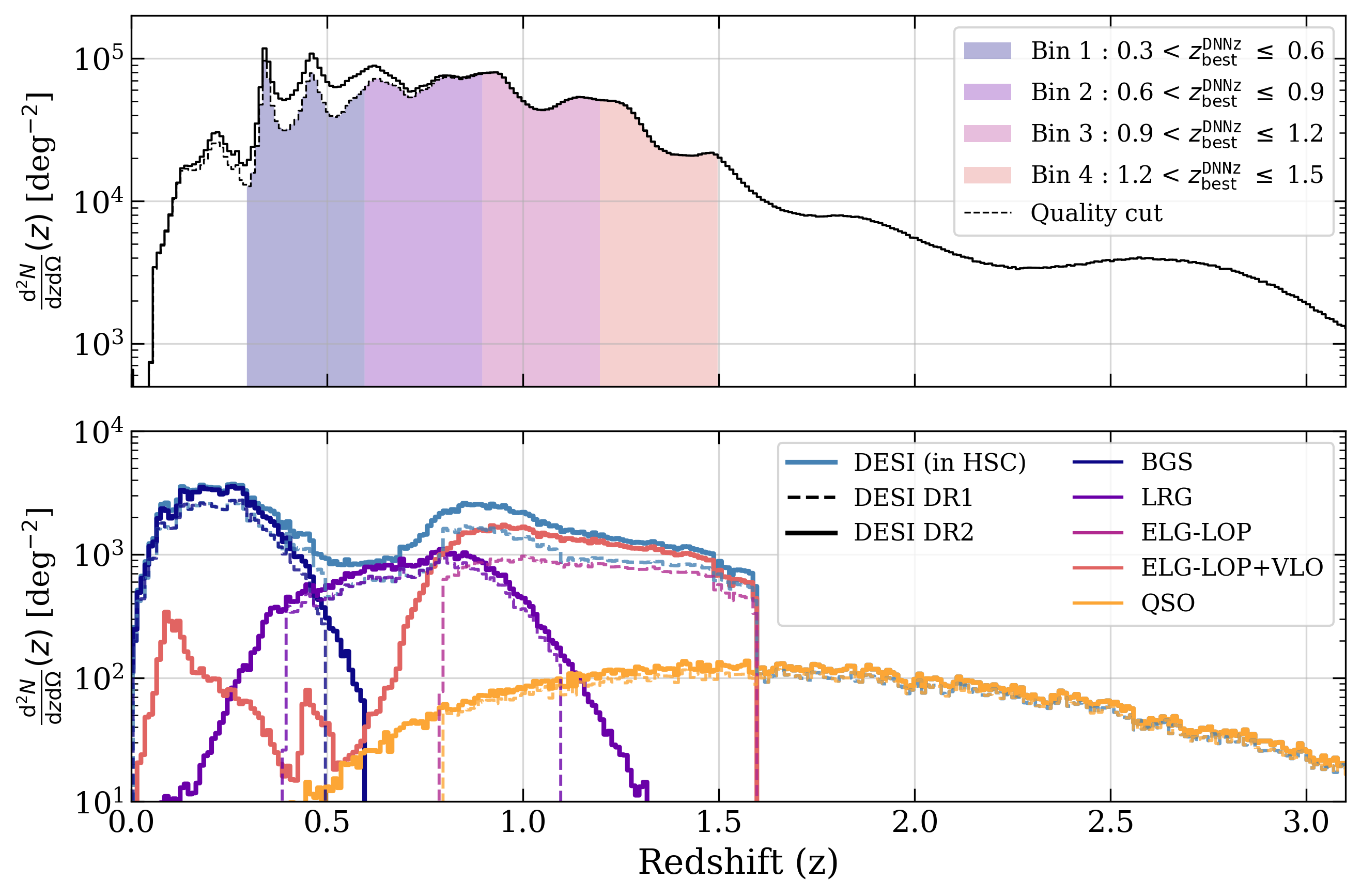}
    \caption{\textit{Upper panel:} N(z) angular density per redshift from the HSC catalog, using photometric redshifts from the \texttt{DNNz} algorithm \cite{DNNZHSCNishizawa2020}. The effect of the calibration cut on HSC is displayed, removing problematic sources up to $z_{\mathrm{best}}^{\mathtt{DNNz}}\leq0.9$ (dotted histogram), as well as HSC's four tomographic bin selections (shaded colors, including the calibration cut). \textit{Lower panel:} N(z) angular density per redshift of the DESI Large Scale Structure clustering catalogs \cite{DESIConstructionLSS2024}, showcasing the combined tracers distribution (light blue) in the HSC footprint and individual tracers (BGS, ELG, LRG and QSOs). DR1 is represented in dotted lines and DR2 is represented in full lines. In this figure, ELGs are distinguished between ELG-LOP (DR1 analysis) and ELG-LOP+VLO (DR2 analysis). Since most of the HSC footprint is already included in DR1, adding DR2 does not provide an improvement as big as one could expect, though there still are density differences as showcased in Figure \ref{fig:footprint}. Per tomographic bin, the effective number density counts of photometric sources are 3.77, 5.07, 4.00 and 2.12 arcmin$^{-2}$ \cite{PSFModellingHSCZhang2023} after applying the calibration cut.}
    \label{fig:density_z}
\end{figure}

This work uses multiple target classes: the BGS \cite{DESI.Selection.BGS.Hahn2023}, the LRGs \cite{DESI.Selection.LRG.Zhou2023}, the ELGs \cite{DESI.Selection.ELG.Raichoor2023} and the QSOs \cite{DESI.Selection.QSO.Chaussidon2023} as described above. 
The target classes' redshift ranges used in this study are described in Table \ref{tab:desi_z_bounds}, per data release, and the redshift density distribution for both data releases over the HSC footprint is showcased in Figure \ref{fig:density_z}. 
DESI uses priority ranking to perform fiber assignment on targets \cite{FiberSystem.Poppett.2024}. For the BGS sample, this work uses the \texttt{BGS\_ANY} sample, which includes both the \texttt{BGS\_BRIGHT} and \texttt{BGS\_FAINT} samples, and hereafter referred to as BGS. More details can be found in the BGS selection description, under section 3 of \cite{DESI.Selection.BGS.Hahn2023}.
For the ELG sample, we decide to not use DR1 ELGs in this work, as their very low completeness interferes with the auto-correlation measurements and provides only little information for the two first tomographic bins due to low redshift overlap. In DR2, as shown by Figure \ref{fig:footprint}, the local density is much higher and we can now measure auto-correlation of ELGs in this context.
For the ELG sample, this analysis uses the LOw-Priority (\texttt{LOP}) sample \cite{DESI.Selection.ELG.Raichoor2023} preferentially selects high redshift ELGs at $1.1\lesssim z\lesssim1.6$ . 
In DR2, in addition to \texttt{LOP}, we use the Very LOw priority (\texttt{VLO}) sample for ELGs, which focuses on targets in $0.6\lesssim z\lesssim1.1$. Indeed, the inclusion of \texttt{VLO} to the LSS catalogs is a new feature of DR2 compared to DR1, as displayed in figure \ref{fig:density_z}.

\begin{table}[h]
\centering
\caption{Redshift ranges as inclusive bin edges used for each data release per tracer.}
\begin{tabular}{c|c|c}
\hline
\textbf{Tracer} & \textbf{DR1} & \textbf{DR2} \\
\hline
BGS & $0-0.5$ & $0-0.6$ \\
LRG & $0.4-1.1$ & $0.3-1.2$ \\
ELG (LOP+VLO) & - & $0.7-1.6$ \\
QSO & $0.8-2.8$ & $0.7-2.8$ \\
\end{tabular}
\label{tab:desi_z_bounds}
\end{table}

Note that for DR2, some sources are expected outside of the bounds described in Table \ref{tab:desi_z_bounds} and shown in Figure \ref{fig:density_z}.
Their lower number densities and negligible overlap with the two highest redshift tomographic bins of HSC motivates our choice for redshift cuts. 
For example, information from low redshift ELGs ($\lesssim0.7$) is not included.

\subsection{HSC-SSP PDR3 Shape Catalog}
\label{sec:data:hsc}

The weak lensing catalog calibrated in this study is the S19A galaxy shape catalog \cite{HSCShapeCataloguePDR3Li2022} provided by the Public Data Release 3 (PDR3) \cite{HSC2022datarelease3} of the Hyper Suprime-Cam Subaru Strategic Program. This catalog builds upon $i_{\mathrm{band}}$ imaging data obtained from 2014 to 2019 in the Wide fields of the HSC survey, at a mean $i_{\mathrm{band}}$ seeing of $0.59''$ \cite{HSCShapeCataloguePDR3Li2022}. A conservative $i_{\mathrm{band}}<24.5$ magnitude cut is applied to the dataset. The catalog itself is split into 6 fields named \texttt{HECTOMAP}, \texttt{XMM}, \texttt{VVDS}, \texttt{GAMA09H}, \texttt{GAMA15H} and \texttt{WIDE12H}. 

As a preliminary step, a \textit{calibration cut} is applied to the first two tomographic bins of the HSC shape catalog. 
It removes $\sim31\%$ of sources in the first bin, and $\sim8\%$ of sources in the second bin. The cut, shown in equation \ref{eq:calib_cut}, is not applied on the third and fourth bin. This selection is performed in order to remove possible outliers in the individual probability density functions from the photo-$z$ algorithm reporting multi-modal distributions with a second, smaller peak around $z\sim3$. The cut depends on the $95\%$ confidence interval boundaries of the \texttt{DNNz} \cite{DNNZHSCNishizawa2020} and \texttt{Mizuki} \cite{MizukiHSCTanaka2015} photometric redshift algorithms, such that:

\begin{equation}\label{eq:calib_cut}
(z^{\texttt{DNNz}}_{95,\,\mathrm{max}}-z^{\texttt{DNNz}}_{95,\,\mathrm{min}})<2.7\;\;\&\;\;(z^{\texttt{Mizuki}}_{95,\,\mathrm{max}}-z^{\texttt{Mizuki}}_{95,\,\mathrm{min}})<2.7
\end{equation}

\noindent Moreover, for the rest of this analysis, when using redshifts outside of the HSC tomographic bins, the calibration cut is applied to all sources with redshifts between $z\sim0$ and $z\sim0.3$. This removes about $\sim16\%$ of the sources within that redshift range. Redshifts outside the fiducial tomographic bins are used to evaluate galaxy bias evolution for HSC, in section \ref{sec:modeling:photoz_gal_bias}. The fiducial redshift value kept to determine if a redshift is included in a bin or not is $z_{\mathrm{best}}^{\texttt{DNNz}}$. The $_\mathrm{best}$ subscript notes the redshift point estimate that minimizes the risk function, $R(z_{\mathrm{p}})=\int P(z)L(\frac{z_{\mathrm{p}}-z}{1+z})\mathrm{d}z$, based on the individual probability density function $P(z)$ obtained through $\texttt{DNNz}$, and with $L(x)$ a loss function. More details can be found in the following works \cite{HSCPHotozTanaka2017, DNNZHSCNishizawa2020}. 

\section{Modeling}
\label{sec:modeling}

This section describes the modeling choices and assumptions used to model the sample redshift distributions in the tomographic bins. We first present the estimator used to infer the photometric sample redshift distribution $n_p(z)$ for each bin, then discuss further corrections to potential systematic effects affecting this measurement, such as correcting for magnification effects and photometric sample galaxy evolution. One can express the observed projected galaxy density contrast at position $\bm{\hat{\theta}}$ as the contribution of clustering and magnification \cite{DES2021Krause, EuclidClusterRedshifts2025}:

\begin{equation}
    \delta_{x}^{\mathrm{obs}}(\bm{\hat{\theta}})=\delta_{x}^{\mathrm{g}}(\bm{\hat{\theta}})+
    \delta_{x}^{\mu}(\bm{\hat{\theta}})
\end{equation}

\noindent where $\delta_{x}^{\mathrm{g}}$ is the line of sight projection of the three dimensional galaxy contrast for galaxy sample $x$ and $\delta_{x}^{\mu}(\bm{\hat{\theta}})$ is the magnification term. Specifically:

\begin{equation}
    \delta_{x}^{\mathrm{g}}(\bm{\hat{\theta}})=\int n_x(z)\delta_{x}^{\mathrm{g}}(z,\,\bm{\hat{\theta}})\mathrm{d}z
\end{equation}

\noindent where $\delta_{x}^{\mathrm{g}}(z,\,\bm{\hat{\theta}})$ is the 3D galaxy contrast and $n_x(z)$ is the mean distribution of galaxies in the considered sample $x$. Under the assumption that the universe is isotropic, $\theta$ will now be treated as a scalar variable. When it comes to measuring the angular two-point correlations, one can write \cite{EuclidClusterRedshifts2025}:

\begin{equation}
    \omega_{x,y}(\theta)=\omega^{\mathrm{g} \times \mathrm{g}}_{x,y}(\theta)
    +\omega^{\mu \times \mathrm{g}}_{x,y}(\theta)+\omega^{\mathrm{g} \times \mu}_{x,y}(\theta)
\end{equation}

\noindent where $\omega^{\mathrm{A} \times \mathrm{B}}_{x,y}=\langle \delta_x^{\mathrm{A}}\delta_y^{\mathrm{B}}\rangle(\theta)$ with $\mathrm{A}, \mathrm{B}$ the respective contributions for sample $x$ and $y$.
The dominant contribution to the angular two-point correlation is the galaxy contrast ($\mathrm{g} \times \mathrm{g}$). Corrections related to magnification effects ($\mu \times \mathrm{g}, \mathrm{g} \times \mu$) are presented in section \ref{sec:modeling:magnification}. Higher order terms, such as $\mu\times\mu$, are not considered due to being subdominant with respect to any contribution including the galaxy density contribution $\mathrm{g}$. Redshift Space Distortions (RSD) are also neglected in our analysis: such contributions were shown to be negligible for wide projections and small scales used here \cite{2010PhRvD..81f3531B}. 

In most clustering redshift analyses, at least one of the samples is considered in sufficiently narrow redshift bins such that the distribution of each of the bins can be approximated as an indicator function of the redshift interval. Our fiducial bin size for the spectroscopic sample is $\dz_{\mathrm{s}}=0.05$, which is sufficient to capture the shape of distributions as large as the HSC tomographic bins \cite{EuclidClusterRedshifts2025}. Furthermore, as explored by \cite{EuclidClusterRedshifts2025}, large bin width conserves validity of the Limber equation and does not need for further corrections on border effects. Validity of such slicing is explored and $\Delta z_s=0.05$ is found to be adapted even for narrower tomographic bins than the ones used in the present work. It is also useful to express the angular two point correlation functions as a function of projected comoving transverse distance $r_p$, where $r_p=\theta\chi(z)$ with the flat-sky approximation and at small angular scales. This is the transverse comoving distance at redshift $z$ for a $\theta$ angular separation, and is equal to the comoving distance if the curvature of the universe $\Omega_k$ is null. 

\begin{figure}
    \centering
    \includegraphics[width=0.95\textwidth]{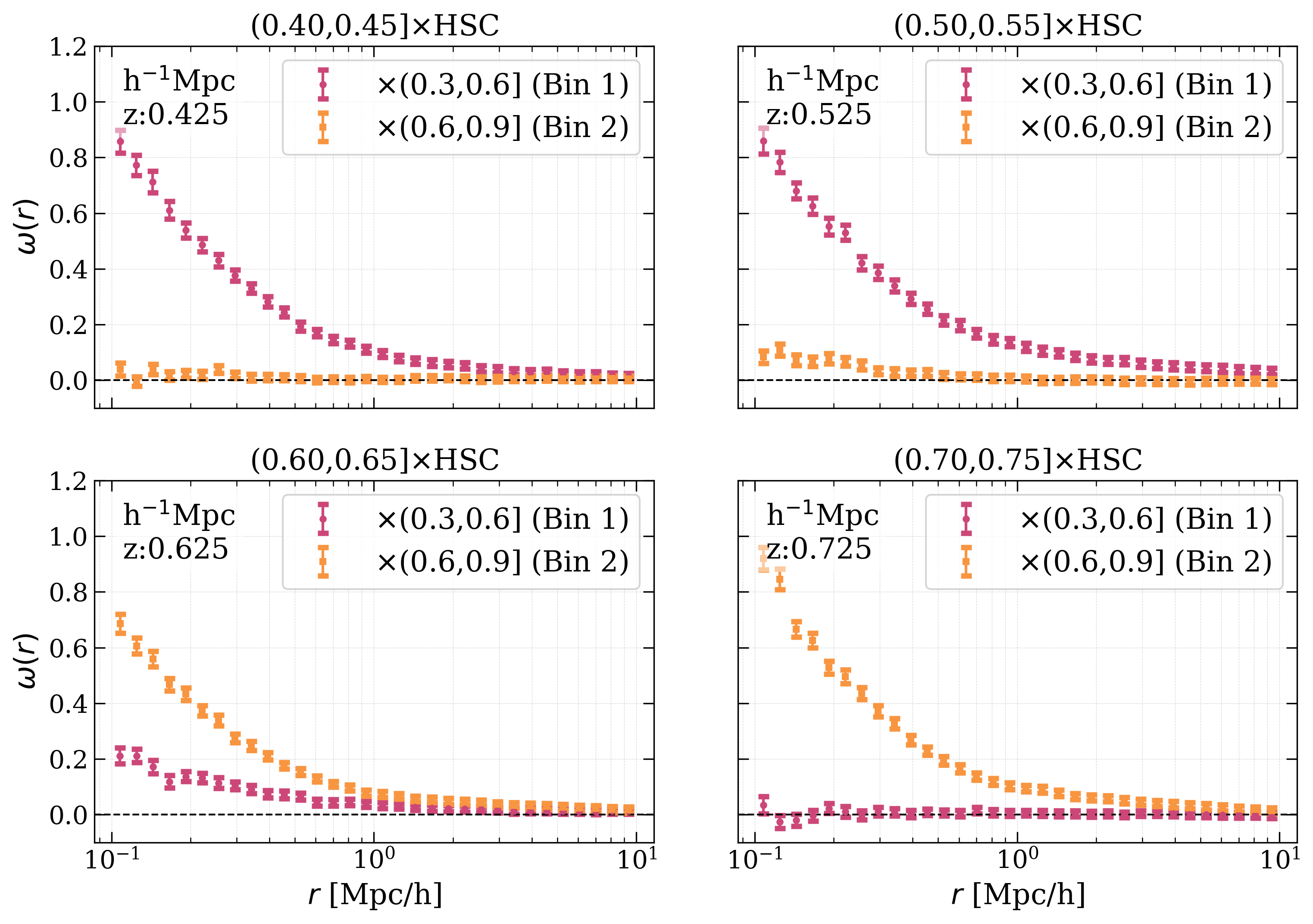}
    \caption{Example cross-correlation measurements for small spectroscopic slices with HSC bins. Here, the spectroscopic sources are DR1 LRGs, cross-correlated with the first two HSC tomographic bins. Comoving distance is computed at the redshift highlighted in the top left corner of each plot. As the redshift increases, one can see the weaker integrated signal for cross correlations with the first tomographic bin (purple), since the $n(z)$ decreases when far away from the photo-$z$ boundaries, and the stronger signal for the second tomographic bin as the redshifts reaches $z\sim0.6-0.9$ (orange).}
    \label{fig:cross-corr}
\end{figure}

For this analysis, the same cosmological parameters as those used by HSC's own clustering redshifts analysis \cite{HSCClusteringRau2023} are used: $\Omega_{\mathrm{DM}}=0.258868$, $\Omega_{\mathrm{b}}=0.048252$, $h=0.6777$, $n_s=0.95$ and $\sigma_8=0.8$\label{cosmology}. For the rest of this work, angular separation will use the projected comoving distance $r_p$ instead of $\theta$ within this cosmology, where the redshift used to compute $r_p$ is the center of the spectroscopic bin used, noted $z_j$ for a bin interval $[z_j-\frac{\dz}{2},z_j+\frac{\dz}{2})$\footnote{For auto-correlations on the photometric dataset used in section \ref{sec:modeling:photoz_gal_bias}, the distances are computed with the mid-point of the tomographic bin in mind.}, with $j$ the index of the bin. We will denote with $\omega_{x,y}(r_p,\,z)$ the projected
correlation function at $z$ over the 
projection width $\dz$. 

\subsection{Integrating over small-scales}
\label{sec:modeling:integration}

Following similar notations to \cite{EuclidClusterRedshifts2025, EuclidClusteringRedshifts2023}, subscript $s$ is used for the spectroscopic target class(es) and $p$ for the photometric sample, replacing the generic $x,y$ used previously (with $x=y$ for auto-correlations). $s$ can be any DESI tracer, such as BGS, LRG, ELG or QSO, or a combination of tracers. The photometric sample from HSC corresponding to the tomographic bin $i$ is noted $p_i$, though in the remainder of this work $p$ is often used to describe any tomographic bin.
The target class selected on that redshift bin is noted $s_j$, and $n_{s_j}(z)=n_s(z_j)$ are used interchangeably. The midpoint $z_j$ is the fiducial redshift for that bin, useful for example to compute comoving scales. Finally, $\bo_{xy}$ is the scale-averaged angular cross correlation vector, such that (in the case of cross-correlation between $s_j$ and $p_i$):

\begin{equation}
    \bo_{s_j, \,p_i}(z_j)=\int_{r_{\mathrm{min}}}^{r_{\mathrm{max}}}W(r)\omega_{s_j,\,p_i}(r,\,z_j)\mathrm{d}r
\end{equation}

\noindent with $W(r)\propto r^\beta$ the scale-dependent weighting function. In \texttt{pycorr}\footnote{\textcolor{blue}{\href{https://github.com/cosmodesi/pycorr}{\texttt{github.com/cosmodesi/pycorr}}}}, two-point correlation functions are expressed in terms of the fiducial angular separation $\theta$: the present work expresses bin edges defined with the transverse comoving distance $r_p$. For each narrow spectroscopic bin, $r_p$ distances are computed with \texttt{astropy} \cite{astropy2023paper3} at the mid-point redshift $z_j$. This sets $z_j$ as a parameter for $\bo$. The weighting function $W(r)=r^\beta/\int_{r_{\mathrm{min}}}^{r_{\mathrm{max}}}r^\beta \mathrm{d}r$ is normalized to unity over integration scales, and $\beta=-1$ is chosen following \cite{ménard2014clusteringbasedredshiftestimationmethod, Schmidt2013ClusteringRedshifts}. This weighting function up-weights the smaller scales, where higher clustering signal is expected. Scale cut choices to apply to our correlation functions are represented by $r_{min}$ and $r_{max}$. In literature, other weighting schemes have been explored, such as weighting directly the numerator and denominator of the two-point estimator (described in section \ref{sec:2pcf}), and integrating independently the numerator and denominator \cite{KidsCrossCorrvdB2020,Gatti2018DESY1ClusteringRedshifts,Schmidt2013ClusteringRedshifts}. Alternative weighting schemes are not explored in this work. The fiducial scale cuts of this analysis are $0.3-3\hMpc$: these cuts are more conservative than the cuts on the first clustering redshifts calibration \cite{HSCClusteringRau2023}, $0.1-1\mathrm{Mpc}$. In section \ref{sec:results}, we explore using an even more conservative scale cut ($1-5\hMpc$), and compare the results and error bars obtained. 
 The effective scales of the two bins correspond 
to about 0.6$\hMpc$ and 2$\hMpc$, 
respectively. One can relate auto and cross-correlations with:

\begin{equation}\label{eq:cross}
     \omega_{x,y}^{\mathrm{g}\times\mathrm{g}}(r,\,z)\approx r_{x,y}(r,\,z)\sqrt{\omega_{x,x}^{\mathrm{g}\times\mathrm{g}}(r,z)\omega_{y,y}^{\mathrm{g}\times\mathrm{g}}(r,z)}
\end{equation}

\noindent Here, $r_{x,y}(r,\, z)$ is the cross-correlation coefficient between the two galaxy samples, which is assumed to remain constant across redshift within a tomographic bin, but not necessarily unity since one can always re-normalize the redshift distribution later. This assumption is weaker than that of linear galaxy bias, which we do not assume, but is still expected to fail on very small scales. Previous work \cite{2010PhRvD..81f3531B,CrossCorrSingh2019} has shown that the cross-correlation coefficient for projected density correlations, with projection lengths larger than $\dz \sim 0.02$, is quite close to unity even down to $1\hMpc$. Since the signal to noise is the largest on small scales, the goal is to find an optimal compromise between the high signal to noise (SNR) obtainable at the very small angular clustering scales, and the breakdown of this assumption. We will show that the results from two scale cuts are very consistent with each other, suggesting that the assumption of the cross-correlation coefficient not evolving with redshift is likely to be valid down to $0.5\hMpc$.
Example cross-correlation measurements between small spectroscopic bins from the DR1 LRGs and the two first tomographic bins of HSC are showcased in figure \ref{fig:cross-corr}.

\subsection{Building the n(z) model}\label{sec:modeling:nz}

For a single spectroscopic redshift bin (here, of size $\dz_\mathrm{s}=0.05$), the $n(z)$ distribution is assumed to be an indicator function of the redshift bin ($n_x(z)=\mathbbm{1}_{[z_j-\dz/2,\,z_j+\dz/2)}(z)/\dz$). 
$n(z)$ may be varying across the bin, 
and may be noisy, which we 
will address later with spline fits to the binned $n(z)$.  
To isolate the $n(z)$ contribution of the
photometric galaxies $p$, the cross-correlation vector is performed over the complete population $p$ (here, of a tomographic bin), while the auto-correlation is performed over a redshift slice $p_j$:

\begin{equation}\label{eq:cross_binned} 
 \bo_{s_jp}^{\mathrm{g}\times \mathrm{g}}(z_j)\approx n_p(z_j)\dz\bo_{s_jp_j}^{\mathrm{g}\times \mathrm{g}}(z_j)
\end{equation}

\noindent Binning by true redshift is not possible due to photometric redshift uncertainties for $p_j$ bins. However, 
for sufficiently narrow photoz bins one can investigate the evolution of photometric clustering across the 
tomographic bin: this is addressed in section \ref{sec:modeling:photoz_gal_bias}, where a proxy expression for $\bo_{p_jp_j}$ is developed. Assuming the above expression \ref{eq:cross_binned}, the $n_p(z)$ distribution can be written as:

\begin{equation}\label{eq:nz}
    \mathrm{A}n_p(z_j)\approx\frac{\bo_{s_jp}^{\mathrm{g}\times \mathrm{g}}(z_j)}{\dz \bar{r}_{s_jp_j}\sqrt{\bo_{s_js_j}^{\mathrm{g}\times \mathrm{g}}(z_j)\bo_{p_jp_j}^{\mathrm{g}\times \mathrm{g}}(z_j)}}
\end{equation}

\noindent Here $\mathrm{A}$ is a normalization constant, which we will determine such that $n(z)$ integrates to unity. $\bar{r}_{s_jp_j}$ is the average cross-correlation coefficient across the scales of our analysis. 
In this context, equation \ref{eq:nz} does not make use of the linear bias approximation. Instead, we measure the evolution of clustering for both spectroscopic and photometric samples (the latter in narrow photometric bins of a given tomographic bin), and we assume that the cross-correlation coefficient is not evolving with redshift. This assumption is almost certainly valid for our conservative choice where the pivot point is around 2$\hMpc$, and is likely to be valid for our standard choice where the pivot point is $0.6\hMpc$.
In the following subsections, further systematic corrections to this expression are described. Section \ref{sec:modeling:photoz_gal_bias} discusses corrections to $\bo_{p_jp_j}$ in order to approximate the photometric sample evolution over redshift, since one cannot obtain accurate $p_j$ bins over photometric redshifts due to photo-$z$ error. In Section \ref{sec:modeling:magnification}, second-order terms from magnification effects to the observed cross-correlation vector $\bo_{sp}$ are considered.

\subsection{Galaxy clustering evolution for the photometric sample}\label{sec:modeling:photoz_gal_bias}

In practice, the true redshifts of the photometric redshift distribution are not known. The photometric galaxy sample evolution has to be taken into account in the $n(z)$ model with the $\bo_{pp}$ term, but the scatter and potential outliers in each of the redshift bins does not allow us to accurately measure the auto-correlation of the photometric sample in small redshift bins $\bo_{p_kp_k}$. Here, $p_k$ is a smaller tomographic bin of width $\dz_{\mathrm{p}}=0.1$ indexed by $k$: the binning is performed on the photometric redshifts. In this work, $p_k$'s are spread linearly from $z_{\mathrm{p}}\sim0$ to $z_{\mathrm{p}}\sim1.6$: probing the galaxy bias further remains difficult due to the low number densities of QSO in higher redshift bins, without help from other tracers.

If the auto-correlation of the photometric tracer using the photometric redshift binning is noted $\bo_{pp}^{\mathrm{photo}-z}$, galaxy bias can be recovered by measuring $\bo_{pp}^{\mathrm{true}}$, that is, measured as if the sample was binned using true redshifts\footnote{Here, these notations are similar terminology to work done on the Dark Energy Survey clustering redshifts analyses \cite{Cawthon2022DESY3Boss}}. The correction employed in this study is close to previous works \cite{Cawthon2022DESY3Boss, EuclidClusterRedshifts2025,KidsCrossCorrvdB2020}. 
Similarly to \cite{Cawthon2022DESY3Boss}, the assumption relies on $\bo_{p_kp_k}$ not significantly evolving in redshift over a single $p_k$ bins. One can further encapsulate nonlinearities by including a correction for non-linear dark matter evolution as well, as in \cite{EuclidClusterRedshifts2025}: we opt to not include this further correction as nonlinear bias and dark matter evolution tend to be competing effects, and discuss this choice in appendix \ref{sec:appendix_pz}.
A correction to $\bo_{pp}^{\mathrm{photo}-z}$ is obtained by measuring $n_{p_k}(z)$, where $n_{p_k}(z)$ is the clustering redshifts measurement of $n(z)$ over the $p_k$ tomographic bin and is obtained with equation \ref{eq:npkz}:
\begin{equation}\label{eq:npkz}
    n_{p_k}(z_j)\propto\frac{\bo_{s,p_k}(z_j)}{\sqrt{\bo_{ss}(z_j)}}
\end{equation} 

\noindent The above expression assumes constant $\bo_{pp}$ over the smaller tomographic bin, only correcting for the spectroscopic sample evolution with auto-correlations in order to obtain the true redshift distribution, with $\dz_{\mathrm{p}}=0.1$ as the small photometric redshift bin interval ranging $z\sim0$ to $z\sim1.6$. While $\dz_{\mathrm{p}}=0.1$ is not a small redshift interval for such an assumption, \cite{EuclidClusterRedshifts2025} explores the recovered bias expression differences with different bin sizes and finds similar corrections for $\dz=0.1$ and $\dz=0.02$. We assess the validity of such an approximation in appendix \ref{sec:appendix_pz}.

\begin{figure}[h!]
    \centering
    \begin{minipage}[t]{0.35\textwidth}
        \vspace{0pt} 
        \includegraphics[width=\textwidth]{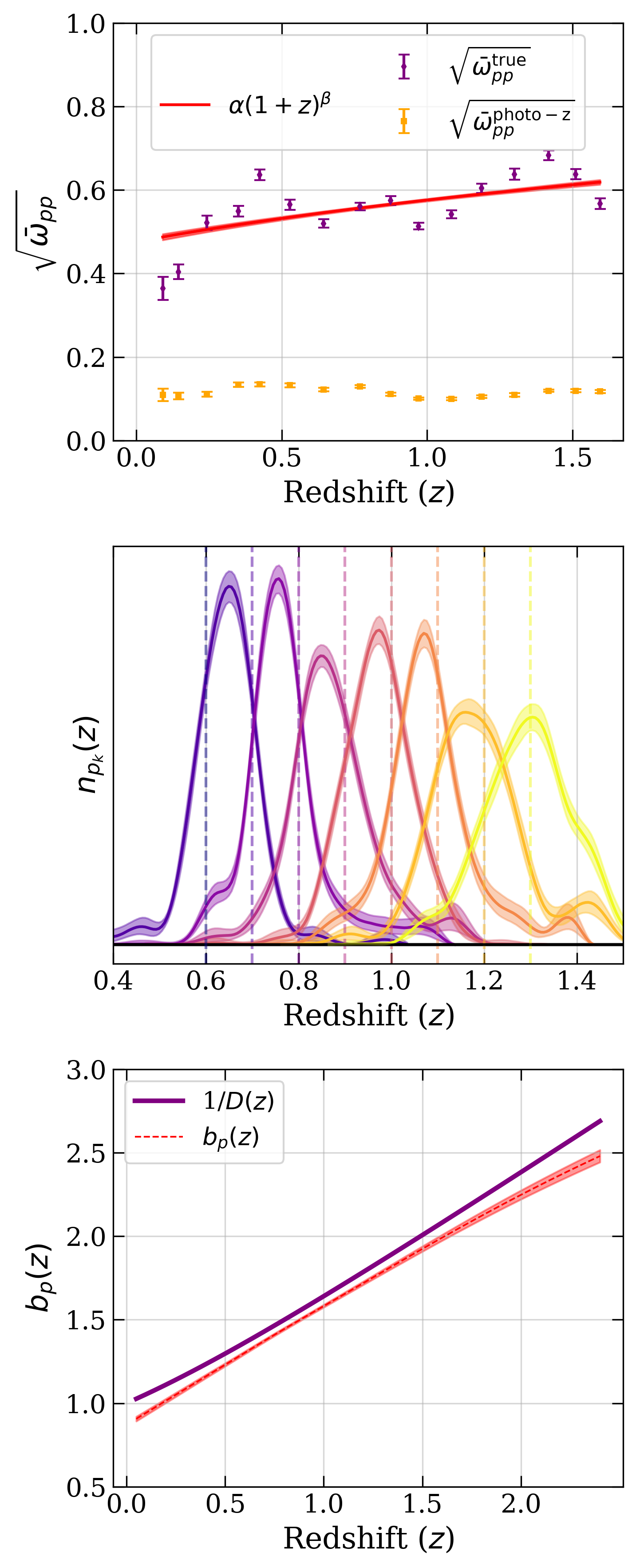}
    \end{minipage}%
    \hfill
    \begin{minipage}[t]{0.58\textwidth}
        \vspace{0pt}
        \captionof{figure}{
            \textit{Top panel}: $\sqrt{\bo_{pp}^{\mathrm{true}}}$ and $\sqrt{\bo_{pp}^{\mathrm{photo}-z}}$ with the power law fit through the corrected measurements. This showcases the importance of the correction, as the $\sqrt{\bo_{pp}^{\mathrm{photo}-z}}$ alone does not capture significant galaxy bias evolution, whereas the correction captures a general evolution trend.
            \\ \vspace{1.3em} \\
            \noindent\textit{Middle panel}: Clustering redshifts measurements for a few of the intermediate small tomographic bins ($n_{p_k}(z)$ distributions), highlighting the tomographic bin ranges with the dotted lines. The $n_{p_k}(z)$ are modeled with splines, described further in section \ref{sec:parametrization:bspline}. Nonetheless, the photometric redshift spread and the gradual expectation shift can be seen increasing with redshift, compared to the midpoint of the bins (between the dotted lines).
           \\ \vspace{0.1em} \\
            \textit{Bottom panel}: Comparing $b_p(z)$ computed with the corrections to $\bo_{pp}$ and a "passive evolution" galaxy sample where the linear bias follows $b_p(z)\propto1/D(z)$ with $D(z)$ the growth factor of the universe. Here, $b_p(z)$ is assumed linear in the context of a comparison to the factor of growth, but not used further in $n(z)$ computations.
            \label{fig:photoz_gal_bias}
        }
    \end{minipage}
\end{figure}

Since $\bo_{pp}$ is assumed to be constant within these bins, the measured distributions are re-normalized and only implement the correction to the spectroscopic bias with $\bo_{ss}$, as shown in equation \ref{eq:npkz}. $n_{p_k}(z)$ is measured with finer spectroscopic redshift bins ($\dz_{\mathrm{s}}=0.025$) as finer tomographic bins need better redshift resolution in order to accurately recover the distributions. Dark matter evolution is also taken into account in the correction, following \cite{EuclidClusterRedshifts2025}: one obtains the following correction to the two-point correlation functions: 

\begin{equation}\label{eq:photoz_corr_correction}
    \bo_{pp}^{\mathrm{true}}(z_k)\approx\bo_{pp}^{\mathrm{photo}-z}(z_k)\frac{\dz_{\mathrm{p}}^{-2}\int^{z_k+\dz_{\mathrm{p}}/2}_{z_k-\dz_{\mathrm{p}}/2}\bo_{mm}(z)\mathrm{d}z}{\int_0^{\infty}n_{p_k}^2(z)\bo_{mm}(z)\mathrm{d}z}
\end{equation}
\noindent Here, $\bo_{mm}(z)$ is the dark matter angular auto-correlation at redshift $z$ over a $\Delta z$ slice around $z$. Thus, equation \ref{eq:photoz_corr_correction} allows to approximate the correction to $\bo_{pp}^{\mathrm{photo}-z}$, therefore obtaining a proxy for the auto-correlation evolution in redshift of the photometric sample. The galaxy bias evolution of the photometric sample $b_p(z)$ expression presented in the bottom plot of Figure \ref{fig:photoz_gal_bias} makes use of the Limber approximation \cite{Limber1953Approximation} and the linear bias approximation ($\bo_{pp}^{\mathrm{true}}(z)\propto\bo_{mm}b^2_p(z)$). While the linear bias approximation does not necessarily hold at very small scales, it is important to note that the quantity $\bo_{pp}^{\mathrm{true}}$ does not depend on the linear bias approximation and therefore can be used in Equation \ref{eq:nz} without loss of generality.

The numerator of the correction corresponds to the influence of the spectroscopic redshift distribution over the tomographic bin, assumed to be flat. The denominator corrects for the photo-$z$ spread due to the chosen algorithm's inaccuracies, as the $n_{p_k}(z)$ distribution cannot be assumed flat over the tomographic bin used. The square root of these corrected auto-correlations is fit to a power-law model (see Figure \ref{fig:photoz_gal_bias}), as done in works from DES\footnote{This analysis includes a free amplitude parameter $\alpha$ in the fit. Though normalization allows one to drop the amplitude term, the error is still included in error propagation.} \cite{Cawthon2022DESY3Boss, Gatti2018DESY1ClusteringRedshifts}, and the fitting errors are propagated to the $p(z)$ measurements.

\begin{equation}\label{eq:photoz_bias_model}
\sqrt{\bo_{pp}^{\mathrm{true}}}\approx \alpha(1+z)^{\beta}
\end{equation}


\noindent The fit gives $\alpha=0.476\pm 0.008$ and $\beta=0.276\pm{0.024}$ for the chosen scale cut $0.3-3\,\hMpc$. Values are fit on the expectations of the small tomographic bins. In figure \ref{fig:photoz_gal_bias}, the measurements of galaxy bias and the power law approximation are shown in the top panel, as well as the clustering redshifts measurements for a few of the intermediate tomographic bins used to compute the corrections and characterize the local photo-$z$ spread in the middle panel. In the bottom panel, the bias evolution of the photometric sample $b_p(z)$ is computed using the dark matter auto-correlation and the linear bias approximation, though this expression is not used in computations of $n(z)$ and is solely derived for comparison purposes. The obtained bias $b_p(z)$ is then compared to a passive evolution type sample that would follow $b(z)\propto 1/D(z)$ where $D(z)$ is the (cosmology-dependent) normalized factor of growth. With this correction, the HSC sample bias evolves slightly faster than the factor of growth but remains comparable. This observed behavior is consistent with measurements of the large-scale galaxy bias of HSC Year 1 \cite{Nicola2020TomoclusteringHSCY1}, where proportionality $b(z)\propto1/D(z)$ is observed for flux-limited samples.

\subsection{Magnification}\label{sec:modeling:magnification}

In this subsection, the impact of second order terms ($\mathrm{g}\times\mu$ and $\mu\times \mathrm{g}$) due to magnification effects are described and modeled. Magnification arises via weak lensing effect, which can magnify or de-magnify galaxy flux, depending on the slope of the luminosity function. This (de)magnification is caused by matter distribution along the line of sight, which is correlated with the galaxy distribution at lower redshifts. In turn this introduces non-zero correlation between high redshift galaxies and low redshift galaxies. The total observed flux of individual galaxies increase caused by magnification affects both the selection function of galaxies \cite{ElvinPoole2023MagnificationModelling} and the photometric redshift algorithm, since both rely on the observed properties of the sample. These biases are dominant when cross-correlating sources that are far apart in redshift. Magnification effects for auto-correlations are neglected. Magnification bias effects are described with the log-number count slope, characterizing the effects on the observed projected number density:

\begin{equation}
    s_\mu=\frac{\mathrm{dlog_{10}}\mathrm{N}(>m)}{\mathrm{d}m}
\end{equation}

\noindent where $m$ is the limiting magnitude of the sample, and $\mathrm{N}$ is the cumulative number count as a function of limiting magnitude. In practice, $s_\mu$ is a partial derivative, since $\mathrm{N}(>m)$ depends on other selection effects such as redshift binning depending on the sample \cite{Nicola2020TomoclusteringHSCY1}. DESI tracers often have complex selection functions \cite{DESI.Selection.BGS.Hahn2023, DESI.Selection.LRG.Zhou2023, DESI.Selection.ELG.Raichoor2023, DESI.Selection.QSO.Chaussidon2023} and therefore cannot be characterized by a simple magnitude cut. A common technique to probe the magnification bias in that case, used for example when computing the LRG magnification bias \cite{DESI.LRGCross.Magnifcation.Zhou2023}, is to shift the magnitudes in all bands by a small amount $\delta m$ and measure the response on $s_\mu$ after reapplying target selection, going cut by cut. The magnification bias can then be written as $\alpha_x=2.5s_\mu-1$\cite{Gatti2021ClusteringDESWLsrcDistrib, Gatti2018DESY1ClusteringRedshifts} \footnote{Other parametrizations are common, such as $\alpha=2.5s_\mu$ \cite{DESI.LRGCross.Magnifcation.Zhou2023, MagnificationBiasComplexSelections, ChoiCFHTLensAngularCross2016}, or $\left.\alpha=\frac{\mathrm{dln}(f_1)}{\mathrm{d}m}\right|_m$ \cite{heydenreich2025lensingbordersmeasurementsgalaxygalaxy}, where $f_1$ is the fraction of galaxies passing the cut.}, where $x$ denotes a galaxy sample, either spectroscopic or photometric. 

The formalism developed in works like \cite{DES2021Krause} and used in \cite{Gatti2018DESY1systematics,Cawthon2022DESY3Boss,Gatti2021ClusteringDESWLsrcDistrib,EuclidClusterRedshifts2025} is adopted to model the magnification correction on the cross-correlation vectors. The correction is theoretically obtained using the following:

\begin{equation}
    W^\mu_x(\chi) = \frac{3 \Omega_m H_0^2}{2 c^2} \int_{\chi}^{\infty} n_x(\chi')\frac{\chi}{a(\chi)}\frac{\chi'-\chi}{\chi'}\mathrm{d}\chi'\quad\text{and}\quad W^\mathrm{g}_x(\chi)=n_x(z)\frac{\mathrm{d}z}{\mathrm{d}\chi}(\chi)
\end{equation}

\noindent where $W^\mu_x(\chi)$ is the tomographic lens efficiency of sample $x$ and $W^\mathrm{g}_x(\chi)$ is the selection function of galaxies in the redshift bin. The expression of the angular cross power-spectrum between fields A, B with the Limber approximation \cite{Limber1953Approximation} gives:

\begin{equation}
    C_{s_jp_i}^{\mathrm{AB}}(\ell) = \int \frac{W^\mathrm{A}_{s_j}(\chi) W^\mathrm{B}_{p_i}(\chi)}{\chi^2}P_{\mathrm{AB}}\left(k = \frac{\ell + 0.5}{\chi},\,z(\chi)\right)\mathrm{d}\chi
\end{equation}

\noindent and the angular power spectra can be converted into angular correlation functions through a Legendre transformation. Here, $P_{\mathrm{AB}}$ is the corresponding 3D power spectrum model for fields A,B. More in-depth demonstrations can be found in works \cite{Cawthon2022DESY3Boss, EuclidClusterRedshifts2025}. The following expression per tomographic bin $p$ and target class $s$, as demonstrated in Appendix B of \cite{EuclidClusterRedshifts2025}, is obtained:

\begin{equation}
    \bar{M}(z_j) = \alpha_s(z_j) \sum_{i<j} b_p(z_i)n_p(z_i) \mathcal{D}(z_i, z_j) 
+ b_s(z_j) \sum_{i>j} \alpha_p(z_i)n_p(z_i)\mathcal{D}(z_j, z_i)
\end{equation}
with $\mathcal{D}(z_x,z_y)$ defined as:
\begin{equation}
    \mathcal{D}(z_x,z_y):=\frac{3H_0^2\Omega_m}{c}\frac{\bo_m(z_j)\dz}{H(z_x)a(z_x)}\frac{\chi(z_y)-\chi(z_x)}{\chi(z_y)}\chi(z_x)
\end{equation}

\noindent In the above expression, $H$ is the Hubble parameter, $H_0=H(z=0)$, $\chi$ is the comoving radial distance, $c$ is the speed of light and $a(z)=1/(1+z)$ is the scale factor of the universe. Our expression differs slightly from \cite{Cawthon2022DESY3Boss} (like \cite{EuclidClusterRedshifts2025}, redshift evolution of $\alpha_s,\alpha_p,b_s,b_p$ is included, though for $\alpha_p$ it will be neglected) and is closer to \cite{EuclidClusterRedshifts2025}, the only difference being that the expression is averaged over comoving scales, removing dependencies on angular variables. The non-linear 2D angular matter power spectrum $\bo_{\mathrm{mm}}(z_j)$ (on a $\Delta z$ redshift slice around $z_j$) is obtained using \texttt{Core Cosmology Library (CCL)}\footnote{\href{https://github.com/LSSTDESC/CCL}{\textcolor{blue}{\texttt{github.com/LSSTDESC/CCL}}}} \cite{CCLChisari2019} with the same fiducial cosmology as previously described and used in HSC's own clustering redshift analysis \cite{HSCClusteringRau2023}, reminded earlier in section \ref{cosmology}. The matter power spectrum is predicted and computed with models from \texttt{halofit} \cite{HalofitTakahashi2012} and \texttt{CLASS}\footnote{\href{https://github.com/lesgourg/class_public}{\textcolor{blue}{\texttt{github.com/lesgourg/class\_public}}}} \cite{CLASSBlas2011}, and any computation of the matter power spectrum in this work uses these predictions. The real space contribution is then computed with the Legendre series transform. Because magnification corrections on the auto correlations are neglected, the magnification correction is a linear system to solve $\mathcal{M}\mathbf{X}=\mathbf{n}$ where $\mathbf{n}$ is the measurement vector, $\mathbf{X}=(n_p(z_j))_{j\in\llbracket1,N_{\mathrm{b}}\rrbracket}$ are the magnification-corrected measurements, $N_\mathrm{b}$ is the number of bins and $\mathcal{M}$ is a "magnification matrix" such that:

\begin{equation}
    \mathcal{M}=(m_{kl})_{k,l\in\llbracket1,N_{\mathrm{b}}\rrbracket^2}=
    \begin{cases}
\frac{1}{\dz\sqrt{\bo_{s_ks_k}\bo_{p_kp_k}}}\alpha_s(z_k)b_p(z_l)\mathcal{D}(z_l,z_k) & \text{if } k < l \text{ ($\mathrm{g}\times\mu $)}\\
1 & \text{if } k = l \text{ ($\mathrm{g}\times \mathrm{g} $)}\\
\frac{1}{\dz\sqrt{\bo_{s_ks_k}\bo_{p_kp_k}}}\alpha_p(z_l)b_s(z_k)\mathcal{D}(z_k,z_l) & \text{if } l > k \text{ ($\mu\times \mathrm{g} $)}
    \end{cases}\label{eq:mag_matrix}
\end{equation}

\noindent The measurement vector $\mathbf{n}$ is:
\begin{equation}
    \mathbf{n}=\left(\frac{\bo_{s_jp_i}}{\dz\sqrt{\bo_{p_jp_j}\bo_{s_js_j}}}\right)_{j\in\llbracket1,N_{\mathrm{b}}\rrbracket}
\end{equation}
In parentheses of equation \ref{eq:mag_matrix}, contributions from different effects in the matrix are highlighted. The dominant term is by far the diagonal term, corresponding to the galaxy density contribution $\mathrm{g}\times \mathrm{g}$.

The galaxy bias expressions can be obtained through auto-correlations, as detailed in section \ref{sec:modeling:nz}. In this context however, we probe auto-correlations at larger scales, where linear bias is a valid approximation, as detailed in \ref{sec:modeling:magnification:bias:gal}. For the photometric tracer, the power law approximation described in section \ref{sec:modeling:photoz_gal_bias} is adopted to recover the bias expression, after computing the dark matter auto-correlation. Therefore, the matrix coefficients become:

\begin{equation}
    m_{kl}=
    \begin{cases}
\frac{b_p(z_l)\alpha_s(z_k)}{b_p(z_k)b_s(z_k)\bo_{mm}(z_k)}\mathcal{D}(z_l,z_k) & \text{if } k < l \\
1 & \text{if } k = l \\
\frac{\alpha_p(z_l)}{b_p(z_k)\bo_{mm}(z_k)}\mathcal{D}(z_k,z_l) & \text{if } k > l
    \end{cases}
\end{equation}
The magnification corrected measurements are obtained with
$\mathbf{X}=(n_p(z_j))_{j\in\llbracket1,N_{\mathrm{b}}\rrbracket}$ and the error bars are propagated assuming: 
\begin{equation}
    \mathcal{M}(X+\delta X)=(\mathbf{n}+\delta\mathbf{n})\Rightarrow\delta X=\mathcal{M}^{-1}\delta\mathbf{n}
\end{equation} 

\noindent Magnification bias and galaxy bias measurements are not marginalized over further, given the amplitude of the correction with respect to measurements. In the data products of this work, distributions and measurements with and without this correction are included.

\label{sec:modeling:magnification:bias}

The next paragraphs describe the galaxy bias and magnification bias measurements used in the context of this analysis for our magnification correction.

\paragraph{Galaxy bias}\label{sec:modeling:magnification:bias:gal} 
For the spectroscopic sample, the galaxy bias used to correct for magnification effects is measured using auto-correlation measurements on the spectroscopic tracer. To model the galaxy bias of any spectroscopic tracer, the parametrization introduced by \cite{Laurent2017QSOBias}, and also used by \cite{Chaussidon2025QSOLRGPNG} is implemented:

\begin{equation}
    b_s(z)=c_s(1+z)^2+d_s
\end{equation}

\noindent The linear galaxy bias is obtained with the monopole of the auto-correlation functions, with the technique described in \cite{Chaussidon2025QSOLRGPNG}. The effect of Kaiser redshift space distortions \cite{KaiserRSD} is accounted for with \texttt{desilike}\footnote{\href{https://desilike.readthedocs.io/en/latest/}{\textcolor{blue}{desilike.readthedocs.io}}. Specifically, we use \texttt{desilike.theories.galaxy\_clustering.full\_shape}, and obtain the linear bias with the \texttt{KaiserTracerCorrelationFunctionMultipoles} class.}. Measurements are fit to each of the target classes\footnote{The measurement of the galaxy bias for BGS is performed on the \texttt{BGS\_BRIGHT-21.35} sample. More details about this selection can be found in the DESI DR2 Baryon Acoustic Oscillation (BAO) work \cite{DESI.DR2.II.BAO}. The linear galaxy bias difference between the two samples (\texttt{BGS\_ANY} and \texttt{BGS\_BRIGHT-21.35)} is assumed to be negligible in the context of the magnification correction.} between scales of $30$ to $65\hMpc$ with the DESI fiducial cosmology (described in, for example, \cite{DESI.2024.III.FiducialCosmology.BAO}). While one could use the existing auto-correlation functions to measure linear galaxy bias in the magnification correction, a more robust bias measurement that overcomes small-scale nonlinearities and projection effects is obtained by measuring the 3D auto-correlation on larger scales. Nonetheless, the galaxy bias measurements obtained with the small scale angular auto-correlations are consistent with the larger scale measurements. The fitted parameters are given in the following Table \ref{tab:galaxy_bias}. 

{\renewcommand{\arraystretch}{1.2}
\begin{table}[h]
\centering
\caption{Galaxy bias model coefficients fitted to the DR2 two-point auto-correlation functions, within the DESI fiducial cosmology presented in \cite{DESI.2024.III.FiducialCosmology.BAO} and reduced $\chi^2$ statistic of the fit. The coefficients are assumed to be the same for DR1.}
\begin{tabular}{c|c|c|c}
\hline
\textbf{Tracer} & $c_s$ & $d_s$ & $\chi^2_\mathrm{r}$\\
\hline
BGS & $0.606\pm{0.073}$ & $0.524\pm{0.129}$ & $0.59$ \\
LRG & $0.236\pm{0.019}$ & $1.346\pm{0.058}$ & $1.39$\\
ELG & $0.155\pm{0.012}$ & $0.595\pm{0.057}$ & $1.53$ \\ 
QSO & $0.252\pm{0.008}$ & $0.710\pm{0.058}$ & $1.62$ \\
\end{tabular}
\label{tab:galaxy_bias}
\end{table}
}

For the photometric sample galaxy bias, the fitted power-law described in \ref{sec:modeling:photoz_gal_bias} as correction to $\sqrt{\bo_{pp}^{\mathrm{photo}-z}}$ can be used to recover the expression of $b_p(z)$ at all redshifts using the power-law fit. In this specific case of galaxy bias for the magnification correction, we assume linear galaxy bias for the photometric sample.

\paragraph{Magnification bias}
\label{sec:modeling:magnification:bias:mag}
The number count slope $s_\mu$ and magnification bias $\alpha_x$ were introduced in section earlier \ref{sec:modeling:magnification}. For HSC, the magnitude cut $i_{\mathrm{band}}-a_i<24.5$ is considered to be the dominant selection function in HSC, where $i_{\mathrm{band}}$ is the HSC i-band \texttt{cModel} magnitude\footnote{Details on the HSC \texttt{cModel} magnitudes can be found in sub-subsection 4.9.9 of \cite{BoschHSCpipelineY1}.}, and $a_i$ is the absorption for $i$ band. Magnification slope coefficients for HSC are assumed constant per tomographic bin. Tomographic bins are affected by two significant selections that could alter the magnification slope: the calibration cut, detailed in section \ref{sec:data:hsc} for the first two tomographic bins, and the selection based on binning with photometric redshifts. In a HSC Y1 analysis performed in \cite{Nicola2020TomoclusteringHSCY1}, only the magnitude selection was considered, hence the same assumption is kept. Following the approach described in that work to estimate $s_\mu$, a fourth order polynomial is fit to the cumulative number counts. The polynomial is then derived, log-scaled and evaluated at the limiting magnitude to obtain $s_\mu$. Measurements can be found in the following table \ref{tab:mag_bias:hsc}.

\begin{table}[h]
\centering
\caption{Magnification bias for each tomographic bin. Redshift evolution within tomographic bins is neglected.}
\begin{tabular}{c|c}
\hline
\textbf{Tomographic bin} & $s_\mu$ \\
\hline
Bin 1 ($z:0.3-0.6$) & 0.002 \\
Bin 2 ($z:0.6-0.9$) & 0.065\\
Bin 3 ($z:0.9-1.2$) & 0.142 \\
Bin 4 ($z:1.2-1.5$) & 0.206 \\
\end{tabular}
\label{tab:mag_bias:hsc}
\end{table}

In the case of the DESI dataset, magnification measurements are obtained from literature \cite{Sailer2025EvolutionStructureMagnificationBiasBGS, DESI.LRGCross.Magnifcation.Zhou2023, debelsunce2025QSOmagCMBLensing} as well as from measurements performed with the same methodology as \cite{heydenreich2025lensingbordersmeasurementsgalaxygalaxy} for ELGs. For ELGs, it is assumed that ELG-\texttt{LOP} and ELG-\texttt{LOP}+\texttt{VLO} have the same magnification bias (combined under the ELG tag), computed exclusively on the ELG-\texttt{LOP} target class\footnote{The largest bias from ELG magnification effects is on Bin 3's high redshift tail. This happens at the higher redshift range of ELGs ($z\gtrsim1.1$), where there are significantly less ELGs from ELG-\texttt{VLO}, as shown in Figure \ref{fig:density_z}.}. The magnification biases are compiled under table \ref{tab:mag_bias:desi}.

\begin{table}[h]
\centering
\caption{Magnification slope for each tracer and effective redshift. For LRGs, the provided $\bar{z}$ redshifts are the mid-point redshift of the photometric redshift bins used to compute the magnification coefficient. The effect of photo-$z$ bias or scatter is not considered further in the measurement.}
\begin{tabular}{c|c|c}
\hline
\textbf{Tracer} & $s_\mu$ & $\bar{z}$ \\
\hline
\multirow{2}{*}{BGS \cite{Sailer2025EvolutionStructureMagnificationBiasBGS}} & 0.81 & 0.211 \\
                    & 0.80 & 0.352 \\
\hline
\multirow{8}{*}{LRG \cite{DESI.LRGCross.Magnifcation.Zhou2023}} 
& 0.954 $\pm$ 0.027 & 0.44 \\
& 0.988 $\pm$ 0.025 & 0.51 \\
& 1.040 $\pm$ 0.021 & 0.59 \\
& 1.047 $\pm$ 0.018 & 0.67 \\
& 0.999 $\pm$ 0.021 & 0.75 \\
& 0.957 $\pm$ 0.017 & 0.82 \\
& 0.914 $\pm$ 0.018 & 0.89 \\
& 1.078 $\pm$ 0.020 & 0.97 \\
\hline 
\multirow{2}{*}{ELG} & 0.546 $\pm$ 0.005 & 0.95 \\
                    & 0.666 $\pm$ 0.005 & 1.35 \\
\hline
\multirow{3}{*}{QSO \cite{debelsunce2025QSOmagCMBLensing}} & 0.099 & 1.44 \\
                    & 0.185 & 2.27 \\
                    & 0.284 & 2.75 \\
\end{tabular}
\label{tab:mag_bias:desi}
\end{table}

Magnification biases are linearly interpolated with these measurements, following choices made by works in DES \cite{Gatti2021ClusteringDESWLsrcDistrib} (for example for the redMaGiC sample). In order to be accurate, magnification biases would require a single measurement per spectroscopic bin $j$, but given the amplitude of the correction with respect to the $\mathrm{g}\times \mathrm{g}$ contribution, interpolation is assumed to be negligible within this correction.

\section{Computing the two-point angular functions}\label{sec:2pcf}

Two point angular correlations are computed with the \texttt{pycorr} software, based on the \texttt{CORRFUNC}\footnote{\textcolor{blue}{\href{https://github.com/manodeep/Corrfunc}{\texttt{github.com/manodeep/Corrfunc}}}} \cite{Sinha2019corrfunc1, Sinha2019corrfunc2} engine. HSC and DESI's survey footprints are extracted using multi-order coverage maps (MOC) \cite{fernique2015moc} made with the \texttt{mocpy}\footnote{\textcolor{blue}{\href{https://github.com/cds-astro/mocpy}{\texttt{github.com/cds-astro/mocpy}}}} python package. The areas of interest for this study represent the intersection of DESI's MOC with HSC's MOC.
The computations of the two-point angular functions are split up into four patches of sky, corresponding to spatially close areas, in order to speed up computing time. Concretely, this corresponds to the \texttt{XMM} field, the \texttt{HECTOMAP} field, the \texttt{VVDS} field, and the joint field composed of \texttt{GAMA09H}, \texttt{GAMA15H} and \texttt{WIDE12H}. We use $N_\mathrm{b}=32$ logarithmically-spaced angular bins, with edges spanning $0.1$ to $10$ $\hMpc$ , in transverse comoving scales.

The DESI angular auto-correlations are obtained by combining the auto-correlation measured in the north galactic cap (NGC) and south galactic cap (SGC) of the dataset used. Two point correlation functions are computed independently per DESI tracer, and include standard weights and the FKP weights \cite{FKPWeights1994}. 
The weighting used is described in \cite{DESIConstructionLSS2024}: $w_\mathrm{tot}=w_\mathrm{comp}w_\mathrm{sys}w_\mathrm{zfail}w_\mathrm{FKP}$. Here, $w_\mathrm{comp}$ is used for completeness, $w_\mathrm{sys}$ accounts for target density fluctuations due to imaging conditions and $w_\mathrm{zfail}$ encompasses the relative redshift success rate. Given the redshift ranges used in this analysis, the FKP weights $w_{\mathrm{FKP}}$ are not strictly necessary, since their purpose is to describe how the number density changes with redshift, and the redshift bins are fine. It is important to note that fiber assignment incompleteness affects our measurements on these small scales, especially for low completeness tracers such as ELGs. 
This effect is all the more prominent for DR1 ELGs, which is why we decide to not use them in this analysis. 
In short, the number of targets who obtain a fiber on a single DESI tile while observing is different from the total number of possible targets in that target class within the tile. 
To mitigate this effect, one could use Pairwise Inverse Probability (PIP, \cite{BianchiPIP2017}) weights when computing two point correlation functions, but this is only doable when performing auto-correlations.
Since cross-correlations are performed with the HSC catalog, PIP weighting is not available, though it is noteworthy that most of HSC's footprint is already within DESI areas with high or maximal completeness, as shown in figure \ref{fig:footprint}. 
This kind of clustering redshift study will benefit from subsequent data releases of DESI, alleviating further this issue. 
To verify this introduces no apparent bias in the measurements, two scale cuts are included in this work: $0.3-3\hMpc$ and $1-5\hMpc$. 
The fiber collisions become an issue below $\theta_\mathrm{fib}=0.05^\circ$ \cite{bianchi2025characterizationdesifiberassignment, Pinon2025FiberAssign}. 
The second scale cut allows for no leakage of fiber collisions in the scale cut up to $z\sim0.45$, with minor contamination for larger redshifts. 
The measurement consistency shown in section \ref{sec:results} supports that no bias is introduces by using measurements at smaller scales.

With the HSC catalog, the lensing weights are included in our two-point estimators, following the same choices as \cite{HSCClusteringRau2023}. 
Jackknife estimates of the covariance matrix are computed to estimate error bars, including the correction given by \cite{Mohammad2022jackknife}, which is the fiducial approach implemented in \texttt{pycorr}. 
Per patch of the sky, $N_{\mathrm{JK}}=100$ "delete-one" jackknife realizations are performed: in each iteration, a random area of sky in the footprint is removed and the two-point angular statistic is computed with that area excluded. 
Removed areas are determined through a KMeans sub-sampler algorithm splitting up the footprint into $N_{\mathrm{JK}}$ areas at \texttt{HEALPix} \cite{Gorski2005HEALPix} nside resolution of $256$ on each of the patches of sky. The covariance matrix obtained on the associated angular two-point correlation functions can be written as such \cite{Mohammad2022jackknife}:

\begin{equation}
    \mathbf{C}_{ij}=\frac{N_{\mathrm{JK}}-1}{N_{\mathrm{JK}}}\sum^{N_{\mathrm{JK}}}_{n=1}(\omega^{(n)}_i-\langle\omega_i\rangle)(\omega^{(n)}_j-\langle\omega_j\rangle) 
\end{equation}

\noindent where $({N_{\mathrm{JK}}-1})/{N_{\mathrm{JK}}}$ is the fraction of the footprint's area used to compute any jackknife realization, $\omega^{(n)}_i$ is the measurement in angular bin $i\in\llbracket 1,N_b \rrbracket$ over the footprint excluding area $n$ and $\langle\omega_i\rangle$ is the mean value over all measurements in bin $i$ from excluded regions. The Davis-Peebles \cite{DavisPeebles1983} estimator is chosen to compute the two-point angular correlation functions:

\begin{equation}
    \hat{\omega}_{xy}(r):=\frac{\mathrm{D}_x\mathrm{D}_y(r)}{\mathrm{R}_x\mathrm{D}_y(r)}-1
\end{equation}

\noindent where $\mathrm{D}_x\mathrm{D}_y$ are normalized data-data pair counts and $\mathrm{D}_x\mathrm{R}_y$ are the normalized data-random pair counts. Although this estimator is not as low bias and variance as the broadly used Landy-Szalay \cite{LandySzalay1993} estimator:

\begin{equation}
    \hat{\omega}^{\mathrm{LS}}_{xy}(r):=\frac{\mathrm{D}_x\mathrm{D}_y(r)-\mathrm{R}_x\mathrm{D}_y(r)-\mathrm{D}_x\mathrm{R}_y(r)}{\mathrm{R}_x\mathrm{R}_y(r)}+1, 
\end{equation}

\noindent the Davis-Peebles estimator allows this study to not rely on the random catalog of the photometric sample, and only rely on the random catalog for our spectroscopic tracers. A random catalog for HSC is still used when computing the auto-correlations on smaller tomographic bins $\bo_{pp}^{\mathrm{photo}-z}$, in the context of photometric galaxy bias mitigation (see section \ref{sec:modeling:photoz_gal_bias}). This random catalog has no redshift assignments, so it is less robust than the DESI catalog, but captures the footprint sufficiently well given our measurement. \footnote{More details about the random points used can be found on the HSC data release 3 portal: \href{https://hsc-release.mtk.nao.ac.jp/doc/index.php/random-points__pdr3/}{\textcolor{blue}{hsc-release.mtk.nao.ac.jp/random-points\_\_pdr3/}}.} Conveniently, the Davis-Peebles estimator is also faster to compute. At least 10 times more randoms from the DESI catalogs than data points are used when computing the two-point correlation functions\footnote{This corresponds to four catalogs of randoms from the Large Scale Structure catalogs of DESI \cite{DESIConstructionLSS2024}. Each catalog corresponds to a number density of $2500$ deg$^{-2}$, so the random catalog number density is $10000$ deg$^{-2}$ before redshift binning.}, and for low-density datasets such as QSOs, this can reach up to 100 times more. 

\section{Parametrization methods}
\label{sec:parametrization}

This section presents our parametrization choices for the tomographic redshift distributions for each bin. The benefit of such modeling is to regularize and smooth out the $n(z)$ distributions, as some measurements statistically fluctuate to negative values. We compare some literature approaches, and introduce a new parametrization based on splines. The hierarchical Bayesian method proposed in \cite{Rau2022compositelikelihood, HSCClusteringRau2023} leveraging both the \textbf{PhotZ} (photometry calibration) and \textbf{WX} (clustering redshifts measurements) distributions is not implemented, due to inconsistencies between the two calibrations: we therefore rely only on the clustering redshift measurements. 

\subsection{Combining tracers}
\label{sec:results:combine_tracers}

Before parameterizing, once the $n(z)$ for each target class is measured and, if necessary, corrected for other systematic effects such as magnification or galaxy bias, only target classes with significant overlap with the bin signal are retained. As such, Bin 1 ($z\sim0.3-0.6$) uses measurements from the BGS, LRGs and ELG-LOP. Bin 2 ($z\sim0.6-0.9$) uses every target class available, including QSOs. In Bin 3 ($z\sim0.9-1.2$) and Bin 4 ($z\sim1.2-1.5$), BGS derived constraints are removed as they present no significant overlap in redshift. Still, cross-correlations between all tracers and the tomographic bins are computed to assert consistency will null measurements on the redshift ranges of excluded tracers\footnote{When performing clustering redshifts measurements on smaller tomographic bins of width $\dz_{\mathrm{p}}$ centered on $z_k$ while correcting for sample galaxy bias (see \ref{sec:modeling:photoz_gal_bias}), we use the available tracers in bins of centers included in $\left[z_k-3\dz_{\mathrm{p}},\,z_k+3\dz_{\mathrm{p}}\right]$ per tomographic bin $k$. Bin $z\sim0.-0.025$ is not taken into account due to very low densities (as reported by Figure \ref{fig:density_z} for the HSC sample) and high variance biasing the spline model}. For each bin, after combining results, the measurements are restricted to the following ranges: in Bin 1, the model is retained between $z\sim0$ and $0.8$; in Bin 2, between $z\sim0.3$ and $1.3$; in Bin 3, between $z\sim0.3$ and $2.1$; and in Bin 4, between $z\sim0.7$ and $2.1$. Once the measurements for each necessary target class within a single tomographic bin are obtained, the target classes are combined into a single measurement using inverse variance weighting\footnote{When combining measurements if retaining only the cross-correlations (without correcting for auto-correlations), the measurements are combined on the two-point correlation counts themselves instead, due to normalization differences between target classes because of the amplitude of the auto-correlations. Hence, the samples are treated as a single "combined" target class.}:

\begin{equation}
   n(z_j)=\left(\sum\limits_{t\in\mathcal{T}_j}\frac{1}{\sigma_t^2}\right)^{-1}\sum\limits_{t\in\mathcal{T}_j}\frac{n_t(z_j)}{\sigma_t^2}\pm\sqrt{\left(\sum\limits_{t\in\mathcal{T}_j}\frac{1}{\sigma_t^2}\right)^{-1}}\quad\forall j\in\llbracket 1,\,N_{b}\rrbracket
\end{equation}

\noindent where $\mathcal{T}_j$ is the set of target classes used at redshift $z_j$, $n_t(z_j)$ is the measurement of $n(z)$ derived from target class $t$ at redshift $z_j$ with the associated error $\sigma_t$. Cross-covariance effects between tracers on overlapping spectroscopic redshifts are not taken into account. This ultimately affects our measurements with some over-confidence on multi-target redshift ranges (for example, between $z\sim0.7$ and $z\sim1.2$, ELGs, LRGs and QSOs overlap and are all used in the analysis): this is not modeled further in the context of this work and could be improved upon in future works. 

\subsection{Suppressed Gaussian Processes}
\label{sec:parametrization:sgp}

A recently developed approach introduced and implemented in Euclid methodological analyses for clustering redshifts \cite{EuclidClusteringRedshifts2023, EuclidClusterRedshifts2025} involves using gaussian processes that are then suppressed in order to model measurements in a non-parametric way. In short, the data is modeled using Gaussian Processes (GP) with a Matérn kernel, setting a slope of $\nu_{\mathrm{GP}}=1.5$ and a fixed length scale $l_{\mathrm{GP}}=\dz$. In this work, GPs are implemented using \texttt{scikit-learn}\footnote{\href{https://github.com/scikit-learn/scikit-learn}{\textcolor{blue}{\texttt{github.com/scikit-learn/scikit-learn}}}} \cite{ScikitLearnPedregosa2011} and the suppression function proposed in \cite{EuclidClusteringRedshifts2023, EuclidClusterRedshifts2025}, varying the damping factor. The suppression function implemented depends on $x$, the convolution of a gaussian window with the local Signal to Noise Ratio (SNR) of the measurements, and a dampening factor $k$. The suppression function is as follows:

\begin{equation}\label{eq:suppression_function}
S(x, k) =
\begin{cases}
0 & \text{if } x < 0 \\
1 - (1 - x)^k\; & \text{if } 0 < x < 1, \\
1 & \text{if } x > 1.
\end{cases}
\end{equation}

\noindent Such a suppression function tends to suppress real, tail like effects in large tomographic bins with important scatter. This is not as worrisome for experiments like LSST, expecting narrower tomographic bins, but poses issues with tomographic bins in other weak lensing surveys such as HSC (the concerning bin being Bin 3 with a long, high-redshift tail, for example) \cite{HSCClusteringRau2023} or DES (for example, Bin 4) \cite{Gatti2021ClusteringDESWLsrcDistrib} that are spread out. Fine-tuning the suppression function could be an option to explore in order to correctly use the suppressed gaussian processes model, but is not explored further in the scope of this paper.

\subsection{B-spline parametrization}
\label{sec:parametrization:bspline}

\begin{figure}[h]
    \centering
    \includegraphics[width=0.96\textwidth]{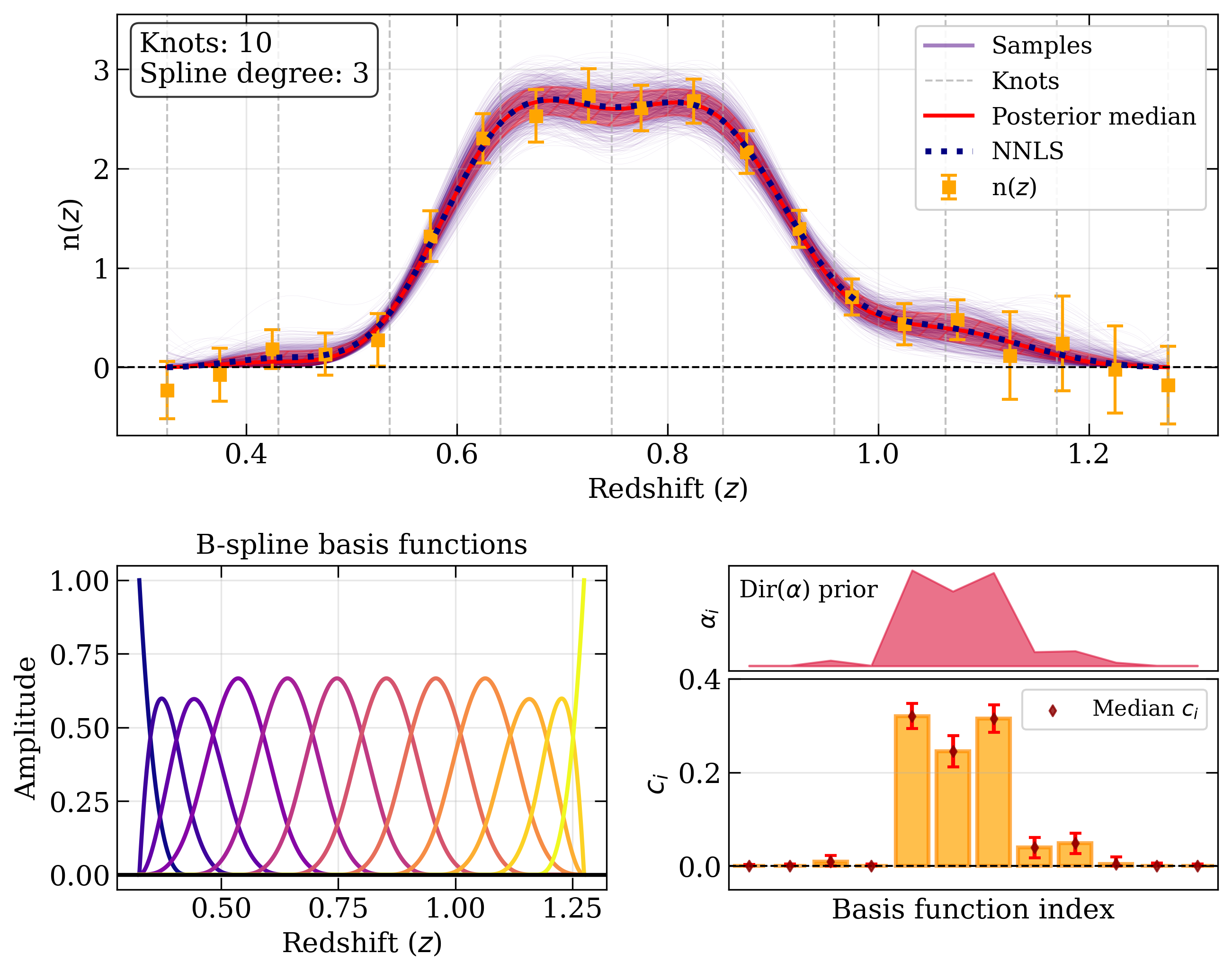}
    \caption{\textit{Top panel:} Bin 2 data (as presented in section \ref{sec:results} with all corrections) underlaid with the spline distribution (red, posterior median and $1\sigma$ contour) and some example samples from the spline modeling (indigo). Position of interior knots are shown in gray dotted lines. The NNLS result of equation \ref{eq:model_nz_bspline} is shown in dashed navy. \textit{Bottom left panel:} B-spline functions $(B_i)_{i\in\llbracket 1,n\rrbracket}$ over the redshift range. \textit{Bottom right panel:} In the bottom plot, we show the posterior median coefficient distribution over the samples with associated standard deviations, where $c_i$ coefficients are constrained to be positive. Above, the $\mathbf{Dir}(\alpha)$ prior is showcased, per expression \ref{eq:dir_prior}.} 
    \label{fig:spline_showcase}
\end{figure}

The $n(z)$ distributions are modeled with univariate basis splines (B-splines) over the redshift range. B-splines are families $(B_i)_{i\in\llbracket 1,n\rrbracket}$ of piece-wise polynomials of order $k\geq1$, with $p=k-1$ being the degree of the polynomials. These are defined by a sequence of knots $t_0 \leq t_1 \leq \dots \leq t_{m}$, with knots $t_0,\dots t_p$ and $t_{m-p},\dots, t_m$ known as external knots, usually repeated in order to support the first and last B-splines of the basis.
The other knots lie on the range one is interested in and are known as internal knots. Knots will now refer to the internal knots. B-splines are defined by recursion using the Cox-de Boor formula:

\begin{equation}
\begin{aligned}
\forall\;p\in\mathbb{N}\quad B_{i,p}(z):=
\begin{cases}
\displaystyle \mathbbm{1}_{[t_i,t_{i+1})}(z) \quad \text{if } p=0, \\
\frac{z - t_i}{t_{i+p} - t_i} \, B_{i,p-1}(z) 
+ \frac{t_{i+p+1} - z}{t_{i+p+1} - t_{i+1}} \, B_{i+1,p-1}(z). \quad \text{if } p > 1
\end{cases}
\end{aligned}
\end{equation}

\noindent where $\mathbbm{1}_{[t_i,t_{i+1})}$ is the indicator function of $[t_i,t_{i+1})$, null when $z\notin [t_i,t_{i+1}[$ or if $t_i=t_{i+1}$. B-splines are by construction non-negative, and an example basis for the distribution in Bin 2 is shown in the bottom left panel of figure \ref{fig:spline_showcase}.
For this model, one can use M-splines, where $B_{i,\,p}(z)$ are normalized to integrate to 1: such a change only affects the prior of the normalization factor $A$ described in the following model \ref{eq:model_nz_bspline}.
To obtain our $n(z)$ distributions with $\hat{n}(z)$, $\hat{n}(z)$ is parametrized as a function of our B-spline basis $(B_{i,\,p})_{i\in\llbracket 1,m\rrbracket}$, free coefficients $(c_i)_{i\in\llbracket 1,m\rrbracket}$, and a free amplitude parameter $A$ to account for re-normalization. Therefore:
\begin{equation}\label{eq:model_nz_bspline}
    \hat{n}(z)=A\sum\limits_{i=1}^{m}c_iB_{i,\,p}(z)
\end{equation}

\noindent The $c_i$ coefficients are subject to a Dirichlet prior \textbf{Dir}$(\bm{\alpha})$, which implies the $c_i$ coefficients lie on the standard $m-1$ simplex: that is, $\sum^m_{i=1}c_i=1$ and $c_i\in[0,1]$ for all $i\in\llbracket1,\,m\rrbracket$. 

In order to refine this prior, prior concentrations defined by the $\bm{\alpha}$ vector on the $c_i$'s are adapted to encourage these coefficients to use values closer to 0 where the measurements are consistent with $0$ (that is, where the cross-correlation signal is consistent with 0, modulo statistical fluctuations), which means little to no contribution from that spline function in the basis. Such refinement of the $\bm{\alpha}$ vector is guessed by solving Non-Negative Least Squares (NNLS) using equation \ref{eq:model_nz_bspline} against the $n(z)$ measurements (and fixing $A$ to 1), obtaining coefficients $c_i^{(\mathrm{NNLS})}$. When solving NNLS in this scenario, varying errors for each measurements are not taken into account, and results are only used as a guess for the prior. Two hyperparameters are defined to tune this prior: the base $a_0$ being the minimal prior concentration to ensure sampling of the Bayesian model and a boost factor $b$ that controls the local prior concentration. Therefore:

\begin{equation}\label{eq:dir_prior}
    \alpha_i=a_0+bc_i^{(\mathrm{NNLS)}} \quad \forall i\in\llbracket1,\,m\rrbracket
\end{equation}

\noindent This analysis chose $b=3$ and $a_0=0.05$. The prior \textbf{Dir}$(\bm{\alpha})$ set on Bin 2 of the tomographic bins is given as example in Figure \ref{fig:spline_showcase} in the bottom right panel. The knots are uniformly distributed on the redshift grid with one knot every two data points: their positions are shown in dashed gray in figure \ref{fig:spline_showcase}. The free amplitude coefficient is subject to a broad gaussian prior over the initial amplitude $A_0$ parameter computed with NNLS, with a standard deviation of $0.25A_0$. The spline models are obtained with Bayesian analysis using the software package \texttt{pymc}\footnote{\href{https://github.com/pymc-devs/pymc}{\textcolor{blue}{github.com/pymc-devs/pymc}}} \cite{Abril-Pla_PyMC_a_modern_2023}.

\section{Results}
\label{sec:results}

This section presents the calibrated $n(z)$ distributions and the expectations we measure on each of the tomographic Bins. The measurements of this work are put in context of the previous clustering redshift analysis \cite{HSCClusteringRau2023} on the HSC dataset, as well as the cosmic shear results \cite{Dalal2023CosmoShearHSC, CosmoShearLi2023HSCY3}, cosmic shear ratios (CSR) calibration \cite{Rana2025HSCY3CosmicShearRatios} and the combination of galaxy clustering with weak lensing (GC-WL) measurements \cite{Zhang2025RedshiftGCPointMass} of the photometric redshift shifts. 
Redshift expectation values are measured as the first moment of the posterior draws from the spline model described in section \ref{sec:parametrization} for each Bin. 
The resulting expectation value for each draw is split into 16$^{th}$, 50$^{th}$ and 84$^{th}$ percentiles in order to obtain error measurements and the center expectation value. 

\begin{figure}
    \centering
    \includegraphics[width=0.95\textwidth]{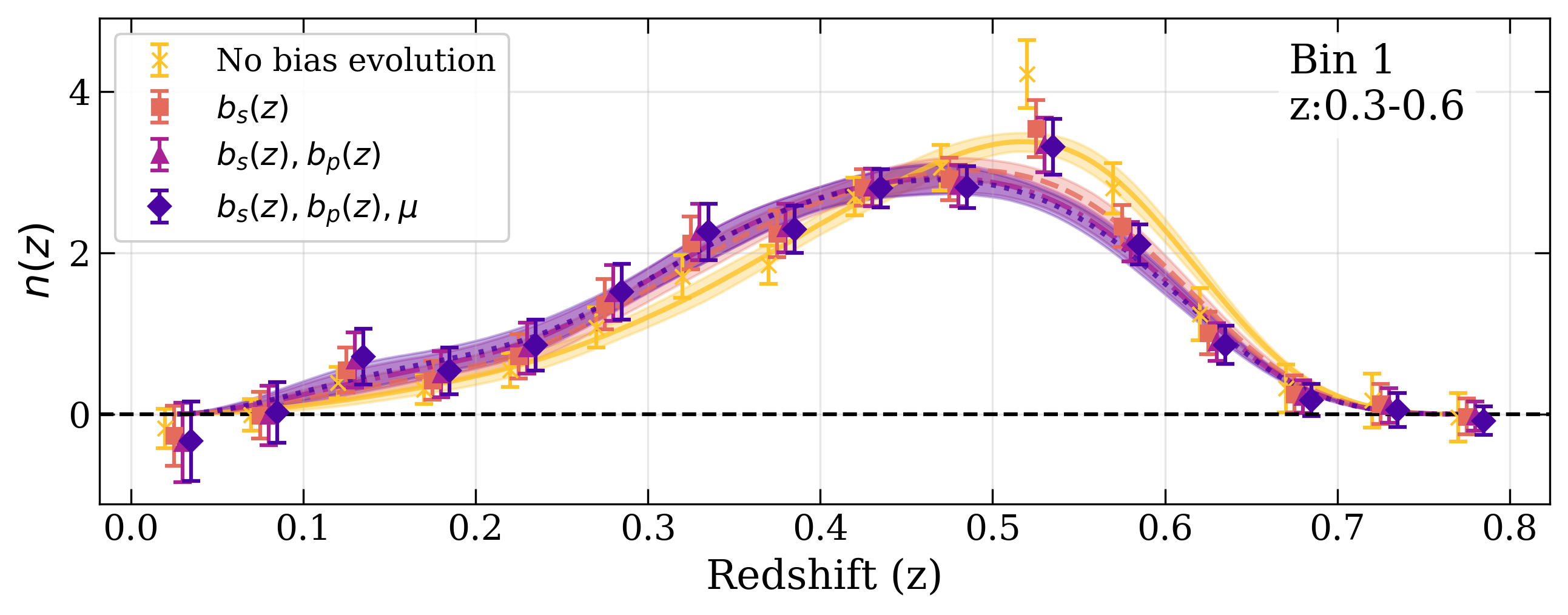}
    \includegraphics[width=0.95\textwidth]{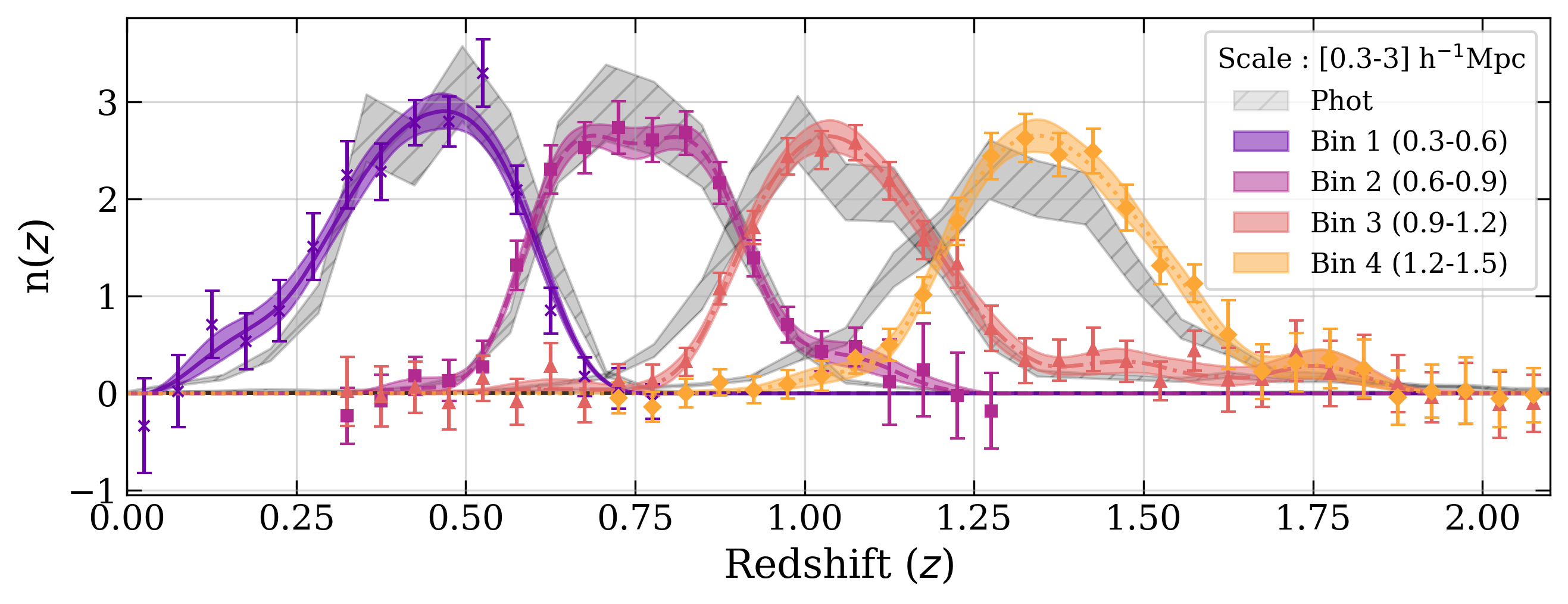}
    \caption{\textit{Upper panel}: Calibrated distributions on tomographic Bin 1 ($z\sim0.3-0.6$), showing the impact of the different corrections on the shape of the distribution. For each color, we include an additional correction. No bias evolution corresponds to only using cross-correlations with dark matter evolution. $b(z)$ corrections account for spectroscopic galaxy bias with auto-correlations ($b_s(z)$), and photometric galaxy bias ($b_p(z)$), as described in section \ref{sec:modeling:photoz_gal_bias}. Finally, the last correction are the magnification corrections $\mu$, described in section \ref{sec:modeling:magnification}. Measurements are offset for readability. We observe similar shifts with these corrections for all bins, as displayed in Table \ref{tab:shifts_results} and Figure \ref{fig:shifts}. \textit{Lower panel}: Model distributions on each of the 4 tomographic bins, with the clustering redshift measurements and the photometry calibrated distributions \textbf{PhotZ} (in hatched gray) replicated from Figure 8 of \cite{HSCClusteringRau2023}. Distributions are shown with all corrections included.}
    \label{fig:distributions}
\end{figure}

\begin{figure}
    \centering
    \includegraphics[width=0.98\textwidth]{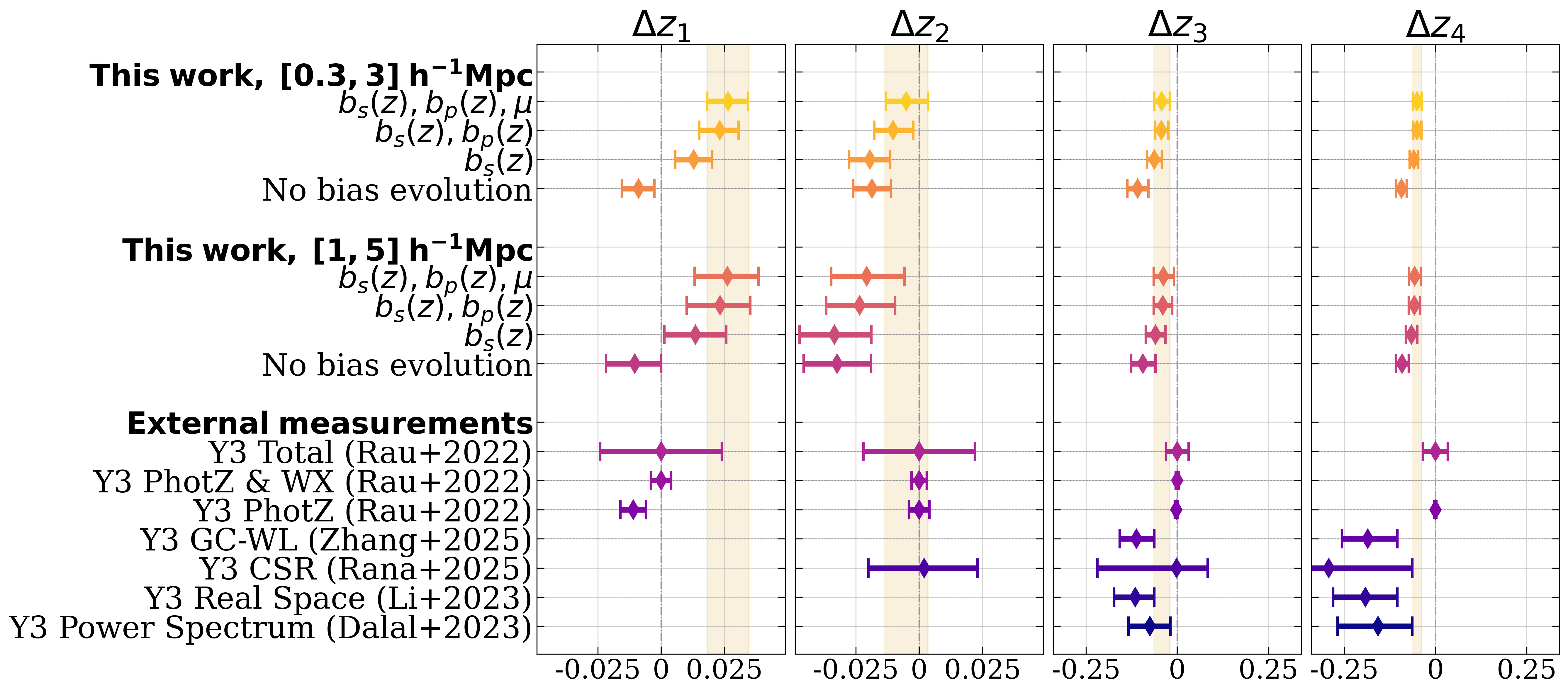}
    \caption{Shift values for each of the 4 tomographic bins in the catalog from posterior draws compared to Y3 Total, Y3 \textbf{PhotZ}, Y3 \textbf{PhotZ \& WX} described in \cite{HSCClusteringRau2023} and replicated in Table \ref{tab:shifts_results}.
    Note that the \textbf{PhotZ} errors do not include systematics, and that \textbf{WX} has incomplete redshift coverage, or is completely absent, for Bin 3 and Bin 4, respectively. The results are compared to Cosmic Shear Ratio (CSR) \cite{Rana2025HSCY3CosmicShearRatios} photo-$z$ shift calibration, combined galaxy clustering and weak lensing (GC-WL) \cite{Zhang2025RedshiftGCPointMass} shift measurements, angular power spectrum cosmic shear \cite{Dalal2023CosmoShearHSC}, and two-point correlation cosmic shear \cite{CosmoShearLi2023HSCY3} posterior modes. We display the effect of the successive corrections to $n(z)$. No bias evolution corresponds to only using the cross-correlations with dark matter evolution over the bin. $b_s(z)$ and $b_p(z)$ respectively correspond to the corrections of the bias evolution for the spectroscopic sample and the photometric sample, discussed in sections \ref{sec:modeling:nz} and \ref{sec:modeling:photoz_gal_bias}. When correcting only for $b_s(z)$, we take into account dark matter evolution. Finally, $\mu$ is the magnification correction detailed in section \ref{sec:modeling:magnification}. We showcase the impact of choosing a different, more conservative scale cut such as $1-5\,\hMpc$. The shaded gold area corresponds to our fiducial result with all calibrations included over scales $0.3-3\,\hMpc$, projected to compare with other measurements.} 
    \label{fig:shifts}
\end{figure}

The impact of the successive corrections made to the clustering redshifts measurements are shown with Bin 1 as an example in the upper panel of Figure \ref{fig:distributions}. 
Similar global shifts are observed on every tomographic Bin due to the successive corrections. 
In the lower panel of Figure \ref{fig:distributions}, measurements with all corrections and the associated spline models are shown. 
For comparison, we replicate the \textbf{PhotZ} photometry calibrated distributions presented in Figure 8 of \cite{HSCClusteringRau2023}. $\dz$ "shift" values are reported in Table \ref{tab:shifts_results} and displayed in Figure \ref{fig:shifts}. These shifts are measured as a difference of mean values with respect to the reported Y3 Total distribution means obtained in \cite{HSCClusteringRau2023}. Per convention, a negative shift implies the considered measurement has a preference for a higher expectation value compared to the Y3 Total measurement.
{\renewcommand{\arraystretch}{1.2}
\begin{table}[h!]
\centering
\caption{\textit{HSC calibration results}: Measurements from the clustering redshift analysis \cite{HSCClusteringRau2023}. The \textbf{PhotZ} calibration measurements correspond to the hatched gray background curves from photometry calibration shown in Figure \ref{fig:distributions} and in Figure 8 of \cite{HSCClusteringRau2023}. \textbf{WX} adds in calibration from the clustering redshifts measurements with the \texttt{CAMIRA} LRG sample. The Bayesian surprise term is the Kullback-Leibler divergence, characterizing the information gained from the addition of cross-correlations over the \textbf{PhotZ} calibration. Values in brackets are the standard deviation for the considered measurement. For the following table sections, measurements displayed correspond to shifts with respect to the "Y3 Total" measurement. A negative shift denotes a shift of the new measurements towards higher redshifts. \textit{Cosmic Shear Results}: For Y3 power spectrum and real space cosmic shear measurements, uniform prior $\mathcal{U}(-1,1)$ was set on the shift parameters: the measurements in the table report the mode of the posterior distribution. Y3 CSR and Y3 GC-WL respectively correspond to the photo-$z$ shift measurements performed using Cosmic Shear Ratios (CSR) \cite{Rana2025HSCY3CosmicShearRatios} and the combined information from Galaxy Clustering and Weak Lensing (GC-WL) \cite{Zhang2025RedshiftGCPointMass}. \textit{This work}: Shifts measured for two scale cuts $0.3-3\,\hMpc$ and $1-5\,\hMpc$, excluding magnification corrections (first row) and including the magnification corrections (second row) for all four tomographic bins. The sections \textit{Cosmic Shear Results} and \textit{This work} showcase the results as shifts with respect to the Y3 Total measurements.}
\begin{tabular}{c|c|c|c|c}
\textbf{Measurement} & \textbf{Bin 1} & \textbf{Bin 2} & \textbf{Bin 3} & \textbf{Bin 4} \\
\hline
\multicolumn{4}{c}{\textbf{HSC calibration results \cite{HSCClusteringRau2023}}} \\ \hline
\textbf{Y3 PhotZ (DNNz)} & 0.463 [0.005] & 0.766 [0.004] & 1.084 [0.004] & 1.330 [0.003] \\
\textbf{Y3 DEMPz} & 0.463 & 0.777 & 1.097 & 1.350 \\
\textbf{Y3 PhotoZ \& WX} & 0.452 [0.004] & 0.766 [0.003] & 1.081 [0.004] & - \\
\textbf{Y3 Bayesian Surprise} & 3.84 & 0.10 & 0.28 & - \\
\textbf{Y3 Total} & 0.452 [0.024] & 0.766 [0.022] & 1.081 [0.031] & 1.330 [0.034] \\
\hline
\multicolumn{4}{c}{\textbf{Cosmic shear results}} \\ \hline
\textbf{Y3 Power Spectrum \cite{Dalal2023CosmoShearHSC}} 
    & - 
    & - 
    & $-0.075^{+0.056}_{-0.059}$ 
    & $-0.157^{+0.094}_{-0.111}$ \\ 
\textbf{Y3 Real Space \cite{CosmoShearLi2023HSCY3}} 
    & - 
    & - 
    & $-0.115^{+0.052}_{-0.058}$ 
    & $-0.192^{+0.088}_{-0.088}$ \\ 
\textbf{Y3 CSR \cite{Rana2025HSCY3CosmicShearRatios}} 
    & - 
    & $+0.002^{+0.021}_{-0.022}$
    & $-0.002^{+0.085}_{-0.217}$ 
    & $-0.292^{+0.229}_{-0.324}$ \\
\textbf{Y3 GC-WL \cite{Zhang2025RedshiftGCPointMass}}
    & - 
    & -
    & $-0.112^{+0.046}_{-0.049}$ 
    & $-0.185^{+0.071}_{-0.081}$ \\
\hline
\multicolumn{4}{c}{\textbf{This work, $\mathbf{[0.3,3]\,\hMpc}$}} \\ \hline
\textbf{B-spline | $b_s(z)$, $b_p(z)$} 
    & $+0.023^{+0.007}_{-0.008}$ 
    & $-0.010^{+0.008}_{-0.007}$ 
    & $-0.043^{+0.019}_{-0.017}$ 
    & $-0.050^{+0.011}_{-0.011}$ \\ 
\textbf{B-spline | $b_s(z)$, $b_p(z)$, $\mu$} 
    & $+0.026^{+0.008}_{-0.008}$ 
    & $-0.005^{+0.009}_{-0.008}$ 
    & $-0.043^{+0.022}_{-0.020}$ 
    & $-0.050^{+0.012}_{-0.012}$ \\ 
\hline
\multicolumn{4}{c}{\textbf{This work, $\mathbf{[1,5]\,\hMpc}$}} \\ \hline
\textbf{B-spline | $b_s(z)$, $b_p(z)$} 
    & $+0.023^{+0.012}_{-0.013}$ 
    & $-0.023^{+0.014}_{-0.013}$ 
    & $-0.040^{+0.026}_{-0.025}$ 
    & $-0.058^{+0.016}_{-0.015}$ \\ 
\textbf{B-spline | $b_s(z)$, $b_p(z)$, $\mu$} 
    & $+0.026^{+0.012}_{-0.013}$ 
    & $-0.021^{+0.015}_{-0.014}$ 
    & $-0.038^{+0.028}_{-0.027}$ 
    & $-0.057^{+0.017}_{-0.016}$ \\ 
\end{tabular}
\label{tab:shifts_results} 
\end{table}}

The shift parameter results offer competitive measurements exclusively from clustering redshifts, as showcased in Figure \ref{fig:shifts}. In this work, we follow the recommendations from the first HSC analysis (see notes (iv) and (v) of Section 7 from \cite{HSCClusteringRau2023}), by modeling the evolution of the galaxy-dark matter bias for both the photometric sample and the spectroscopic sample, as well as corrections due to magnification effects. Moreover, the calibration sample we use here has redshifts $z\gtrsim1.2$ with QSO and ELG sources, allowing for complete calibration of Bin 3 and Bin 4, and the calibration sample is entirely composed of spectroscopic redshifts. These measurements are new with respect to the original analysis. In particular, quasars offer significant potential for high redshift calibration of source galaxies. In appendix \ref{sec:appendix_qso}, we explore their potential and sensitivity by showcasing clustering redshifts over a toy population at redshifts $z_\mathrm{phot}\in[1.8,2.0)$ at raw density $\sim0.1\;\mathrm{arcmin}^{-2}$.

As such, in the upper panel of Figure \ref{fig:distributions}, we can notice the relative impact of each of the calibrations on tomographic Bin 1. 
This work hints at biases in Bin 1 calibration, as displayed in Figure \ref{fig:shifts}. 
We explain this discrepancy with the successive corrections to correct galaxy bias with the auto-correlation measurements. 
With the "No bias evolution" correction, the measurement is $n_p(z_j)\propto \bo_{sp}(z_j)/\bo_{mm}(z_j)$ and is more in-line with reported measurements of \textbf{WX} from the CAMIRA LRG sample. Moreover, as reported by Table \ref{tab:galaxy_bias}, the BGS galaxies ($z\sim0-0.5$) have strong bias evolution, hence provoking significant shifts towards lower redshift expectations. 
This linear galaxy bias measurement is further corroborated by two-point angular measurements used in the auto-correlation correction. 
Note that clustering redshifts measurements of $n(z)$ that do not take into account galaxy bias evolution are not necessarily comparable due to evolution rate differences from different calibration samples, hence partially explaining the mild difference we see between \textbf{PhotZ \& WX} and the measurement with no bias evolution made in this work.

In Bin 2, we observe full consistency with the HSC Y3 results for both scale cuts.
This result is consistent with the Cosmic Shear Ratio photo-$z$ \cite{Rana2025HSCY3CosmicShearRatios} calibration. 
Bin 2 has the highest relative magnification correction: still, magnification has a minor effect with respect to the other corrections made to the $n(z)$ measurement. 
In Bin 2, we decide to not use ELGs from DR1 as they are subject to significant fiber collisions: about $\sim35\%$ of ELGs were observed in DR1, resulting in approximately $\sim50\%$ missing small-scale pairs \cite{Pinon2025FiberAssign}.
This effect is much less significant at larger scales $1-5\,\hMpc$ as we leave the regime of small-scale fiber assignment issues, especially for the higher redshift range. 
The consistency showcased between our measurements at different scales furthermore supports that fiber assignment systematics do not introduce bias in the measurements, as detailed in section \ref{sec:2pcf}. 
We notice Bin 2 presents minor shifts with respect to the smaller scale measurements, however this seems entirely driven by a single redshift bin and subsequently unrelated to fiber assignment issues. 
In contrast, the large scale incompleteness is corrected by the completeness weights included in the measurements. 
With subsequent data releases for DESI, the ELG completeness and fiber assignment issues are expected to significantly improve. 
In that regard, further improving these measurements is left for future work.

Bin 3 and Bin 4 leverage the DESI DR2 dataset for the calibrated $n(z)$. 
Due to their high galaxy bias, the measurements for the QSO sample offer very good insights in calibrating high-redshift ranges, despite the low number densities reported in Figure \ref{fig:density_z}. 
We measure a shift towards higher redshifts: however, the shift is less strong that the shifts predicted by GC-WL and cosmic shear analyses. CSR analysis for Bin 3 shows better consistency with the shifts inferred in the present work, though the larger error bars allow the measurement to remain consistent with the remaining calibrations. 
Moreover, in Bin 3, the \textbf{PhotZ} calibration predicted significant tail-like effects on the high redshift range $z\sim1.3-1.8$. 
While Bin 3 has partial clustering redshifts measurements to calibrate the distributions in \cite{HSCClusteringRau2023}, the tail-like effect is not probed by cross-correlations due to the limited redshift range of the CAMIRA LRG sample ($z\lesssim1.2$).
The tail contribution is measured in this analysis and we find it has stronger amplitude than initially expected by the \textbf{PhotZ} distribution, as shown in Figure \ref{fig:distributions}. 
This can be used to explain our noticeable shift in the tomographic mean with respect to the calibration provided in the first analysis, where clustering measurements spanned up to $z\sim1.2$ and therefore did not have clustering insights for the tail effects.
This result is more in-line with the expected possible shift in $\dz_3$ reported by the cosmic shear analyses \cite{Dalal2023CosmoShearHSC, CosmoShearLi2023HSCY3}, though still lower than reported by these analyses. 
In the lower panel of Figure \ref{fig:distributions}, we show the $n(z)$ distributions for the four tomographic bins on the same figure. We underlay the figure with the hatched distributions from \textbf{PhotZ} calibration for each of the bins. 
When it comes to cosmology dependence of these results, they are two-fold: cosmology impacts the scale cuts we define for this analysis, and cosmology intervenes in the dark matter auto-correlation $\bo_{mm}$ function. In any correction that does not include both $b_s(z)$ and $b_p(z)$, we compensate our measurements with $\bo_{mm}(z)$. However, if one corrects for both $b_s(z)$ and $b_p(z)$ with the measured auto-correlations (for example, any measurement showcased in Table \ref{tab:shifts_results}), then $\bo_{mm}$ is encompassed in the auto-correlation vectors, and therefore we do not introduce cosmology dependence from the computation, except from chosen scale cuts. As reported by \cite{HSCClusteringRau2023}, for fairly extreme cosmologies at the $2\sigma$ contour of Stage III surveys, one can see up to $\sim10\%$ effects in the redshift scaling of $\bo_{mm}$. In this work, in the context of fiducial results, this dependence only affects the magnification corrections presented in section \ref{sec:modeling:magnification} as they rely on the theoretical dark matter auto-correlation: this scaling effect is not considered in this work with respect to the other assumptions relevant to the magnification correction.

\section{Conclusion}
\label{sec:conclusion}

In this work, we provide a full calibration of the four HSC tomographic bins, including constraints on the two higher redshift tomographic bins. 
With the arrival of Stage-IV weak lensing surveys such as LSST, Euclid and the Roman Space Telescope, and further spectroscopic data programs available such as the next data releases of DESI and the data acquisition of the Prime Focus Spectrograph (PFS) \cite{2016PFSspectro} program, systematic mitigation will become significantly more important. 
Therefore, this work explored reducing some of the common systematics in clustering redshifts, such as galaxy bias and magnification effects. 
The fiducial scale cut choice of this analysis is $0.3-3\,\hMpc$. Consistency between the two presented scale cuts ($0.3-3\,\hMpc$ and $1-5\,\hMpc$) alleviates concerns that bias modeling complexities have substantial impacts on the showcased results in section \ref{sec:results}. Moreover, we show that one can relax the assumption of linear bias commonly used in clustering redshift works in the $n(z)$ model to the assumption that the cross-correlation coefficient does not evolve in redshift over the chosen scales.


Our calibration of the tomographic distributions suggests smaller shifts than expected by the cosmic shear analyses from HSC Year 3 and the shear calibration ratios analysis in Bin 3 and Bin 4. 
This analysis also hints at biases present in Bin 1 and a mild bias present in Bin 2, both towards lower redshifts. 
In companion paper \cite{ChoppinDeJanvry2025b}, we present the cosmological parameter inference on cosmic shear in Fourier space with the calibrated redshift distributions from this analysis.

The code used for this work and all the data to replicate the figures of this paper is made publicly available at \href{https://github.com/JeanCHDJdev/desi-y3-hsc/}{\textcolor{blue}{github.com/jeanchdjdev/desi-y3-hsc}}.

\acknowledgments
\label{sec:acknowledgements}

\hspace{0.85cm}The authors thank Quentin Lavier for pipeline improvements and debugging.
JCdJ acknowledges support from Fondation CentraleSupélec and Fondation Ailes de France. SGG acknowledges that this work was performed in part at the Aspen Center for Physics, which is supported by National Science Foundation grant PHY-2210452 and a grant from the Alfred P Sloan Foundation (G-2024-22395). US is supported by NSF CDSE grant number AST-2408026 and NASA TCAN grant number 80NSSC24K0101. TZ is supported by Schmidt Sciences.
 
This research used data obtained with the Dark Energy Spectroscopic Instrument (DESI). DESI construction and operations is managed by the Lawrence Berkeley National Laboratory. This material is based upon work supported by the U.S. Department of Energy, Office of Science, Office of High-Energy Physics, under Contract No. DE–AC02–05CH11231, and by the National Energy Research Scientific Computing Center, a DOE Office of Science User Facility under the same contract. Additional support for DESI was provided by the U.S. National Science Foundation (NSF), Division of Astronomical Sciences under Contract No. AST-0950945 to the NSF’s National Optical-Infrared Astronomy Research Laboratory; the Science and Technology Facilities Council of the United Kingdom; the Gordon and Betty Moore Foundation; the Heising-Simons Foundation; the French Alternative Energies and Atomic Energy Commission (CEA); the National Council of Humanities, Science and Technology of Mexico (CONAHCYT); the Ministry of Science and Innovation of Spain (MICINN), and by the DESI Member Institutions: \url{www.desi.lbl.gov/collaborating-institutions}. The DESI collaboration is honored to be permitted to conduct scientific research on I’oligam Du’ag (Kitt Peak), a mountain with particular significance to the Tohono O’odham Nation. Any opinions, findings, and conclusions or recommendations expressed in this material are those of the author(s) and do not necessarily reflect the views of the U.S. National Science Foundation, the U.S. Department of Energy, or any of the listed funding agencies.

The Hyper Suprime-Cam (HSC) collaboration includes the astronomical communities of Japan and Taiwan, and Princeton University. The HSC instrumentation and software were developed by the National Astronomical Observatory of Japan (NAOJ), the Kavli Institute for the Physics and Mathematics of the Universe (Kavli IPMU), the University of Tokyo, the High Energy Accelerator Research Organization (KEK), the Academia Sinica Institute for Astronomy and Astrophysics in Taiwan (ASIAA), and Princeton University. Funding was contributed by the FIRST program from the Japanese Cabinet Office, the Ministry of Education, Culture, Sports, Science and Technology (MEXT), the Japan Society for the Promotion of Science (JSPS), Japan Science and Technology Agency (JST), the Toray Science Foundation, NAOJ, Kavli IPMU, KEK, ASIAA, and Princeton University. This paper makes use of software developed for Vera C. Rubin Observatory. We thank the Rubin Observatory for making their code available as free software at \url{http://pipelines.lsst.io/}. This paper is based on data collected at the Subaru Telescope and retrieved from the HSC data archive system, which is operated by the Subaru Telescope and Astronomy Data Center (ADC) at NAOJ. Data analysis was in part carried out with the cooperation of Center for Computational Astrophysics (CfCA), NAOJ. We are honored and grateful for the opportunity of observing the Universe from Maunakea, which has the cultural, historical and natural significance in Hawai'i. 

This analysis used the following software packages: \texttt{astropy} \cite{astropy2013paper1, astropy2018paper2, astropy2023paper3}, \texttt{CCL} \cite{CCLChisari2019}, \texttt{Corrfunc} \cite{Sinha2019corrfunc1, Sinha2019corrfunc2} (with \texttt{cosmodesi/pycorr}), \texttt{mocpy} \cite{fernique2015moc}, \texttt{numpy} \cite{harris2020numpy}, \texttt{SciPy} \cite{2020SciPyNMeth}, \texttt{matplotlib} \cite{Hunter2007Matplotlib}, \texttt{pyMC} \cite{Abril-Pla_PyMC_a_modern_2023}, \texttt{scikit-learn} \cite{ScikitLearnPedregosa2011}, \texttt{fitsio}.

\bibliographystyle{JHEP}
\bibliography{src/desi,src/hsc,src/othersurveys,src/software,src/methods,src/clusterz}

@article{KidsCrossCorrvdB2020,
   title={Testing KiDS cross-correlation redshifts with simulations},
   volume={642},
   ISSN={1432-0746},
   url={http://dx.doi.org/10.1051/0004-6361/202038835},
   DOI={10.1051/0004-6361/202038835},
   journal={\aap},
   publisher={EDP Sciences},
   author={van den Busch, J. L. and Hildebrandt, H. and Wright, A. H. and Morrison, C. B. and Blake, C. and Joachimi, B. and Erben, T. and Heymans, C. and Kuijken, K. and Taylor, E. N.},
   year={2020},
   month=oct, pages={A200} 
}

@article{EuclidClusteringRedshifts2023,
   title={Euclid: Calibrating photometric redshifts with spectroscopic cross-correlations},
   volume={670},
   ISSN={1432-0746},
   url={http://dx.doi.org/10.1051/0004-6361/202244795},
   DOI={10.1051/0004-6361/202244795},
   journal={\aap},
   publisher={EDP Sciences},
   author={Naidoo, K. and Johnston, H. and Joachimi, B. and van den Busch, J. L. and Hildebrandt, H. and Ilbert, O. and Lahav, O. and Aghanim, N. and Altieri, B. and Amara, A. and Baldi, M. and Bender, R. and Bodendorf, C. and Branchini, E. and Brescia, M. and Brinchmann, J. and Camera, S. and Capobianco, V. and Carbone, C. and Carretero, J. and Castander, F. J. and Castellano, M. and Cavuoti, S. and Cimatti, A. and Cledassou, R. and Congedo, G. and Conselice, C. J. and Conversi, L. and Copin, Y. and Corcione, L. and Courbin, F. and Cropper, M. and Da Silva, A. and Degaudenzi, H. and Dinis, J. and Dubath, F. and Dupac, X. and Dusini, S. and Farrens, S. and Ferriol, S. and Fosalba, P. and Frailis, M. and Franceschi, E. and Franzetti, P. and Fumana, M. and Galeotta, S. and Garilli, B. and Gillard, W. and Gillis, B. and Giocoli, C. and Grazian, A. and Grupp, F. and Haugan, S. V. H. and Holmes, W. and Hormuth, F. and Hornstrup, A. and Jahnke, K. and Kümmel, M. and Kiessling, A. and Kilbinger, M. and Kitching, T. and Kohley, R. and Kurki-Suonio, H. and Ligori, S. and Lilje, P. B. and Lloro, I. and Maiorano, E. and Mansutti, O. and Marggraf, O. and Markovic, K. and Marulli, F. and Massey, R. and Maurogordato, S. and Meneghetti, M. and Merlin, E. and Meylan, G. and Moresco, M. and Moscardini, L. and Munari, E. and Nakajima, R. and Niemi, S. M. and Padilla, C. and Paltani, S. and Pasian, F. and Pedersen, K. and Percival, W. J. and Pettorino, V. and Pires, S. and Polenta, G. and Poncet, M. and Popa, L. and Pozzetti, L. and Raison, F. and Rebolo, R. and Renzi, A. and Rhodes, J. and Riccio, G. and Romelli, E. and Rosset, C. and Rossetti, E. and Saglia, R. and Sapone, D. and Sartoris, B. and Schneider, P. and Secroun, A. and Seidel, G. and Sirignano, C. and Sirri, G. and Starck, J.-L. and Surace, C. and Tallada-Crespí, P. and Taylor, A. N. and Tereno, I. and Toledo-Moreo, R. and Torradeflot, F. and Tutusaus, I. and Valentijn, E. A. and Valenziano, L. and Vassallo, T. and Wang, Y. and Weller, J. and Wetzstein, M. and Zacchei, A. and Zamorani, G. and Zoubian, J. and Andreon, S. and Maino, D. and Scottez, V. and Wright, A. H.},
   year={2023},
   month=feb, pages={A149} }

@article{EuclidClusterRedshifts2025,
      title={Euclid: Photometric redshift calibration with the clustering redshifts technique}, 
      author={W. d'Assignies Doumerg and M. Manera and C. Padilla and O. Ilbert and H. Hildebrandt and L. Reynolds and J. Chaves-Montero and A. H. Wright and P. Tallada-Crespí and M. Eriksen and J. Carretero and W. Roster and Y. Kang and K. Naidoo and R. Miquel and B. Altieri and A. Amara and S. Andreon and N. Auricchio and C. Baccigalupi and D. Bagot and M. Baldi and A. Balestra and S. Bardelli and P. Battaglia and A. Biviano and E. Branchini and M. Brescia and S. Camera and V. Capobianco and C. Carbone and V. F. Cardone and S. Casas and F. J. Castander and M. Castellano and G. Castignani and S. Cavuoti and K. C. Chambers and A. Cimatti and C. Colodro-Conde and G. Congedo and C. J. Conselice and L. Conversi and Y. Copin and F. Courbin and H. M. Courtois and M. Crocce and A. Da Silva and H. Degaudenzi and S. de la Torre and G. De Lucia and M. Douspis and X. Dupac and A. Ealet and S. Escoffier and M. Farina and F. Faustini and S. Ferriol and F. Finelli and P. Fosalba and S. Fotopoulou and M. Frailis and E. Franceschi and M. Fumana and S. Galeotta and K. George and B. Gillis and C. Giocoli and P. Gómez-Alvarez and J. Gracia-Carpio and A. Grazian and F. Grupp and W. Holmes and I. M. Hook and A. Hornstrup and K. Jahnke and M. Jhabvala and B. Joachimi and E. Keihänen and S. Kermiche and A. Kiessling and B. Kubik and M. Kümmel and M. Kunz and H. Kurki-Suonio and O. Lahav and A. M. C. Le Brun and S. Ligori and P. B. Lilje and V. Lindholm and I. Lloro and G. Mainetti and D. Maino and E. Maiorano and O. Mansutti and S. Marcin and O. Marggraf and K. Markovic and M. Martinelli and N. Martinet and F. Marulli and R. Massey and D. C. Masters and E. Medinaceli and S. Mei and M. Melchior and Y. Mellier and M. Meneghetti and E. Merlin and G. Meylan and A. Mora and M. Moresco and L. Moscardini and C. Neissner and S. -M. Niemi and S. Paltani and F. Pasian and K. Pedersen and V. Pettorino and S. Pires and G. Polenta and M. Poncet and L. A. Popa and L. Pozzetti and F. Raison and R. Rebolo and A. Renzi and J. Rhodes and G. Riccio and E. Romelli and M. Roncarelli and E. Rossetti and R. Saglia and Z. Sakr and D. Sapone and B. Sartoris and J. A. Schewtschenko and P. Schneider and T. Schrabback and A. Secroun and E. Sefusatti and G. Seidel and M. Seiffert and S. Serrano and P. Simon and C. Sirignano and G. Sirri and A. Spurio Mancini and L. Stanco and J. Steinwagner and D. Tavagnacco and A. N. Taylor and H. I. Teplitz and I. Tereno and N. Tessore and S. Toft and R. Toledo-Moreo and F. Torradeflot and A. Tsyganov and I. Tutusaus and L. Valenziano and J. Valiviita and T. Vassallo and G. Verdoes Kleijn and Y. Wang and J. Weller and G. Zamorani and E. Zucca and M. Bolzonella and C. Burigana and L. Gabarra and J. Martín-Fleitas and I. Risso and V. Scottez and M. Viel},
      year={2025},
      eprint={2505.10416},
      archivePrefix={arXiv},
      primaryClass={astro-ph.CO},
      url={https://arxiv.org/abs/2505.10416}, 
}

@article{ménard2014clusteringbasedredshiftestimationmethod,
      title={Clustering-based redshift estimation: method and application to data}, 
      author={Brice Ménard and Ryan Scranton and Samuel Schmidt and Chris Morrison and Donghui Jeong and Tamas Budavari and Mubdi Rahman},
      year={2014},
      eprint={1303.4722},
      archivePrefix={arXiv},
      primaryClass={astro-ph.CO},
      url={https://arxiv.org/abs/1303.4722}, 
}

@article{Cawthon2022DESY3Boss,
    author = {Cawthon, R and Elvin-Poole, J and Porredon, A and Crocce, M and Giannini, G and Gatti, M and Ross, A J and Rykoff, E S and Carnero Rosell, A and DeRose, J and Lee, S and Rodriguez-Monroy, M and Amon, A and Bechtol, K and De Vicente, J and Gruen, D and Morgan, R and Sanchez, E and Sanchez, J and Sevilla-Noarbe, I and Abbott, T M C and Aguena, M and Allam, S and Annis, J and Avila, S and Bacon, D and Bertin, E and Brooks, D and Burke, D L and Carrasco Kind, M and Carretero, J and Castander, F J and Choi, A and Costanzi, M and da Costa, L N and Pereira, M E S and Dawson, K and Desai, S and Diehl, H T and Eckert, K and Everett, S and Ferrero, I and Fosalba, P and Frieman, J and García-Bellido, J and Gaztanaga, E and Gruendl, R A and Gschwend, J and Gutierrez, G and Hinton, S R and Hollowood, D L and Honscheid, K and Huterer, D and James, D J and Kim, A G and Kneib, J-P and Kuehn, K and Kuropatkin, N and Lahav, O and Lima, M and Lin, H and Maia, M A G and Melchior, P and Menanteau, F and Miquel, R and Mohr, J J and Muir, J and Myles, J and Palmese, A and Pandey, S and Paz-Chinchón, F and Percival, W J and Plazas, A A and Roodman, A and Rossi, G and Scarpine, V and Serrano, S and Smith, M and Soares-Santos, M and Suchyta, E and Swanson, M E C and Tarle, G and To, C and Troxel, M A and Wilkinson, R D and (DES Collaboration)},
    title = {Dark Energy Survey Year 3 results: calibration of lens sample redshift distributions using clustering redshifts with BOSS/eBOSS},
    journal = {Monthly Notices of the Royal Astronomical Society},
    volume = {513},
    number = {4},
    pages = {5517-5539},
    year = {2022},
    month = {05},
    issn = {0035-8711},
    doi = {10.1093/mnras/stac1160},
    url = {https://doi.org/10.1093/mnras/stac1160},
    eprint = {https://academic.oup.com/mnras/article-pdf/513/4/5517/43865105/stac1160.pdf},
}

@article{CrossCorrSingh2019,
   title={Cosmological constraints from galaxy–lensing cross-correlations using BOSS galaxies with SDSS and CMB lensing},
   volume={491},
   ISSN={1365-2966},
   url={http://dx.doi.org/10.1093/mnras/stz2922},
   DOI={10.1093/mnras/stz2922},
   number={1},
   journal={Monthly Notices of the Royal Astronomical Society},
   publisher={Oxford University Press (OUP)},
   author={Singh, Sukhdeep and Mandelbaum, Rachel and Seljak, Uroš and Rodríguez-Torres, Sergio and Slosar, Anže},
   year={2019},
   month=oct, pages={51–68} 
}

@article{Newman2008ClusterZ,
    doi = {10.1086/589982},
    url = {https://dx.doi.org/10.1086/589982},
    year = {2008},
    month = {sep},
    publisher = {},
    volume = {684},
    number = {1},
    pages = {88},
    author = {Newman, Jeffrey A.},
    title = {Calibrating Redshift Distributions beyond Spectroscopic Limits with Cross-Correlations},
    journal = {The Astrophysical Journal}
}

@article{Morrison2017TheWiZZClusterZ,
   title={the-wizz: clustering redshift estimation for everyone},
   volume={467},
   ISSN={1365-2966},
   url={http://dx.doi.org/10.1093/mnras/stx342},
   DOI={10.1093/mnras/stx342},
   number={3},
   journal={Monthly Notices of the Royal Astronomical Society},
   publisher={Oxford University Press (OUP)},
   author={Morrison, C. B. and Hildebrandt, H. and Schmidt, S. J. and Baldry, I. K. and Bilicki, M. and Choi, A. and Erben, T. and Schneider, P.},
   year={2017},
   month=feb, pages={3576–3589} 
}

@article{wright2025kidslegacyClusterZ,
      title={KiDS-Legacy: Redshift distributions and their calibration}, 
      author={Angus H. Wright and Hendrik Hildebrandt and Jan Luca van den Busch and Maciej Bilicki and Catherine Heymans and Benjamin Joachimi and Constance Mahony and Robert Reischke and Benjamin Stölzner and Anna Wittje and Marika Asgari and Nora Elisa Chisari and Andrej Dvornik and Christos Georgiou and Benjamin Giblin and Henk Hoekstra and Priyanka Jalan and Anjitha John William and Shahab Joudaki and Konrad Kuijken and Giorgio Francesco Lesci and Shun-Sheng Li and Laila Linke and Arthur Loureiro and Matteo Maturi and Lauro Moscardin and Lucas Porth and Mario Radovich and Tilman Tröster and Maximilian von Wietersheim-Kramsta and Ziang Yan and Mijin Yoon and Yun-Hao Zhang},
      year={2025},
      eprint={2503.19440},
      archivePrefix={arXiv},
      primaryClass={astro-ph.CO},
      url={https://arxiv.org/abs/2503.19440}, 
}

@article{Schmidt2013ClusteringRedshifts,
   title={Recovering redshift distributions with cross-correlations: pushing the boundaries},
   volume={431},
   ISSN={0035-8711},
   url={http://dx.doi.org/10.1093/mnras/stt410},
   DOI={10.1093/mnras/stt410},
   number={4},
   journal={Monthly Notices of the Royal Astronomical Society},
   publisher={Oxford University Press (OUP)},
   author={Schmidt, Samuel J. and Ménard, Brice and Scranton, Ryan and Morrison, Christopher and McBride, Cameron K.},
   year={2013},
   month=apr, 
   pages={3307–3318} 
}

@article{McQuinnClusterZ2013,
   title={On using angular cross-correlations to determine source redshift distributions},
   volume={433},
   ISSN={1365-2966},
   url={http://dx.doi.org/10.1093/mnras/stt914},
   DOI={10.1093/mnras/stt914},
   number={4},
   journal={Monthly Notices of the Royal Astronomical Society},
   publisher={Oxford University Press (OUP)},
   author={McQuinn, M. and White, M.},
   year={2013},
   month=jul, pages={2857–2883} }

@ARTICLE{DESI2016b.Instr,
       author = {{DESI Collaboration} and {Aghamousa}, Amir and {Aguilar}, Jessica and {Ahlen}, Steve and {Alam}, Shadab and {Allen}, Lori E. and {Allende Prieto}, Carlos and {Annis}, James and {Bailey}, Stephen and {Balland}, Christophe and {Ballester}, Otger and {Baltay}, Charles and {Beaufore}, Lucas and {Bebek}, Chris and {Beers}, Timothy C. and {Bell}, Eric F. and {Bernal}, Jos{\'e} Luis and {Besuner}, Robert and {Beutler}, Florian and {Blake}, Chris and {Bleuler}, Hannes and {Blomqvist}, Michael and {Blum}, Robert and {Bolton}, Adam S. and {Briceno}, Cesar and {Brooks}, David and {Brownstein}, Joel R. and {Buckley-Geer}, Elizabeth and {Burden}, Angela and {Burtin}, Etienne and {Busca}, Nicolas G. and {Cahn}, Robert N. and {Cai}, Yan-Chuan and {Cardiel-Sas}, Laia and {Carlberg}, Raymond G. and {Carton}, Pierre-Henri and {Casas}, Ricard and {Castander}, Francisco J. and {Cervantes-Cota}, Jorge L. and {Claybaugh}, Todd M. and {Close}, Madeline and {Coker}, Carl T. and {Cole}, Shaun and {Comparat}, Johan and {Cooper}, Andrew P. and {Cousinou}, M. -C. and {Crocce}, Martin and {Cuby}, Jean-Gabriel and {Cunningham}, Daniel P. and {Davis}, Tamara M. and {Dawson}, Kyle S. and {de la Macorra}, Axel and {De Vicente}, Juan and {Delubac}, Timoth{\'e}e and {Derwent}, Mark and {Dey}, Arjun and {Dhungana}, Govinda and {Ding}, Zhejie and {Doel}, Peter and {Duan}, Yutong T. and {Ealet}, Anne and {Edelstein}, Jerry and {Eftekharzadeh}, Sarah and {Eisenstein}, Daniel J. and {Elliott}, Ann and {Escoffier}, St{\'e}phanie and {Evatt}, Matthew and {Fagrelius}, Parker and {Fan}, Xiaohui and {Fanning}, Kevin and {Farahi}, Arya and {Farihi}, Jay and {Favole}, Ginevra and {Feng}, Yu and {Fernandez}, Enrique and {Findlay}, Joseph R. and {Finkbeiner}, Douglas P. and {Fitzpatrick}, Michael J. and {Flaugher}, Brenna and {Flender}, Samuel and {Font-Ribera}, Andreu and {Forero-Romero}, Jaime E. and {Fosalba}, Pablo and {Frenk}, Carlos S. and {Fumagalli}, Michele and {Gaensicke}, Boris T. and {Gallo}, Giuseppe and {Garcia-Bellido}, Juan and {Gaztanaga}, Enrique and {Pietro Gentile Fusillo}, Nicola and {Gerard}, Terry and {Gershkovich}, Irena and {Giannantonio}, Tommaso and {Gillet}, Denis and {Gonzalez-de-Rivera}, Guillermo and {Gonzalez-Perez}, Violeta and {Gott}, Shelby and {Graur}, Or and {Gutierrez}, Gaston and {Guy}, Julien and {Habib}, Salman and {Heetderks}, Henry and {Heetderks}, Ian and {Heitmann}, Katrin and {Hellwing}, Wojciech A. and {Herrera}, David A. and {Ho}, Shirley and {Holland}, Stephen and {Honscheid}, Klaus and {Huff}, Eric and {Hutchinson}, Timothy A. and {Huterer}, Dragan and {Hwang}, Ho Seong and {Illa Laguna}, Joseph Maria and {Ishikawa}, Yuzo and {Jacobs}, Dianna and {Jeffrey}, Niall and {Jelinsky}, Patrick and {Jennings}, Elise and {Jiang}, Linhua and {Jimenez}, Jorge and {Johnson}, Jennifer and {Joyce}, Richard and {Jullo}, Eric and {Juneau}, St{\'e}phanie and {Kama}, Sami and {Karcher}, Armin and {Karkar}, Sonia and {Kehoe}, Robert and {Kennamer}, Noble and {Kent}, Stephen and {Kilbinger}, Martin and {Kim}, Alex G. and {Kirkby}, David and {Kisner}, Theodore and {Kitanidis}, Ellie and {Kneib}, Jean-Paul and {Koposov}, Sergey and {Kovacs}, Eve and {Koyama}, Kazuya and {Kremin}, Anthony and {Kron}, Richard and {Kronig}, Luzius and {Kueter-Young}, Andrea and {Lacey}, Cedric G. and {Lafever}, Robin and {Lahav}, Ofer and {Lambert}, Andrew and {Lampton}, Michael and {Landriau}, Martin and {Lang}, Dustin and {Lauer}, Tod R. and {Le Goff}, Jean-Marc and {Le Guillou}, Laurent and {Le Van Suu}, Auguste and {Lee}, Jae Hyeon and {Lee}, Su-Jeong and {Leitner}, Daniela and {Lesser}, Michael and {Levi}, Michael E. and {L'Huillier}, Benjamin and {Li}, Baojiu and {Liang}, Ming and {Lin}, Huan and {Linder}, Eric and {Loebman}, Sarah R. and {Luki{\'c}}, Zarija and {Ma}, Jun and {MacCrann}, Niall and {Magneville}, Christophe and {Makarem}, Laleh and {Manera}, Marc and {Manser}, Christopher J. and {Marshall}, Robert and {Martini}, Paul and {Massey}, Richard and {Matheson}, Thomas and {McCauley}, Jeremy and {McDonald}, Patrick and {McGreer}, Ian D. and {Meisner}, Aaron and {Metcalfe}, Nigel and {Miller}, Timothy N. and {Miquel}, Ramon and {Moustakas}, John and {Myers}, Adam and {Naik}, Milind and {Newman}, Jeffrey A. and {Nichol}, Robert C. and {Nicola}, Andrina and {Nicolati da Costa}, Luiz and {Nie}, Jundan and {Niz}, Gustavo and {Norberg}, Peder and {Nord}, Brian and {Norman}, Dara and {Nugent}, Peter and {O'Brien}, Thomas and {Oh}, Minji and {Olsen}, Knut A.~G. and {Padilla}, Cristobal and {Padmanabhan}, Hamsa and {Padmanabhan}, Nikhil and {Palanque-Delabrouille}, Nathalie and {Palmese}, Antonella and {Pappalardo}, Daniel and {P{\^a}ris}, Isabelle and {Park}, Changbom and {Patej}, Anna and {Peacock}, John A. and {Peiris}, Hiranya V. and {Peng}, Xiyan and {Percival}, Will J. and {Perruchot}, Sandrine and {Pieri}, Matthew M. and {Pogge}, Richard and {Pollack}, Jennifer E. and {Poppett}, Claire and {Prada}, Francisco and {Prakash}, Abhishek and {Probst}, Ronald G. and {Rabinowitz}, David and {Raichoor}, Anand and {Ree}, Chang Hee and {Refregier}, Alexandre and {Regal}, Xavier and {Reid}, Beth and {Reil}, Kevin and {Rezaie}, Mehdi and {Rockosi}, Constance M. and {Roe}, Natalie and {Ronayette}, Samuel and {Roodman}, Aaron and {Ross}, Ashley J. and {Ross}, Nicholas P. and {Rossi}, Graziano and {Rozo}, Eduardo and {Ruhlmann-Kleider}, Vanina and {Rykoff}, Eli S. and {Sabiu}, Cristiano and {Samushia}, Lado and {Sanchez}, Eusebio and {Sanchez}, Javier and {Schlegel}, David J. and {Schneider}, Michael and {Schubnell}, Michael and {Secroun}, Aur{\'e}lia and {Seljak}, Uros and {Seo}, Hee-Jong and {Serrano}, Santiago and {Shafieloo}, Arman and {Shan}, Huanyuan and {Sharples}, Ray and {Sholl}, Michael J. and {Shourt}, William V. and {Silber}, Joseph H. and {Silva}, David R. and {Sirk}, Martin M. and {Slosar}, Anze and {Smith}, Alex and {Smoot}, George F. and {Som}, Debopam and {Song}, Yong-Seon and {Sprayberry}, David and {Staten}, Ryan and {Stefanik}, Andy and {Tarle}, Gregory and {Sien Tie}, Suk and {Tinker}, Jeremy L. and {Tojeiro}, Rita and {Valdes}, Francisco and {Valenzuela}, Octavio and {Valluri}, Monica and {Vargas-Magana}, Mariana and {Verde}, Licia and {Walker}, Alistair R. and {Wang}, Jiali and {Wang}, Yuting and {Weaver}, Benjamin A. and {Weaverdyck}, Curtis and {Wechsler}, Risa H. and {Weinberg}, David H. and {White}, Martin and {Yang}, Qian and {Yeche}, Christophe and {Zhang}, Tianmeng and {Zhao}, Gong-Bo and {Zheng}, Yi and {Zhou}, Xu and {Zhou}, Zhimin and {Zhu}, Yaling and {Zou}, Hu and {Zu}, Ying},
        title = "{The DESI Experiment Part II: Instrument Design}",
      journal = {arXiv e-prints},
         year = 2016,
        month = oct,
          eid = {arXiv:1611.00037},
        pages = {arXiv:1611.00037},
archivePrefix = {arXiv},
       eprint = {1611.00037},
 primaryClass = {astro-ph.IM},
       adsurl = {https://ui.adsabs.harvard.edu/abs/2016arXiv161100037D}
}

@ARTICLE{DESI2022.KP1.Instr,
       author = {{DESI Collaboration} and {Abareshi}, B. and {Aguilar}, J. and {Ahlen}, S. and {Alam}, Shadab and {Alexander}, David M. and {Alfarsy}, R. and {Allen}, L. and {Allende Prieto}, C. and {Alves}, O. and {Ameel}, J. and {Armengaud}, E. and {Asorey}, J. and {Aviles}, Alejandro and {Bailey}, S. and {Balaguera-Antol{\'\i}nez}, A. and {Ballester}, O. and {Baltay}, C. and {Bault}, A. and {Beltran}, S.~F. and {Benavides}, B. and {BenZvi}, S. and {Berti}, A. and {Besuner}, R. and {Beutler}, Florian and {Bianchi}, D. and {Blake}, C. and {Blanc}, P. and {Blum}, R. and {Bolton}, A. and {Bose}, S. and {Bramall}, D. and {Brieden}, S. and {Brodzeller}, A. and {Brooks}, D. and {Brownewell}, C. and {Buckley-Geer}, E. and {Cahn}, R.~N. and {Cai}, Z. and {Canning}, R. and {Capasso}, R. and {Carnero Rosell}, A. and {Carton}, P. and {Casas}, R. and {Castander}, F.~J. and {Cervantes-Cota}, J.~L. and {Chabanier}, S. and {Chaussidon}, E. and {Chuang}, C. and {Circosta}, C. and {Cole}, S. and {Cooper}, A.~P. and {da Costa}, L. and {Cousinou}, M. -C. and {Cuceu}, A. and {Davis}, T.~M. and {Dawson}, K. and {de la Cruz-Noriega}, R. and {de la Macorra}, A. and {de Mattia}, A. and {Della Costa}, J. and {Demmer}, P. and {Derwent}, M. and {Dey}, A. and {Dey}, B. and {Dhungana}, G. and {Ding}, Z. and {Dobson}, C. and {Doel}, P. and {Donald-McCann}, J. and {Donaldson}, J. and {Douglass}, K. and {Duan}, Y. and {Dunlop}, P. and {Edelstein}, J. and {Eftekharzadeh}, S. and {Eisenstein}, D.~J. and {Enriquez-Vargas}, M. and {Escoffier}, S. and {Evatt}, M. and {Fagrelius}, P. and {Fan}, X. and {Fanning}, K. and {Fawcett}, V.~A. and {Ferraro}, S. and {Ereza}, J. and {Flaugher}, B. and {Font-Ribera}, A. and {Forero-Romero}, J.~E. and {Frenk}, C.~S. and {Fromenteau}, S. and {G{\"a}nsicke}, B.~T. and {Garcia-Quintero}, C. and {Garrison}, L. and {Gazta{\~n}aga}, E. and {Gerardi}, F. and {Gil-Mar{\'\i}n}, H. and {Gontcho a Gontcho}, S. and {Gonzalez-Morales}, Alma X. and {Gonzalez-de-Rivera}, G. and {Gonzalez-Perez}, V. and {Gordon}, C. and {Graur}, O. and {Green}, D. and {Grove}, C. and {Gruen}, D. and {Gutierrez}, G. and {Guy}, J. and {Hahn}, C. and {Harris}, S. and {Herrera}, D. and {Herrera-Alcantar}, Hiram K. and {Honscheid}, K. and {Howlett}, C. and {Huterer}, D. and {Ir{\v{s}}i{\v{c}}}, V. and {Ishak}, M. and {Jelinsky}, P. and {Jiang}, L. and {Jimenez}, J. and {Jing}, Y.~P. and {Joyce}, R. and {Jullo}, E. and {Juneau}, S. and {Kara{\c{c}}ayl{\i}}, N.~G. and {Karamanis}, M. and {Karcher}, A. and {Karim}, T. and {Kehoe}, R. and {Kent}, S. and {Kirkby}, D. and {Kisner}, T. and {Kitaura}, F. and {Koposov}, S.~E. and {Kov{\'a}cs}, A. and {Kremin}, A. and {Krolewski}, Alex and {L'Huillier}, B. and {Lahav}, O. and {Lambert}, A. and {Lamman}, C. and {Lan}, Ting-Wen and {Landriau}, M. and {Lane}, S. and {Lang}, D. and {Lange}, J.~U. and {Lasker}, J. and {Le Guillou}, L. and {Leauthaud}, A. and {Le Van Suu}, A. and {Levi}, Michael E. and {Li}, T.~S. and {Magneville}, C. and {Manera}, M. and {Manser}, Christopher J. and {Marshall}, B. and {Martini}, Paul and {McCollam}, W. and {McDonald}, P. and {Meisner}, Aaron M. and {Mena-Fern{\'a}ndez}, J. and {Meneses-Rizo}, J. and {Mezcua}, M. and {Miller}, T. and {Miquel}, R. and {Montero-Camacho}, P. and {Moon}, J. and {Moustakas}, J. and {Mueller}, E. and {Mu{\~n}oz-Guti{\'e}rrez}, Andrea and {Myers}, Adam D. and {Nadathur}, S. and {Najita}, J. and {Napolitano}, L. and {Neilsen}, E. and {Newman}, Jeffrey A. and {Nie}, J.~D. and {Ning}, Y. and {Niz}, G. and {Norberg}, P. and {Noriega}, Hern{\'a}n E. and {O'Brien}, T. and {Obuljen}, A. and {Palanque-Delabrouille}, N. and {Palmese}, A. and {Zhiwei}, P. and {Pappalardo}, D. and {PENG}, X. and {Percival}, W.~J. and {Perruchot}, S. and {Pogge}, R. and {Poppett}, C. and {Porredon}, A. and {Prada}, F. and {Prochaska}, J. and {Pucha}, R. and {P{\'e}rez-Fern{\'a}ndez}, A. and {P{\'e}rez-R{\`a}fols}, I. and {Rabinowitz}, D. and {Raichoor}, A. and {Ramirez-Solano}, S. and {Ram{\'\i}rez-P{\'e}rez}, C{\'e}sar and {Ravoux}, C. and {Reil}, K. and {Rezaie}, M. and {Rocher}, A. and {Rockosi}, C. and {Roe}, N.~A. and {Roodman}, A. and {Ross}, A.~J. and {Rossi}, G. and {Ruggeri}, R. and {Ruhlmann-Kleider}, V. and {Sabiu}, C.~G. and {Safonova}, S. and {Said}, K. and {Saintonge}, A. and {Salas Catonga}, Javier and {Samushia}, L. and {Sanchez}, E. and {Saulder}, C. and {Schaan}, E. and {Schlafly}, E. and {Schlegel}, D. and {Schmoll}, J. and {Scholte}, D. and {Schubnell}, M. and {Secroun}, A. and {Seo}, H. and {Serrano}, S. and {Sharples}, Ray M. and {Sholl}, Michael J. and {Silber}, Joseph Harry and {Silva}, D.~R. and {Sirk}, M. and {Siudek}, M. and {Smith}, A. and {Sprayberry}, D. and {Staten}, R. and {Stupak}, B. and {Tan}, T. and {Tarl{\'e}}, Gregory and {Tie}, Suk Sien and {Tojeiro}, R. and {Ure{\~n}a-L{\'o}pez}, L.~A. and {Valdes}, F. and {Valenzuela}, O. and {Valluri}, M. and {Vargas-Maga{\~n}a}, M. and {Verde}, L. and {Walther}, M. and {Wang}, B. and {Wang}, M.~S. and {Weaver}, B.~A. and {Weaverdyck}, C. and {Wechsler}, R. and {Wilson}, Michael J. and {Yang}, J. and {Yu}, Y. and {Yuan}, S. and {Y{\`e}che}, Christophe and {Zhang}, H. and {Zhang}, K. and {Zhao}, Cheng and {Zhou}, Rongpu and {Zhou}, Zhimin and {Zou}, H. and {Zou}, J. and {Zou}, S. and {Zu}, Y. and {DESI Collaboration}},
        title = "{Overview of the Instrumentation for the Dark Energy Spectroscopic Instrument}",
      journal = {\aj},
         year = 2022,
        month = nov,
       volume = {164},
       number = {5},
          eid = {207},
        pages = {207},
          doi = {10.3847/1538-3881/ac882b},
archivePrefix = {arXiv},
       eprint = {2205.10939},
 primaryClass = {astro-ph.IM},
       adsurl = {https://ui.adsabs.harvard.edu/abs/2022AJ....164..207A}
}

@ARTICLE{Corrector.Miller.2023,
       author = {{Miller}, Timothy N. and {Doel}, Peter and {Gutierrez}, Gaston and {Besuner}, Robert and {Brooks}, David and {Gallo}, Giuseppe and {Heetderks}, Henry and {Jelinsky}, Patrick and {Kent}, Stephen M. and {Lampton}, Michael and {Levi}, Michael E. and {Liang}, Ming and {Meisner}, Aaron and {Sholl}, Michael J. and {Silber}, Joseph Harry and {Sprayberry}, David and {Aguilar}, Jessica Nicole and {de la Macorra}, Axel and {Eisenstein}, Daniel and {Fanning}, Kevin and {Font-Ribera}, Andreu and {Gazta{\~n}aga}, Enrique and {Gontcho A Gontcho}, Satya and {Honscheid}, Klaus and {Jimenez}, Jorge and {Joyce}, Dick and {Kehoe}, Robert and {Kisner}, Theodore and {Kremin}, Anthony and {Landriau}, Martin and {Le Guillou}, Laurent and {Magneville}, Christophe and {Martini}, Paul and {Miquel}, Ramon and {Moustakas}, John and {Nie}, Jundan and {Percival}, Will and {Poppett}, Claire and {Prada}, Francisco and {Rossi}, Graziano and {Schlegel}, David and {Schubnell}, Michael and {Seo}, Hee-Jong and {Sharples}, Ray and {Tarl{\'e}}, Gregory and {Vargas-Maga{\~n}a}, Mariana and {Zhou}, Zhimin and {the DESI Collaboration}},
        title = "{The Optical Corrector for the Dark Energy Spectroscopic Instrument}",
      journal = {\aj},
         year = 2024,
        month = aug,
       volume = {168},
       number = {2},
          eid = {95},
        pages = {95},
          doi = {10.3847/1538-3881/ad45fe},
archivePrefix = {arXiv},
       eprint = {2306.06310},
 primaryClass = {astro-ph.IM},
       adsurl = {https://ui.adsabs.harvard.edu/abs/2024AJ....168...95M}
}

@ARTICLE{FiberSystem.Poppett.2024,
       author = {{Poppett}, Claire and {Tyas}, Luke and {Aguilar}, J. and {Bebek}, Christopher and {Bramall}, D. and {Claybaugh}, T. and {Edelstein}, J. and {Fagrelius}, P. and {Heetderks}, H. and {Jelinsky}, P. and {Jelinsky}, S. and {Lafever}, Robin and {Lambert}, A. and {Lampton}, M. and {Levi}, Michael E. and {Martini}, P. and {Rockosi}, C. and {Schmoll}, J. and {Sharples}, Ray M. and {Sirk}, Martin and {Wishnow}, Edward and {Yu}, Jiaxi and {Ahlen}, S. and {Bault}, A. and {BenZvi}, S. and {Brooks}, D. and {Cole}, S. and {de la Macorra}, A. and {Dey}, Arjun and {Doel}, P. and {Fanning}, K. and {Font-Ribera}, A. and {Forero-Romero}, J.~E. and {Gazta{\~n}aga}, E. and {Gontcho A Gontcho}, S. and {Gonzalez-Morales}, A.~X. and {Hahn}, C. and {Honscheid}, K. and {Jimenez}, J. and {Juneau}, S. and {Kirkby}, D. and {Kremin}, A. and {Landriau}, M. and {Le Guillou}, L. and {Manera}, M. and {Meisner}, A. and {Miquel}, R. and {Moustakas}, J. and {Mueller}, E. and {Mu{\~n}oz-Guti{\'e}rrez}, A. and {Myers}, A.~D. and {Nie}, J. and {Niz}, G. and {Palanque-Delabrouille}, N. and {Percival}, W.~J. and {Prada}, F. and {Rabinowitz}, D. and {Rezaie}, M. and {Rossi}, G. and {Sanchez}, E. and {Schlafly}, Edward F. and {Schlegel}, D. and {Schubnell}, M. and {Seo}, H. and {Sprayberry}, D. and {Tarl{\'e}}, G. and {Vargas-Maga{\~n}a}, M. and {Weaver}, B.~A. and {Zhou}, R.},
        title = "{Overview of the Fiber System for the Dark Energy Spectroscopic Instrument}",
      journal = {\aj},
         year = 2024,
        month = dec,
       volume = {168},
       number = {6},
          eid = {245},
        pages = {245},
          doi = {10.3847/1538-3881/ad76a4},
       adsurl = {https://ui.adsabs.harvard.edu/abs/2024AJ....168..245P}
}

@ARTICLE{Spectro.Pipeline.Guy.2023,
       author = {{Guy}, J. and {Bailey}, S. and {Kremin}, A. and {Alam}, Shadab and {Alexander}, D.~M. and {Allende Prieto}, C. and {BenZvi}, S. and {Bolton}, A.~S. and {Brooks}, D. and {Chaussidon}, E. and {Cooper}, A.~P. and {Dawson}, K. and {de la Macorra}, A. and {Dey}, A. and {Dey}, Biprateep and {Dhungana}, G. and {Eisenstein}, D.~J. and {Font-Ribera}, A. and {Forero-Romero}, J.~E. and {Gazta{\~n}aga}, E. and {Gontcho A Gontcho}, S. and {Green}, D. and {Honscheid}, K. and {Ishak}, M. and {Kehoe}, R. and {Kirkby}, D. and {Kisner}, T. and {Koposov}, Sergey E. and {Lan}, Ting-Wen and {Landriau}, M. and {Le Guillou}, L. and {Levi}, Michael E. and {Magneville}, C. and {Manser}, Christopher J. and {Martini}, P. and {Meisner}, Aaron M. and {Miquel}, R. and {Moustakas}, J. and {Myers}, Adam D. and {Newman}, Jeffrey A. and {Nie}, Jundan and {Palanque-Delabrouille}, N. and {Percival}, W.~J. and {Poppett}, C. and {Prada}, F. and {Raichoor}, A. and {Ravoux}, C. and {Ross}, A.~J. and {Schlafly}, E.~F. and {Schlegel}, D. and {Schubnell}, M. and {Sharples}, Ray M. and {Tarl{\'e}}, Gregory and {Weaver}, B.~A. and {Y{\'e}che}, Christophe and {Zhou}, Rongpu and {Zhou}, Zhimin and {Zou}, H.},
        title = "{The Spectroscopic Data Processing Pipeline for the Dark Energy Spectroscopic Instrument}",
      journal = {\aj},
         year = 2023,
        month = apr,
       volume = {165},
       number = {4},
          eid = {144},
        pages = {144},
          doi = {10.3847/1538-3881/acb212},
archivePrefix = {arXiv},
       eprint = {2209.14482},
 primaryClass = {astro-ph.IM},
       adsurl = {https://ui.adsabs.harvard.edu/abs/2023AJ....165..144G}
}

@ARTICLE{SurveyOps.Schlafly.2023,
       author = {{Schlafly}, Edward F. and {Kirkby}, David and {Schlegel}, David J. and {Myers}, Adam D. and {Raichoor}, Anand and {Dawson}, Kyle and {Aguilar}, Jessica and {Allende Prieto}, Carlos and {Bailey}, Stephen and {BenZvi}, Segev and {Bermejo-Climent}, Jose and {Brooks}, David and {de la Macorra}, Axel and {Dey}, Arjun and {Doel}, Peter and {Fanning}, Kevin and {Font-Ribera}, Andreu and {Forero-Romero}, Jaime E. and {Garc{\'\i}a-Bellido}, Juan and {Gontcho A Gontcho}, Satya and {Guy}, Julien and {Hahn}, ChangHoon and {Honscheid}, Klaus and {Ishak}, Mustapha and {Juneau}, St{\'e}phanie and {Kehoe}, Robert and {Kisner}, Theodore and {Kremin}, Anthony and {Landriau}, Martin and {Lang}, Dustin A. and {Lasker}, James and {Levi}, Michael E. and {Magneville}, Christophe and {Manser}, Christopher J. and {Martini}, Paul and {Meisner}, Aaron M. and {Miquel}, Ramon and {Moustakas}, John and {Newman}, Jeffrey A. and {Nie}, Jundan and {Palanque-Delabrouille}, Nathalie. and {Percival}, Will J. and {Poppett}, Claire and {Rockosi}, Constance and {Ross}, Ashley J. and {Rossi}, Graziano and {Tarl{\'e}}, Gregory and {Weaver}, Benjamin A. and {Y{\`e}che}, Christophe and {Zhou}, Rongpu and {DESI Collaboration}},
        title = "{Survey Operations for the Dark Energy Spectroscopic Instrument}",
      journal = {\aj},
         year = 2023,
        month = dec,
       volume = {166},
       number = {6},
          eid = {259},
        pages = {259},
          doi = {10.3847/1538-3881/ad0832},
archivePrefix = {arXiv},
       eprint = {2306.06309},
 primaryClass = {astro-ph.CO},
       adsurl = {https://ui.adsabs.harvard.edu/abs/2023AJ....166..259S}
}

@ARTICLE{DESI.DR1.I.Presentation,
       author = {{DESI Collaboration} and {Abdul-Karim}, M. and {Adame}, A.~G. and {Aguado}, D. and {Aguilar}, J. and {Ahlen}, S. and {Alam}, S. and {Aldering}, G. and {Alexander}, D.~M. and {Alfarsy}, R. and {Allen}, L. and {Allende Prieto}, C. and {Alves}, O. and {Anand}, A. and {Andrade}, U. and {Armengaud}, E. and {Avila}, S. and {Aviles}, A. and {Awan}, H. and {Bailey}, S. and {Baleato Lizancos}, A. and {Ballester}, O. and {Bault}, A. and {Bautista}, J. and {BenZvi}, S. and {Beraldo e Silva}, L. and {Bermejo-Climent}, J.~R. and {Beutler}, F. and {Bianchi}, D. and {Blake}, C. and {Blum}, R. and {Bolton}, A.~S. and {Bonici}, M. and {Brieden}, S. and {Brodzeller}, A. and {Brooks}, D. and {Buckley-Geer}, E. and {Burtin}, E. and {Canning}, R. and {Carnero Rosell}, A. and {Carr}, A. and {Carrilho}, P. and {Casas}, L. and {Castander}, F.~J. and {Cereskaite}, R. and {Cervantes-Cota}, J.~L. and {Chaussidon}, E. and {Chaves-Montero}, J. and {Chen}, S. and {Chen}, X. and {Claybaugh}, T. and {Cole}, S. and {Cooper}, A.~P. and {Cousinou}, M. -C. and {Cuceu}, A. and {Davis}, T.~M. and {Dawson}, K.~S. and {de Belsunce}, R. and {de la Cruz}, R. and {de la Macorra}, A. and {de Mattia}, A. and {Deiosso}, N. and {Della Costa}, J. and {Demina}, R. and {Demirbozan}, U. and {DeRose}, J. and {Dey}, A. and {Dey}, B. and {Ding}, J. and {Ding}, Z. and {Doel}, P. and {Douglass}, K. and {Dowicz}, M. and {Ebina}, H. and {Edelstein}, J. and {Eisenstein}, D.~J. and {Elbers}, W. and {Emas}, N. and {Escoffier}, S. and {Fagrelius}, P. and {Fan}, X. and {Fanning}, K. and {Fawcett}, V.~A. and {Fern\textbackslash'andez-Garc\textbackslash'ia}, E. and {Ferraro}, S. and {Findlay}, N. and {Font-Ribera}, A. and {Forero-Romero}, J.~E. and {Forero-S\textbackslash'anchez}, D. and {Frenk}, C.~S. and {G\textbackslash''ansicke}, B.~T. and {Galbany}, L. and {Garc\textbackslash'ia-Bellido}, J. and {Garcia-Quintero}, C. and {Garrison}, L.~H. and {Gazta\textbackslash\raisebox{-0.5ex}\textasciitildenaga}, E. and {Gil-Mar\textbackslash'in}, H. and {Gnedin}, O.~Y. and {Gontcho}, S. Gontcho A and {Gonzalez-Morales}, A.~X. and {Gonzalez-Perez}, V. and {Gordon}, C. and {Graur}, O. and {Green}, D. and {Gruen}, D. and {Gsponer}, R. and {Guandalin}, C. and {Gutierrez}, G. and {Guy}, J. and {Hahn}, C. and {Han}, J.~J. and {Han}, J. and {He}, S. and {Herrera-Alcantar}, H.~K. and {Honscheid}, K. and {Hou}, J. and {Howlett}, C. and {Huterer}, D. and {Ir\textbackslashv\{s\}i\textbackslashv\{c\}}, V. and {Ishak}, M. and {Jacques}, A. and {Jimenez}, J. and {Jing}, Y.~P. and {Joachimi}, B. and {Joudaki}, S. and {Joyce}, R. and {Jullo}, E. and {Juneau}, S. and {Kara\textbackslashc\{c\}ayl\{\textbackslashi\}}, N.~G. and {Karim}, T. and {Kehoe}, R. and {Kent}, S. and {Khederlarian}, A. and {Kirkby}, D. and {Kisner}, T. and {Kitaura}, F. -S. and {Kizhuprakkat}, N. and {Kong}, H. and {Koposov}, S.~E. and {Kremin}, A. and {Krolewski}, A. and {Lahav}, O. and {Lai}, Y. and {Lamman}, C. and {Lan}, T. -W. and {Landriau}, M. and {Lang}, D. and {Lange}, J.~U. and {Lasker}, J. and {Le Goff}, J.~M. and {Le Guillou}, L. and {Leauthaud}, A. and {Levi}, M.~E. and {Li}, S. and {Li}, T.~S. and {Lodha}, K. and {Lokken}, M. and {Luo}, Y. and {Magneville}, C. and {Manera}, M. and {Manser}, C.~J. and {Margala}, D. and {Martini}, P. and {Maus}, M. and {McCullough}, J. and {McDonald}, P. and {Medina}, G.~E. and {Medina-Varela}, L. and {Meisner}, A. and {Mena-Fern\textbackslash'andez}, J. and {Menegas}, A. and {Mezcua}, M. and {Miquel}, R. and {Montero-Camacho}, P. and {Moon}, J. and {Moustakas}, J. and {Mu\textbackslash\raisebox{-0.5ex}\textasciitildenoz-Guti\textbackslash'errez}, A. and {Mu\textbackslash\raisebox{-0.5ex}\textasciitildenoz-Santos}, D. and {Myers}, A.~D. and {Myles}, J. and {Nadathur}, S. and {Najita}, J. and {Napolitano}, L. and {Newman}, J.~A. and {Nikakhtar}, F. and {Nikutta}, R. and {Niz}, G. and {Noriega}, H.~E. and {Padmanabhan}, N. and {Paillas}, E. and {Palanque-Delabrouille}, N. and {Palmese}, A. and {Pan}, J. and {Pan}, Z. and {Parkinson}, D. and {Peacock}, J. and {Percival}, W.~J. and {P\textbackslash'erez-Fern\textbackslash'andez}, A. and {P\textbackslash'erez-R\textbackslash`afols}, I. and {Peterson}, P.},
        title = "{Data Release 1 of the Dark Energy Spectroscopic Instrument}",
      journal = {arXiv e-prints},
         year = 2025,
        month = mar,
          eid = {arXiv:2503.14745},
        pages = {arXiv:2503.14745},
          doi = {10.48550/arXiv.2503.14745},
archivePrefix = {arXiv},
       eprint = {2503.14745},
 primaryClass = {astro-ph.CO},
       adsurl = {https://ui.adsabs.harvard.edu/abs/2025arXiv250314745D}
}

@unpublished{DESI.DR2.Presentation.InPrep,
      author    = {{DESI Collaboration}},
      title     = "{Data Release 2 of the Dark Energy Spectroscopic Instrument}",
}

@ARTICLE{DESI2024.VII.KP7B,
       author = {{DESI Collaboration} and {Adame}, A.~G. and {Aguilar}, J. and {Ahlen}, S. and {Alam}, S. and {Alexander}, D.~M. and {Allende Prieto}, C. and {Alvarez}, M. and {Alves}, O. and {Anand}, A. and {Andrade}, U. and {Armengaud}, E. and {Avila}, S. and {Aviles}, A. and {Awan}, H. and {Bahr-Kalus}, B. and {Bailey}, S. and {Baltay}, C. and {Bault}, A. and {Behera}, J. and {BenZvi}, S. and {Beutler}, F. and {Bianchi}, D. and {Blake}, C. and {Blum}, R. and {Bonici}, M. and {Brieden}, S. and {Brodzeller}, A. and {Brooks}, D. and {Buckley-Geer}, E. and {Burtin}, E. and {Calderon}, R. and {Canning}, R. and {Carnero Rosell}, A. and {Cereskaite}, R. and {Cervantes-Cota}, J.~L. and {Chabanier}, S. and {Chaussidon}, E. and {Chaves-Montero}, J. and {Chebat}, D. and {Chen}, S. and {Chen}, X. and {Claybaugh}, T. and {Cole}, S. and {Cuceu}, A. and {Davis}, T.~M. and {Dawson}, K. and {de la Macorra}, A. and {de Mattia}, A. and {Deiosso}, N. and {Dey}, A. and {Dey}, B. and {Ding}, Z. and {Doel}, P. and {Edelstein}, J. and {Eftekharzadeh}, S. and {Eisenstein}, D.~J. and {Elbers}, W. and {Elliott}, A. and {Fagrelius}, P. and {Fanning}, K. and {Ferraro}, S. and {Ereza}, J. and {Findlay}, N. and {Flaugher}, B. and {Font-Ribera}, A. and {Forero-S{\'a}nchez}, D. and {Forero-Romero}, J.~E. and {Frenk}, C.~S. and {Garcia-Quintero}, C. and {Garrison}, L.~H. and {Gazta{\~n}aga}, E. and {Gil-Mar{\'\i}n}, H. and {Gontcho}, S. Gontcho A and {Gonzalez-Morales}, A.~X. and {Gonzalez-Perez}, V. and {Gordon}, C. and {Green}, D. and {Gruen}, D. and {Gsponer}, R. and {Gutierrez}, G. and {Guy}, J. and {Hadzhiyska}, B. and {Hahn}, C. and {Hanif}, M.~M. S and {Herrera-Alcantar}, H.~K. and {Honscheid}, K. and {Howlett}, C. and {Huterer}, D. and {Ir{\v{s}}i{\v{c}}}, V. and {Ishak}, M. and {Joyce}, R. and {Juneau}, S. and {Kara{\c{c}}ayl{\i}}, N.~G. and {Kehoe}, R. and {Kent}, S. and {Kirkby}, D. and {Kong}, H. and {Koposov}, S.~E. and {Kremin}, A. and {Krolewski}, A. and {Lahav}, O. and {Lai}, Y. and {Lan}, T. -W. and {Landriau}, M. and {Lang}, D. and {Lasker}, J. and {Le Goff}, J.~M. and {Le Guillou}, L. and {Leauthaud}, A. and {Levi}, M.~E. and {Li}, T.~S. and {Lodha}, K. and {Magneville}, C. and {Manera}, M. and {Margala}, D. and {Martini}, P. and {Matthewson}, W. and {Maus}, M. and {McDonald}, P. and {Medina-Varela}, L. and {Meisner}, A. and {Mena-Fern{\'a}ndez}, J. and {Miquel}, R. and {Moon}, J. and {Moore}, S. and {Moustakas}, J. and {Mudur}, N. and {Mueller}, E. and {Mu{\~n}oz-Guti{\'e}rrez}, A. and {Myers}, A.~D. and {Nadathur}, S. and {Napolitano}, L. and {Neveux}, R. and {Newman}, J.~A. and {Nguyen}, N.~M. and {Nie}, J. and {Niz}, G. and {Noriega}, H.~E. and {Padmanabhan}, N. and {Paillas}, E. and {Palanque-Delabrouille}, N. and {Pan}, J. and {Penmetsa}, S. and {Percival}, W.~J. and {Pieri}, M.~M. and {Pinon}, M. and {Poppett}, C. and {Porredon}, A. and {Prada}, F. and {P{\'e}rez-Fern{\'a}ndez}, A. and {P{\'e}rez-R{\`a}fols}, I. and {Rabinowitz}, D. and {Raichoor}, A. and {Ram{\'\i}rez-P{\'e}rez}, C. and {Ramirez-Solano}, S. and {Rashkovetskyi}, M. and {Ravoux}, C. and {Rezaie}, M. and {Rich}, J. and {Rocher}, A. and {Rockosi}, C. and {Roe}, N.~A. and {Rosado-Marin}, A. and {Ross}, A.~J. and {Rossi}, G. and {Ruggeri}, R. and {Ruhlmann-Kleider}, V. and {Samushia}, L. and {Sanchez}, E. and {Saulder}, C. and {Schlafly}, E.~F. and {Schlegel}, D. and {Schubnell}, M. and {Seo}, H. and {Shafieloo}, A. and {Sharples}, R. and {Silber}, J. and {Slosar}, A. and {Smith}, A. and {Sprayberry}, D. and {Tan}, T. and {Tarl{\'e}}, G. and {Taylor}, P. and {Trusov}, S. and {Vaisakh}, R. and {Valcin}, D. and {Valdes}, F. and {Valogiannis}, G. and {Vargas-Maga{\~n}a}, M. and {Verde}, L. and {Walther}, M. and {Wang}, B. and {Wang}, M.~S. and {Weaver}, B.~A. and {Weaverdyck}, N. and {Wechsler}, R.~H. and {Weinberg}, D.~H. and {White}, M. and {Wilson}, M.~J.},
        title = "{DESI 2024 VII: Cosmological Constraints from the Full-Shape Modeling of Clustering Measurements}",
      journal = {arXiv e-prints},
         year = 2024,
        month = nov,
          eid = {arXiv:2411.12022},
        pages = {arXiv:2411.12022},
          doi = {10.48550/arXiv.2411.12022},
archivePrefix = {arXiv},
       eprint = {2411.12022},
 primaryClass = {astro-ph.CO},
       adsurl = {https://ui.adsabs.harvard.edu/abs/2024arXiv241112022D}
}

@ARTICLE{DESI.DR2.II.BAO,
       author = {{DESI Collaboration} and {Abdul-Karim}, M. and {Aguilar}, J. and {Ahlen}, S. and {Alam}, S. and {Allen}, L. and {Allende Prieto}, C. and {Alves}, O. and {Anand}, A. and {Andrade}, U. and {Armengaud}, E. and {Aviles}, A. and {Bailey}, S. and {Baltay}, C. and {Bansal}, P. and {Bault}, A. and {Behera}, J. and {BenZvi}, S. and {Bianchi}, D. and {Blake}, C. and {Brieden}, S. and {Brodzeller}, A. and {Brooks}, D. and {Buckley-Geer}, E. and {Burtin}, E. and {Calderon}, R. and {Canning}, R. and {Carnero Rosell}, A. and {Carrilho}, P. and {Casas}, L. and {Castander}, F.~J. and {Cereskaite}, R. and {Charles}, M. and {Chaussidon}, E. and {Chaves-Montero}, J. and {Chebat}, D. and {Chen}, X. and {Claybaugh}, T. and {Cole}, S. and {Cooper}, A.~P. and {Cuceu}, A. and {Dawson}, K.~S. and {de la Macorra}, A. and {de Mattia}, A. and {Deiosso}, N. and {Della Costa}, J. and {Demina}, R. and {Dey}, A. and {Dey}, B. and {Ding}, Z. and {Doel}, P. and {Edelstein}, J. and {Eisenstein}, D.~J. and {Elbers}, W. and {Fagrelius}, P. and {Fanning}, K. and {Fern\textbackslash'andez-Garc\textbackslash'ia}, E. and {Ferraro}, S. and {Font-Ribera}, A. and {Forero-Romero}, J.~E. and {Frenk}, C.~S. and {Garcia-Quintero}, C. and {Garrison}, L.~H. and {Gazta\textbackslash\raisebox{-0.5ex}\textasciitildenaga}, E. and {Gil-Mar\textbackslash'in}, H. and {Gontcho}, S. Gontcho A and {Gonzalez}, D. and {Gonzalez-Morales}, A.~X. and {Gordon}, C. and {Green}, D. and {Gutierrez}, G. and {Guy}, J. and {Hadzhiyska}, B. and {Hahn}, C. and {He}, S. and {Herbold}, M. and {Herrera-Alcantar}, H.~K. and {Ho}, M. and {Honscheid}, K. and {Howlett}, C. and {Huterer}, D. and {Ishak}, M. and {Juneau}, S. and {Kamble}, N.~V. and {Kara\textbackslashc\{c\}ayl\{\textbackslashi\}}, N.~G. and {Kehoe}, R. and {Kent}, S. and {Kim}, A.~G. and {Kirkby}, D. and {Kisner}, T. and {Koposov}, S.~E. and {Kremin}, A. and {Krolewski}, A. and {Lahav}, O. and {Lamman}, C. and {Landriau}, M. and {Lang}, D. and {Lasker}, J. and {Le Goff}, J.~M. and {Le Guillou}, L. and {Leauthaud}, A. and {Levi}, M.~E. and {Li}, Q. and {Li}, T.~S. and {Lodha}, K. and {Lokken}, M. and {Lozano-Rodr\textbackslash'iguez}, F. and {Magneville}, C. and {Manera}, M. and {Martini}, P. and {Matthewson}, W.~L. and {Meisner}, A. and {Mena-Fern\textbackslash'andez}, J. and {Menegas}, A. and {Mergulh\textbackslash\raisebox{-0.5ex}\textasciitildeao}, T. and {Miquel}, R. and {Moustakas}, J. and {Mu\textbackslash\raisebox{-0.5ex}\textasciitildenoz-Guti\textbackslash'errez}, A. and {Mu\textbackslash\raisebox{-0.5ex}\textasciitildenoz-Santos}, D. and {Myers}, A.~D. and {Nadathur}, S. and {Naidoo}, K. and {Napolitano}, L. and {Newman}, J.~A. and {Niz}, G. and {Noriega}, H.~E. and {Paillas}, E. and {Palanque-Delabrouille}, N. and {Pan}, J. and {Peacock}, J. and {Pellejero Ibanez}, Marcos and {Percival}, W.~J. and {P\textbackslash'erez-Fern\textbackslash'andez}, A. and {P\textbackslash'erez-R\textbackslash`afols}, I. and {Pieri}, M.~M. and {Poppett}, C. and {Prada}, F. and {Rabinowitz}, D. and {Raichoor}, A. and {Ram\textbackslash'irez-P\textbackslash'erez}, C. and {Rashkovetskyi}, M. and {Ravoux}, C. and {Rich}, J. and {Rocher}, A. and {Rockosi}, C. and {Rohlf}, J. and {Rom\textbackslash'an-Herrera}, J.~O. and {Ross}, A.~J. and {Rossi}, G. and {Ruggeri}, R. and {Ruhlmann-Kleider}, V. and {Samushia}, L. and {Sanchez}, E. and {Sanders}, N. and {Schlegel}, D. and {Schubnell}, M. and {Seo}, H. and {Shafieloo}, A. and {Sharples}, R. and {Silber}, J. and {Sinigaglia}, F. and {Sprayberry}, D. and {Tan}, T. and {Tarl\textbackslash'e}, G. and {Taylor}, P. and {Turner}, W. and {Ure\textbackslash\raisebox{-0.5ex}\textasciitildena-L\textbackslash'opez}, L.~A. and {Vaisakh}, R. and {Valdes}, F. and {Valogiannis}, G. and {Vargas-Maga\textbackslash\raisebox{-0.5ex}\textasciitildena}, M. and {Verde}, L. and {Walther}, M. and {Weaver}, B.~A. and {Weinberg}, D.~H. and {White}, M. and {Wolfson}, M. and {Y\textbackslash`eche}, C. and {Yu}, J. and {Zaborowski}, E.~A. and {Zarrouk}, P. and {Zhai}, Z. and {Zhang}, H. and {Zhao}, C. and {Zhao}, G.~B. and {Zhou}, R. and {Zou}, H.},
        title = "{DESI DR2 Results II: Measurements of Baryon Acoustic Oscillations and Cosmological Constraints}",
      journal = {arXiv e-prints},
         year = 2025,
        month = mar,
          eid = {arXiv:2503.14738},
        pages = {arXiv:2503.14738},
          doi = {10.48550/arXiv.2503.14738},
archivePrefix = {arXiv},
       eprint = {2503.14738},
 primaryClass = {astro-ph.CO},
       adsurl = {https://ui.adsabs.harvard.edu/abs/2025arXiv250314738D}
}

@ARTICLE{DESIConstructionLSS2024,
       author = {{Ross}, A.~J. and {Aguilar}, J. and {Ahlen}, S. and {Alam}, S. and {Anand}, A. and {Bailey}, S. and {Bianchi}, D. and {Brieden}, S. and {Brooks}, D. and {Burtin}, E. and {Carnero Rosell}, A. and {Chaussidon}, E. and {Claybaugh}, T. and {Cole}, S. and {Dawson}, K. and {de la Macorra}, A. and {de Mattia}, A. and {Dey}, A. and {Dey}, B. and {Doel}, P. and {Fanning}, K. and {Ferraro}, S. and {Ereza}, J. and {Font-Ribera}, A. and {Forero-Romero}, J.~E. and {Gazta{\~n}aga}, E. and {Gil-Mar{\'\i}n}, H. and {Gontcho A Gontcho}, S. and {Gonzalez-Morales}, A.~X. and {Guy}, J. and {Hahn}, C. and {Heydenreich}, S. and {Honscheid}, K. and {Howlett}, C. and {Ishak}, M. and {Karim}, T. and {Kirkby}, D. and {Kisner}, T. and {Kong}, H. and {Kremin}, A. and {Krolewski}, A. and {Lambert}, A. and {Landriau}, M. and {Lasker}, J. and {Guillou}, L.~L. and {Levi}, M.~E. and {Manera}, M. and {Martini}, P. and {McDonald}, P. and {Meisner}, A. and {Miquel}, R. and {Moon}, J. and {Moustakas}, J. and {Mu{\~n}oz-Guti{\'e}rrez}, A. and {Myers}, A.~D. and {Nadathur}, S. and {Napolitano}, L. and {Newman}, J.~A. and {Nie}, J. and {Niz}, G. and {Palanque-Delabrouille}, N. and {Percival}, W.~J. and {Poppett}, C. and {Prada}, F. and {Raichoor}, A. and {Ravoux}, C. and {Rezaie}, M. and {Rosado-Marin}, A. and {Rossi}, G. and {Samushia}, L. and {Sanchez}, E. and {Schlafly}, E.~F. and {Schlegel}, D. and {Seo}, H. and {Smith}, A. and {Sprayberry}, D. and {Tarl{\'e}}, G. and {Valcin}, D. and {Vargas-Maga{\~n}a}, M. and {Weaver}, B.~A. and {Wilson}, M.~J. and {Yu}, J. and {Zarrouk}, P. and {Zhao}, C. and {Zhou}, R. and {Zou}, H.},
        title = "{The construction of large-scale structure catalogs for the Dark Energy Spectroscopic Instrument}",
      journal = {\jcap},
         year = 2025,
        month = jan,
       volume = {2025},
       number = {1},
          eid = {125},
        pages = {125},
          doi = {10.1088/1475-7516/2025/01/125},
archivePrefix = {arXiv},
       eprint = {2405.16593},
 primaryClass = {astro-ph.CO},
       adsurl = {https://ui.adsabs.harvard.edu/abs/2025JCAP...01..125R}
}

@ARTICLE{Chaussidon2025QSOLRGPNG,
      title={Constraining primordial non-Gaussianity with DESI 2024 LRG and QSO samples}, 
      author={E. Chaussidon and C. Yèche and A. de Mattia and C. Payerne and P. McDonald and A. J. Ross and S. Ahlen and D. Bianchi and D. Brooks and E. Burtin and T. Claybaugh and A. de la Macorra and P. Doel and S. Ferraro and A. Font-Ribera and J. E. Forero-Romero and E. Gaztañaga and H. Gil-Marín and S. Gontcho A Gontcho and G. Gutierrez and J. Guy and K. Honscheid and C. Howlett and D. Huterer and R. Kehoe and D. Kirkby and T. Kisner and A. Kremin and L. Le Guillou and M. E. Levi and M. Manera and A. Meisner and R. Miquel and J. Moustakas and J. A. Newman and G. Niz and N. Palanque-Delabrouille and W. J. Percival and F. Prada and I. Pérez-Ràfols and C. Ravoux and G. Rossi and E. Sanchez and D. Schlegel and M. Schubnell and H. Seo and D. Sprayberry and G. Tarlé and M. Vargas-Magaña and B. A. Weaver and C. Zhao and H. Zou},
      year={2025},
      eprint={2411.17623},
      archivePrefix={arXiv},
      primaryClass={astro-ph.CO},
      url={https://arxiv.org/abs/2411.17623}, 
}

@ARTICLE{DESI.Selection.BGS.Hahn2023,
   title={The DESI Bright Galaxy Survey: Final Target Selection, Design, and Validation},
   volume={165},
   ISSN={1538-3881},
   url={http://dx.doi.org/10.3847/1538-3881/accff8},
   DOI={10.3847/1538-3881/accff8},
   number={6},
   journal={\aj},
   publisher={American Astronomical Society},
   author={Hahn, ChangHoon and Wilson, Michael J. and Ruiz-Macias, Omar and Cole, Shaun and Weinberg, David H. and Moustakas, John and Kremin, Anthony and Tinker, Jeremy L. and Smith, Alex and Wechsler, Risa H. and Ahlen, Steven and Alam, Shadab and Bailey, Stephen and Brooks, David and Cooper, Andrew P. and Davis, Tamara M. and Dawson, Kyle and Dey, Arjun and Dey, Biprateep and Eftekharzadeh, Sarah and Eisenstein, Daniel J. and Fanning, Kevin and Forero-Romero, Jaime E. and Frenk, Carlos S. and Gaztañaga, Enrique and A Gontcho, Satya Gontcho and Guy, Julien and Honscheid, Klaus and Ishak, Mustapha and Juneau, Stéphanie and Kehoe, Robert and Kisner, Theodore and Lan, Ting-Wen and Landriau, Martin and Le Guillou, Laurent and Levi, Michael E. and Magneville, Christophe and Martini, Paul and Meisner, Aaron and Myers, Adam D. and Nie, Jundan and Norberg, Peder and Palanque-Delabrouille, Nathalie and Percival, Will J. and Poppett, Claire and Prada, Francisco and Raichoor, Anand and Ross, Ashley J. and Gaines, Sasha and Saulder, Christoph and Schlafly, Eddie and Schlegel, David and Sierra-Porta, David and Tarle, Gregory and Weaver, Benjamin A. and Yèche, Christophe and Zarrouk, Pauline and Zhou, Rongpu and Zhou, Zhimin and Zou, Hu},
   year={2023},
   month=may, pages={253} }

@ARTICLE{DESI.Selection.LRG.Zhou2023,
   title={Target Selection and Validation of DESI Luminous Red Galaxies},
   volume={165},
   ISSN={1538-3881},
   url={http://dx.doi.org/10.3847/1538-3881/aca5fb},
   DOI={10.3847/1538-3881/aca5fb},
   number={2},
   journal={\aj},
   publisher={American Astronomical Society},
   author={Zhou, Rongpu and Dey, Biprateep and Newman, Jeffrey A. and Eisenstein, Daniel J. and Dawson, K. and Bailey, S. and Berti, A. and Guy, J. and Lan, Ting-Wen and Zou, H. and Aguilar, J. and Ahlen, S. and Alam, Shadab and Brooks, D. and de la Macorra, A. and Dey, A. and Dhungana, G. and Fanning, K. and Font-Ribera, A. and Gontcho, S. Gontcho A. and Honscheid, K. and Ishak, Mustapha and Kisner, T. and Kovács, A. and Kremin, A. and Landriau, M. and Levi, Michael E. and Magneville, C. and Manera, Marc and Martini, P. and Meisner, Aaron M. and Miquel, R. and Moustakas, J. and Myers, Adam D. and Nie, Jundan and Palanque-Delabrouille, N. and Percival, W. J. and Poppett, C. and Prada, F. and Raichoor, A. and Ross, A. J. and Schlafly, E. and Schlegel, D. and Schubnell, M. and Tarlé, Gregory and Weaver, B. A. and Wechsler, R. H. and Yèche, Christophe and Zhou, Zhimin},
   year={2023},
   month=jan, pages={58} }

@ARTICLE{DESI.Selection.ELG.Raichoor2023,
   title={Target Selection and Validation of DESI Emission Line Galaxies},
   volume={165},
   ISSN={1538-3881},
   url={http://dx.doi.org/10.3847/1538-3881/acb213},
   DOI={10.3847/1538-3881/acb213},
   number={3},
   journal={\aj},
   publisher={American Astronomical Society},
   author={Raichoor, A. and Moustakas, J. and Newman, Jeffrey A. and Karim, T. and Ahlen, S. and Alam, Shadab and Bailey, S. and Brooks, D. and Dawson, K. and de la Macorra, A. and de Mattia, A. and Dey, A. and Dey, Biprateep and Dhungana, G. and Eftekharzadeh, S. and Eisenstein, D. J. and Fanning, K. and Font-Ribera, A. and García-Bellido, J. and Gaztañaga, E. and A Gontcho, S. Gontcho and Guy, J. and Honscheid, K. and Ishak, M. and Kehoe, R. and Kisner, T. and Kremin, Anthony and Lan, Ting-Wen and Landriau, M. and Le Guillou, L. and Levi, Michael E. and Magneville, C. and Manera, M. and Martini, P. and Meisner, Aaron M. and Myers, Adam D. and Nie, Jundan and Palanque-Delabrouille, N. and Percival, W. J. and Poppett, C. and Prada, F. and Ross, A. J. and Ruhlmann-Kleider, V. and Sabiu, C. G. and Schlafly, E. F. and Schlegel, D. and Tarlé, Gregory and Weaver, B. A. and Yèche, Christophe and Zhou, Rongpu and Zhou, Zhimin and Zou, H.},
   year={2023},
   month=feb, pages={126} }

@ARTICLE{DESI.Selection.QSO.Chaussidon2023,
   title={Target Selection and Validation of DESI Quasars},
   volume={944},
   ISSN={1538-4357},
   url={http://dx.doi.org/10.3847/1538-4357/acb3c2},
   DOI={10.3847/1538-4357/acb3c2},
   number={1},
   journal={The Astrophysical Journal},
   publisher={American Astronomical Society},
   author={Chaussidon, Edmond and Yèche, Christophe and Palanque-Delabrouille, Nathalie and Alexander, David M. and Yang, Jinyi and Ahlen, Steven and Bailey, Stephen and Brooks, David and Cai, Zheng and Chabanier, Solène and Davis, Tamara M. and Dawson, Kyle and de laMacorra, Axel and Dey, Arjun and Dey, Biprateep and Eftekharzadeh, Sarah and Eisenstein, Daniel J. and Fanning, Kevin and Font-Ribera, Andreu and Gaztañaga, Enrique and A Gontcho, Satya Gontcho and Gonzalez-Morales, Alma X. and Guy, Julien and Herrera-Alcantar, Hiram K. and Honscheid, Klaus and Ishak, Mustapha and Jiang, Linhua and Juneau, Stephanie and Kehoe, Robert and Kisner, Theodore and Kovács, Andras and Kremin, Anthony and Lan, Ting-Wen and Landriau, Martin and Le Guillou, Laurent and Levi, Michael E. and Magneville, Christophe and Martini, Paul and Meisner, Aaron M. and Moustakas, John and Muñoz-Gutiérrez, Andrea and Myers, Adam D. and Newman, Jeffrey A. and Nie, Jundan and Percival, Will J. and Poppett, Claire and Prada, Francisco and Raichoor, Anand and Ravoux, Corentin and Ross, Ashley J. and Schlafly, Edward and Schlegel, David and Tan, Ting and Tarlé, Gregory and Zhou, Rongpu and Zhou, Zhimin and Zou, Hu},
   year={2023},
   month=feb, pages={107} }

@ARTICLE{DESI.LRGCross.Magnifcation.Zhou2023,
   title={DESI luminous red galaxy samples for cross-correlations},
   volume={2023},
   ISSN={1475-7516},
   url={http://dx.doi.org/10.1088/1475-7516/2023/11/097},
   DOI={10.1088/1475-7516/2023/11/097},
   number={11},
   journal={Journal of Cosmology and Astroparticle Physics},
   publisher={IOP Publishing},
   author={Zhou, Rongpu and Ferraro, Simone and White, Martin and DeRose, Joseph and Sailer, Noah and Aguilar, Jessica and Ahlen, Steven and Bailey, Stephen and Brooks, David and Claybaugh, Todd and Dawson, Kyle and de la Macorra, Axel and Dey, Biprateep and Doel, Peter and Font-Ribera, Andreu and Forero-Romero, Jaime E. and Gontcho A Gontcho, Satya and Guy, Julien and Kremin, Anthony and Lambert, Andrew and Le Guillou, Laurent and Levi, Michael and Magneville, Christophe and Manera, Marc and Meisner, Aaron and Miquel, Ramon and Moustakas, John and Myers, Adam D. and Newman, Jeffrey A. and Nie, Jundan and Percival, Will and Rezaie, Mehdi and Rossi, Graziano and Sanchez, Eusebio and Schlegel, David and Schubnell, Michael and Seo, Hee-Jong and Tarlé, Gregory and Zhou, Zhimin},
   year={2023},
   month=nov, pages={097} }

@ARTICLE{Sailer2025EvolutionStructureMagnificationBiasBGS,
      title={Evolution of structure growth during dark energy domination: Insights from the cross-correlation of DESI galaxies with CMB lensing and galaxy magnification}, 
      author={Noah Sailer and Joseph DeRose and Simone Ferraro and Shi-Fan Chen and Rongpu Zhou and Martin White and Joshua Kim and Mathew Madhavacheril},
      year={2025},
      eprint={2503.24385},
      archivePrefix={arXiv},
      primaryClass={astro-ph.CO},
      url={https://arxiv.org/abs/2503.24385}, 
}

@ARTICLE{DESI.2024.II.SampleDef,
   title={DESI 2024 II: sample definitions, characteristics, and two-point clustering statistics},
   volume={2025},
   ISSN={1475-7516},
   url={http://dx.doi.org/10.1088/1475-7516/2025/07/017},
   DOI={10.1088/1475-7516/2025/07/017},
   number={07},
   journal={Journal of Cosmology and Astroparticle Physics},
   publisher={IOP Publishing},
   author={Adame, A.G. and Aguilar, J. and Ahlen, S. and Alam, S. and Alexander, D.M. and Alvarez, M. and Alves, O. and Anand, A. and Andrade, U. and Armengaud, E. and Avila, S. and Aviles, A. and Awan, H. and Bailey, S. and Baltay, C. and Bault, A. and Behera, J. and BenZvi, S. and Beutler, F. and Bianchi, D. and Blake, C. and Blum, R. and Brieden, S. and Brodzeller, A. and Brooks, D. and Brown, Z. and Buckley-Geer, E. and Burtin, E. and Calderon, R. and Canning, R. and Carnero Rosell, A. and Cereskaite, R. and Cervantes-Cota, J.L. and Chabanier, S. and Chaussidon, E. and Chaves-Montero, J. and Chen, S. and Chen, X. and Claybaugh, T. and Cole, S. and Cuceu, A. and Davis, T.M. and Dawson, K. and de la Macorra, A. and de Mattia, A. and Deiosso, N. and Demina, R. and Dey, A. and Dey, B. and Ding, Z. and Doel, P. and Edelstein, J. and Eftekharzadeh, S. and Eisenstein, D.J. and Elliott, A. and Fagrelius, P. and Fanning, K. and Ferraro, S. and Ereza, J. and Findlay, N. and Flaugher, B. and Font-Ribera, A. and Forero-Sánchez, D. and Forero-Romero, J.E. and Frenk, C.S. and Garcia-Quintero, C. and Gaztañaga, E. and Gil-Marín, H. and Gontcho, S.Gontcho A. and Gonzalez-Morales, A.X. and Gonzalez-Perez, V. and Gordon, C. and Green, D. and Gruen, D. and Gsponer, R. and Gutierrez, G. and Guy, J. and Hadzhiyska, B. and Hahn, C. and Hanif, M.M.S. and Herrera-Alcantar, H.K. and Honscheid, K. and Hou, J. and Howlett, C. and Huterer, D. and Iršič, V. and Ishak, M. and Juneau, S. and Karaçaylı, N.G. and Kehoe, R. and Kent, S. and Kirkby, D. and Kitaura, F.-S. and Kong, H. and Kremin, A. and Krolewski, A. and Lai, Y. and Lan, T.-W. and Landriau, M. and Lang, D. and Lasker, J. and Le Goff, J.M. and Le Guillou, L. and Leauthaud, A. and Levi, M.E. and Li, T.S. and Lodha, K. and Magneville, C. and Manera, M. and Margala, D. and Martini, P. and Maus, M. and McDonald, P. and Medina-Varela, L. and Meisner, A. and Mena-Fernández, J. and Miquel, R. and Moon, J. and Moore, S. and Moustakas, J. and Mudur, N. and Mueller, E. and Muñoz-Gutiérrez, A. and Myers, A.D. and Nadathur, S. and Napolitano, L. and Neveux, R. and Newman, J.A. and Nguyen, N.M. and Nie, J. and Niz, G. and Noriega, H.E. and Padmanabhan, N. and Paillas, E. and Palanque-Delabrouille, N. and Pan, J. and Penmetsa, S. and Percival, W.J. and Pieri, M.M. and Pinon, M. and Poppett, C. and Porredon, A. and Prada, F. and Pérez-Fernández, A. and Pérez-Ràfols, I. and Rabinowitz, D. and Raichoor, A. and Ramírez-Pérez, C. and Ramirez-Solano, S. and Rashkovetskyi, M. and Ravoux, C. and Rezaie, M. and Rich, J. and Rocher, A. and Rockosi, C. and Roe, N.A. and Rosado-Marin, A. and Ross, A.J. and Rossi, G. and Ruggeri, R. and Ruhlmann-Kleider, V. and Samushia, L. and Sanchez, E. and Saulder, C. and Schlafly, E.F. and Schlegel, D. and Scholte, D. and Schubnell, M. and Seo, H. and Sharples, R. and Silber, J. and Slosar, A. and Smith, A. and Sprayberry, D. and Tan, T. and Tarlé, G. and Trusov, S. and Vaisakh, R. and Valcin, D. and Valdes, F. and Vargas-Magaña, M. and Verde, L. and Walther, M. and Wang, B. and Wang, M.S. and Weaver, B.A. and Weaverdyck, N. and Wechsler, R.H. and Weinberg, D.H. and White, M. and Wilson, M.J. and Yu, J. and Yu, Y. and Yuan, S. and Yèche, C. and Zaborowski, E.A. and Zarrouk, P. and Zhang, H. and Zhao, C. and Zhao, R. and Zhou, R. and Zou, H.},
   year={2025},
   month=jul, 
   pages={017} 
}

@ARTICLE{heydenreich2025lensingbordersmeasurementsgalaxygalaxy,
      title={Lensing Without Borders: Measurements of galaxy-galaxy lensing and projected galaxy clustering in DESI DR1}, 
      author={S. Heydenreich and A. Leauthaud and C. Blake and Z. Sun and J. U. Lange and T. Zhang and M. DeMartino and A. J. Ross and J. Aguilar and S. Ahlen and D. Bianchi and D. Brooks and F. J. Castander and T. Claybaugh and A. Cuceu and A. de la Macorra and J. DeRose and Arjun Dey and Biprateep Dey and P. Doel and N. Emas and S. Ferraro and A. Font-Ribera and J. E. Forero-Romero and C. Garcia-Quintero and E. Gaztañaga and S. Gontcho A Gontcho and G. Gutierrez and B. Hadzhiyska and K. Honscheid and D. Huterer and M. Ishak and N. Jeffrey and S. Joudaki and E. Jullo and S. Juneau and D. Kirkby and T. Kisner and A. Kremin and A. Krolewski and O. Lahav and C. Lamman and M. Landriau and L. Le Guillou and M. Manera and A. Meisner and R. Miquel and S. Nadathur and N. Palanque-Delabrouille and W. J. Percival and A. Porredon and F. Prada and I. Pérez-Ràfols and G. Rossi and R. Ruggeri and E. Sanchez and C. Saulder and D. Schlegel and A. Semenaite and J. Silber and D. Sprayberry and G. Tarlé and B. A. Weaver and S. Yuan and P. Zarrouk and R. Zhou and H. Zou},
      year={2025},
      eprint={2506.21677},
      archivePrefix={arXiv},
      primaryClass={astro-ph.CO},
      url={https://arxiv.org/abs/2506.21677}, 
}

@ARTICLE{debelsunce2025QSOmagCMBLensing,
      title={Cosmology from Planck CMB Lensing and DESI DR1 Quasar Tomography}, 
      author={R. de Belsunce and A. Krolewski and S. Chiarenza and E. Chaussidon and S. Ferraro and B. Hadzhiyska and C. Ravoux and N. Sailer and G. Farren and A. Tamone and J. Aguilar and S. Ahlen and D. Bianchi and D. Brooks and T. Claybaugh and A. Cuceu and A. de la Macorra and J. Della Costa and Biprateep Dey and P. Doel and A. Font-Ribera and J. E. Forero-Romero and E. Gaztañaga and S. Gontcho A Gontcho and G. Gutierrez and J. Guy and H. K. Herrera-Alcantar and K. Honscheid and M. Ishak and R. Joyce and S. Juneau and R. Kehoe and D. Kirkby and T. Kisner and A. Kremin and O. Lahav and A. Lambert and C. Lamman and M. Landriau and L. Le Guillou and M. E. Levi and M. Manera and P. Martini and A. Meisner and R. Miquel and S. Nadathur and G. Niz and N. Palanque-Delabrouille and W. J. Percival and F. Prada and I. Pérez-Ràfols and A. J. Ross and G. Rossi and E. Sanchez and D. Schlegel and M. Schubnell and H. Seo and J. Silber and D. Sprayberry and G. Tarlé and B. A. Weaver and R. Zhou and H. Zou},
      year={2025},
      eprint={2506.22416},
      archivePrefix={arXiv},
      primaryClass={astro-ph.CO},
      url={https://arxiv.org/abs/2506.22416}, 
}

@article{ChoppinDeJanvry2025b,
      title={Cosmic Shear constraints from HSC Year 3 with clustering calibration of the tomographic redshift distributions from DESI}, 
      author={J. Choppin de Janvry and B. Dai and S. Gontcho A Gontcho and U. Seljak and T. Zhang},
      year={2025},
      eprint={2511.18134},
      archivePrefix={arXiv},
      primaryClass={astro-ph.CO},
      url={https://arxiv.org/abs/2511.18134}, 
}

@article{DESI.2024.III.FiducialCosmology.BAO,
   title={DESI 2024 III: baryon acoustic oscillations from galaxies and quasars},
   volume={2025},
   ISSN={1475-7516},
   url={http://dx.doi.org/10.1088/1475-7516/2025/04/012},
   DOI={10.1088/1475-7516/2025/04/012},
   number={04},
   journal={Journal of Cosmology and Astroparticle Physics},
   publisher={IOP Publishing},
   author={Adame, A.G. and Aguilar, J. and Ahlen, S. and Alam, S. and Alexander, D.M. and Alvarez, M. and Alves, O. and Anand, A. and Andrade, U. and Armengaud, E. and Avila, S. and Aviles, A. and Awan, H. and Bailey, S. and Baltay, C. and Bault, A. and Behera, J. and BenZvi, S. and Beutler, F. and Bianchi, D. and Blake, C. and Blum, R. and Brieden, S. and Brodzeller, A. and Brooks, D. and Buckley-Geer, E. and Burtin, E. and Calderon, R. and Canning, R. and Carnero Rosell, A. and Cereskaite, R. and Cervantes-Cota, J.L. and Chabanier, S. and Chaussidon, E. and Chaves-Montero, J. and Chen, S. and Chen, X. and Claybaugh, T. and Cole, S. and Cuceu, A. and Davis, T.M. and Dawson, K. and de la Macorra, A. and de Mattia, A. and Deiosso, N. and Dey, A. and Dey, B. and Ding, Z. and Doel, P. and Edelstein, J. and Eftekharzadeh, S. and Eisenstein, D.J. and Elliott, A. and Fagrelius, P. and Fanning, K. and Ferraro, S. and Ereza, J. and Findlay, N. and Flaugher, B. and Font-Ribera, A. and Forero-Sánchez, D. and Forero-Romero, J.E. and Garcia-Quintero, C. and Gaztañaga, E. and Gil-Marín, H. and Gontcho, S.Gontcho A. and Gonzalez-Morales, A.X. and Gonzalez-Perez, V. and Gordon, C. and Green, D. and Gruen, D. and Gsponer, R. and Gutierrez, G. and Guy, J. and Hadzhiyska, B. and Hahn, C. and Hanif, M.M.S. and Herrera-Alcantar, H.K. and Honscheid, K. and Howlett, C. and Huterer, D. and Iršič, V. and Ishak, M. and Juneau, S. and Karaçaylı, N.G. and Kehoe, R. and Kent, S. and Kirkby, D. and Kong, H. and Kremin, A. and Krolewski, A. and Lai, Y. and Lan, T.-W. and Landriau, M. and Lang, D. and Lasker, J. and Le Goff, J.M. and Le Guillou, L. and Leauthaud, A. and Levi, M.E. and Li, T.S. and Linder, E. and Lodha, K. and Magneville, C. and Manera, M. and Margala, D. and Martini, P. and Maus, M. and McDonald, P. and Medina-Varela, L. and Meisner, A. and Mena-Fernández, J. and Miquel, R. and Moon, J. and Moore, S. and Moustakas, J. and Mueller, E. and Muñoz-Gutiérrez, A. and Myers, A.D. and Nadathur, S. and Napolitano, L. and Neveux, R. and Newman, J.A. and Nguyen, N.M. and Nie, J. and Niz, G. and Noriega, H.E. and Padmanabhan, N. and Paillas, E. and Palanque-Delabrouille, N. and Pan, J. and Penmetsa, S. and Percival, W.J. and Pieri, M.M. and Pinon, M. and Poppett, C. and Porredon, A. and Prada, F. and Pérez-Fernández, A. and Pérez-Ràfols, I. and Rabinowitz, D. and Raichoor, A. and Ramírez-Pérez, C. and Ramirez-Solano, S. and Rashkovetskyi, M. and Ravoux, C. and Rezaie, M. and Rich, J. and Rocher, A. and Rockosi, C. and Roe, N.A. and Rosado-Marin, A. and Ross, A.J. and Rossi, G. and Ruggeri, R. and Ruhlmann-Kleider, V. and Samushia, L. and Sanchez, E. and Saulder, C. and Schlafly, E.F. and Schlegel, D. and Schubnell, M. and Seo, H. and Sharples, R. and Silber, J. and Slosar, A. and Smith, A. and Sprayberry, D. and Swanson, J. and Tan, T. and Tarlé, G. and Trusov, S. and Vaisakh, R. and Valcin, D. and Valdes, F. and Vargas-Magaña, M. and Verde, L. and Walther, M. and Wang, B. and Wang, M.S. and Weaver, B.A. and Weaverdyck, N. and Wechsler, R.H. and Weinberg, D.H. and White, M. and Wilson, M.J. and Yu, J. and Yu, Y. and Yuan, S. and Yèche, C. and Zaborowski, E.A. and Zarrouk, P. and Zhang, H. and Zhao, C. and Zhao, R. and Zhou, R. and Zou, H.},
   year={2025},
   month=apr, pages={012} }

@article{Pinon2025FiberAssign,
doi = {10.1088/1475-7516/2025/01/131},
url = {https://dx.doi.org/10.1088/1475-7516/2025/01/131},
year = {2025},
month = {jan},
publisher = {IOP Publishing},
volume = {2025},
number = {01},
pages = {131},
author = {Pinon, M. and de Mattia, A. and McDonald, P. and Burtin, E. and Ruhlmann-Kleider, V. and White, M. and Bianchi, D. and Ross, A.J. and Aguilar, J. and Ahlen, S. and Brooks, D. and Cahn, R.N. and Chaussidon, E. and Claybaugh, T. and Cole, S. and de la Macorra, A. and Dey, B. and Doel, P. and Fanning, K. and Forero-Romero, J.E. and Gaztañaga, E. and Gontcho A Gontcho, S. and Howlett, C. and Kirkby, D. and Kisner, T. and Kremin, A. and Lambert, A. and Landriau, M. and Lasker, J. and Le Guillou, L. and Levi, M.E. and Manera, M. and Martini, P. and Meisner, A. and Miquel, R. and Moustakas, J. and Myers, A.D. and Niz, G. and Palanque-Delabrouille, N. and Percival, W.J. and Poppett, C. and Rossi, G. and Sanchez, E. and Schlegel, D. and Schubnell, M. and Seo, H. and Sprayberry, D. and Tarlé, G. and Vargas-Magaña, M. and Weaver, B.A. and Zarrouk, P. and Zhou, R. and Zou, H.},
title = {Mitigation of DESI fiber assignment incompleteness effect on two-point clustering with small angular scale truncated estimators},
journal = {Journal of Cosmology and Astroparticle Physics}
}

@article{bianchi2025characterizationdesifiberassignment,
      title={Characterization of DESI fiber assignment incompleteness effect on 2-point clustering and mitigation methods for DR1 analysis}, 
      author={D. Bianchi and M. M. S Hanif and A. Carnero Rosell and J. Lasker and A. J. Ross and M. Pinon and A. de Mattia and M. White and S. Ahlen and S. Bailey and D. Brooks and E. Burtin and E. Chaussidon and T. Claybaugh and S. Cole and A. de la Macorra and S. Ferraro and A. Font-Ribera and J. E. Forero-Romero and E. Gaztañaga and S. Gontcho A Gontcho and G. Gutierrez and J. Guy and C. Hahn and K. Honscheid and C. Howlett and S. Juneau and D. Kirkby and T. Kisner and A. Kremin and M. Landriau and L. Le Guillou and M. E. Levi and P. McDonald and A. Meisner and R. Miquel and J. Moustakas and N. Palanque-Delabrouille and W. J. Percival and F. Prada and I. Pérez-Ràfols and A. Raichoor and G. Rossi and E. Sanchez and D. Schlegel and M. Schubnell and R. Sharples and J. Silber and D. Sprayberry and G. Tarlé and M. Vargas-Magaña and B. A. Weaver and P. Zarrouk and R. Zhou and H. Zou},
      year={2025},
      eprint={2411.12025},
      archivePrefix={arXiv},
      primaryClass={astro-ph.CO},
      url={https://arxiv.org/abs/2411.12025}, 
journal = {Journal of Cosmology and Astroparticle Physics},
}

@article{HSCClusteringRau2023,
   title={Weak lensing tomographic redshift distribution inference for the Hyper Suprime-Cam Subaru Strategic Program three-year shape catalogue},
   volume={524},
   ISSN={1365-2966},
   url={http://dx.doi.org/10.1093/mnras/stad1962},
   DOI={10.1093/mnras/stad1962},
   number={4},
   journal={Monthly Notices of the Royal Astronomical Society},
   publisher={Oxford University Press (OUP)},
   author={Rau, Markus Michael and Dalal, Roohi and Zhang, Tianqing and Li, Xiangchong and Nishizawa, Atsushi J and More, Surhud and Mandelbaum, Rachel and Miyatake, Hironao and Strauss, Michael A and Takada, Masahiro},
   year={2023},
   month=jul, pages={5109–5131} }

@article{MizukiHSCTanaka2015,
   title={PHOTOMETRIC REDSHIFT WITH BAYESIAN PRIORS ON PHYSICAL PROPERTIES OF GALAXIES},
   volume={801},
   ISSN={1538-4357},
   url={http://dx.doi.org/10.1088/0004-637X/801/1/20},
   DOI={10.1088/0004-637x/801/1/20},
   number={1},
   journal={The Astrophysical Journal},
   publisher={American Astronomical Society},
   author={Tanaka, Masayuki},
   year={2015},
   month=feb, pages={20} }

@article{HSCPHotozTanaka2017,
   title={Photometric redshifts for Hyper Suprime-Cam Subaru Strategic Program Data Release 1},
   volume={70},
   ISSN={2053-051X},
   url={http://dx.doi.org/10.1093/pasj/psx077},
   DOI={10.1093/pasj/psx077},
   number={SP1},
   journal={Publications of the Astronomical Society of Japan},
   publisher={Oxford University Press (OUP)},
   author={Tanaka, Masayuki and Coupon, Jean and Hsieh, Bau-Ching and Mineo, Sogo and Nishizawa, Atsushi J and Speagle, Joshua and Furusawa, Hisanori and Miyazaki, Satoshi and Murayama, Hitoshi},
   year={2017},
   month=oct }

@article{DNNZHSCNishizawa2020,
      title={Photometric Redshifts for the Hyper Suprime-Cam Subaru Strategic Program Data Release 2}, 
      author={Atsushi J. Nishizawa and Bau-Ching Hsieh and Masayuki Tanaka and Tadafumi Takata},
      year={2020},
      eprint={2003.01511},
      archivePrefix={arXiv},
      primaryClass={astro-ph.GA},
      url={https://arxiv.org/abs/2003.01511}, 
}

@article{HSC2022datarelease3,
   title={Third data release of the Hyper Suprime-Cam Subaru Strategic Program},
   volume={74},
   ISSN={2053-051X},
   url={http://dx.doi.org/10.1093/pasj/psab122},
   DOI={10.1093/pasj/psab122},
   number={2},
   journal={Publications of the Astronomical Society of Japan},
   publisher={Oxford University Press (OUP)},
   author={Aihara, Hiroaki and AlSayyad, Yusra and Ando, Makoto and Armstrong, Robert and Bosch, James and Egami, Eiichi and Furusawa, Hisanori and Furusawa, Junko and Harasawa, Sumiko and Harikane, Yuichi and Hsieh, Bau-Ching and Ikeda, Hiroyuki and Ito, Kei and Iwata, Ikuru and Kodama, Tadayuki and Koike, Michitaro and Kokubo, Mitsuru and Komiyama, Yutaka and Li, Xiangchong and Liang, Yongming and Lin, Yen-Ting and Lupton, Robert H and Lust, Nate B and MacArthur, Lauren A and Mawatari, Ken and Mineo, Sogo and Miyatake, Hironao and Miyazaki, Satoshi and More, Surhud and Morishima, Takahiro and Murayama, Hitoshi and Nakajima, Kimihiko and Nakata, Fumiaki and Nishizawa, Atsushi J and Oguri, Masamune and Okabe, Nobuhiro and Okura, Yuki and Ono, Yoshiaki and Osato, Ken and Ouchi, Masami and Pan, Yen-Chen and Plazas Malagón, Andrés A and Price, Paul A and Reed, Sophie L and Rykoff, Eli S and Shibuya, Takatoshi and Simunovic, Mirko and Strauss, Michael A and Sugimori, Kanako and Suto, Yasushi and Suzuki, Nao and Takada, Masahiro and Takagi, Yuhei and Takata, Tadafumi and Takita, Satoshi and Tanaka, Masayuki and Tang, Shenli and Taranu, Dan S and Terai, Tsuyoshi and Toba, Yoshiki and Turner, Edwin L and Uchiyama, Hisakazu and Vijarnwannaluk, Bovornpratch and Waters, Christopher Z and Yamada, Yoshihiko and Yamamoto, Naoaki and Yamashita, Takuji},
   year={2022},
   month=feb, pages={247–272} 
}

@article{PSFModellingHSCZhang2023,
   title={A general framework for removing point-spread function additive systematics in cosmological weak lensing analysis},
   volume={525},
   ISSN={1365-2966},
   url={http://dx.doi.org/10.1093/mnras/stad1801},
   DOI={10.1093/mnras/stad1801},
   number={2},
   journal={Monthly Notices of the Royal Astronomical Society},
   publisher={Oxford University Press (OUP)},
   author={Zhang, Tianqing and Li, Xiangchong and Dalal, Roohi and Mandelbaum, Rachel and Strauss, Michael A and Kannawadi, Arun and Miyatake, Hironao and Nicola, Andrina and Malagón, Andrés A Plazas and Shirasaki, Masato and Sugiyama, Sunao and Takada, Masahiro and More, Surhud},
   year={2023},
   month=jun, pages={2441–2471} }

@article{HSCShapeCataloguePDR3Li2022,
   title={The three-year shear catalog of the Subaru Hyper Suprime-Cam SSP Survey},
   volume={74},
   ISSN={2053-051X},
   url={http://dx.doi.org/10.1093/pasj/psac006},
   DOI={10.1093/pasj/psac006},
   number={2},
   journal={Publications of the Astronomical Society of Japan},
   publisher={Oxford University Press (OUP)},
   author={Li, Xiangchong and Miyatake, Hironao and Luo, Wentao and More, Surhud and Oguri, Masamune and Hamana, Takashi and Mandelbaum, Rachel and Shirasaki, Masato and Takada, Masahiro and Armstrong, Robert and Kannawadi, Arun and Takita, Satoshi and Miyazaki, Satoshi and Nishizawa, Atsushi J and Plazas Malagon, Andres A and Strauss, Michael A and Tanaka, Masayuki and Yoshida, Naoki},
   year={2022},
   month=mar, pages={421–459} }

@article{CosmoShearLi2023HSCY3,
      title={Hyper Suprime-Cam Year 3 Results: Cosmology from Cosmic Shear Two-point Correlation Functions}, 
      author={Xiangchong Li and Tianqing Zhang and Sunao Sugiyama and Roohi Dalal and Ryo Terasawa and Markus M. Rau and Rachel Mandelbaum and Masahiro Takada and Surhud More and Michael A. Strauss and Hironao Miyatake and Masato Shirasaki and Takashi Hamana and Masamune Oguri and Wentao Luo and Atsushi J. Nishizawa and Ryuichi Takahashi and Andrina Nicola and Ken Osato and Arun Kannawadi and Tomomi Sunayama and Robert Armstrong and James Bosch and Yutaka Komiyama and Robert H. Lupton and Nate B. Lust and Lauren A. MacArthur and Satoshi Miyazaki and Hitoshi Murayama and Takahiro Nishimichi and Yuki Okura and Paul A. Price and Philip J. Tait and Masayuki Tanaka and Shiang-Yu Wang},
      year={2023},
      eprint={2304.00702},
      archivePrefix={arXiv},
      primaryClass={astro-ph.CO},
      url={https://arxiv.org/abs/2304.00702}, 
}

@ARTICLE{Nicola2020TomoclusteringHSCY1,
       author = {{Nicola}, Andrina and {Alonso}, David and {S{\'a}nchez}, Javier and {Slosar}, An{\v{z}}e and {Awan}, Humna and {Broussard}, Adam and {Dunkley}, Jo and {Gawiser}, Eric and {Gomes}, Zahra and {Mandelbaum}, Rachel and {Miyatake}, Hironao and {Newman}, Jeffrey A. and {Sevilla-Noarbe}, Ignacio and {Skinner}, Sarah and {Wagoner}, Erika L.},
        title = "{Tomographic galaxy clustering with the Subaru Hyper Suprime-Cam first year public data release}",
      journal = {\jcap},
         year = 2020,
        month = mar,
       volume = {2020},
       number = {3},
          eid = {044},
        pages = {044},
          doi = {10.1088/1475-7516/2020/03/044},
archivePrefix = {arXiv},
       eprint = {1912.08209},
 primaryClass = {astro-ph.CO},
       adsurl = {https://ui.adsabs.harvard.edu/abs/2020JCAP...03..044N}
}

@article{BoschHSCpipelineY1,
    author = {Bosch, James and Armstrong, Robert and Bickerton, Steven and Furusawa, Hisanori and Ikeda, Hiroyuki and Koike, Michitaro and Lupton, Robert and Mineo, Sogo and Price, Paul and Takata, Tadafumi and Tanaka, Masayuki and Yasuda, Naoki and AlSayyad, Yusra and Becker, Andrew C and Coulton, William and Coupon, Jean and Garmilla, Jose and Huang, Song and Krughoff, K Simon and Lang, Dustin and Leauthaud, Alexie and Lim, Kian-Tat and Lust, Nate B and MacArthur, Lauren A and Mandelbaum, Rachel and Miyatake, Hironao and Miyazaki, Satoshi and Murata, Ryoma and More, Surhud and Okura, Yuki and Owen, Russell and Swinbank, John D and Strauss, Michael A and Yamada, Yoshihiko and Yamanoi, Hitomi},
    title = {The Hyper Suprime-Cam software pipeline},
    journal = {Publications of the Astronomical Society of Japan},
    volume = {70},
    number = {SP1},
    pages = {S5},
    year = {2017},
    month = {10},
    issn = {0004-6264},
    doi = {10.1093/pasj/psx080},
    url = {https://doi.org/10.1093/pasj/psx080},
    eprint = {https://academic.oup.com/pasj/article-pdf/70/SP1/S5/54675369/pasj\_70\_sp1\_s5.pdf},
}

@article{Dalal2023CosmoShearHSC,
   title={Hyper Suprime-Cam Year 3 results: Cosmology from cosmic shear power spectra},
   volume={108},
   ISSN={2470-0029},
   url={http://dx.doi.org/10.1103/PhysRevD.108.123519},
   DOI={10.1103/physrevd.108.123519},
   number={12},
   journal={Physical Review D},
   publisher={American Physical Society (APS)},
   author={Dalal, Roohi and Li, Xiangchong and Nicola, Andrina and Zuntz, Joe and Strauss, Michael A. and Sugiyama, Sunao and Zhang, Tianqing and Rau, Markus M. and Mandelbaum, Rachel and Takada, Masahiro and More, Surhud and Miyatake, Hironao and Kannawadi, Arun and Shirasaki, Masato and Taniguchi, Takanori and Takahashi, Ryuichi and Osato, Ken and Hamana, Takashi and Oguri, Masamune and Nishizawa, Atsushi J. and Malagón, Andrés A. Plazas and Sunayama, Tomomi and Alonso, David and Slosar, Anže and Luo, Wentao and Armstrong, Robert and Bosch, James and Hsieh, Bau-Ching and Komiyama, Yutaka and Lupton, Robert H. and Lust, Nate B. and MacArthur, Lauren A. and Miyazaki, Satoshi and Murayama, Hitoshi and Nishimichi, Takahiro and Okura, Yuki and Price, Paul A. and Tait, Philip J. and Tanaka, Masayuki and Wang, Shiang-Yu},
   year={2023},
   month=dec }

@article{Rau2022compositelikelihood,
    author = {Rau, M M and Morrison, C B and Schmidt, S J and Wilson, S and Mandelbaum, R and Mao, Y-Y and LSST Dark Energy Science Collaboration },
    title = {A composite likelihood approach for inference under photometric redshift uncertainty},
    journal = {Monthly Notices of the Royal Astronomical Society},
    volume = {509},
    number = {4},
    pages = {4886-4907},
    year = {2021},
    month = {11},
    issn = {0035-8711},
    doi = {10.1093/mnras/stab3290},
    url = {https://doi.org/10.1093/mnras/stab3290},
    eprint = {https://academic.oup.com/mnras/article-pdf/509/4/4886/41715550/stab3290.pdf},
}

@article{Zhang2022PhotometricRedshiftShifts,
   title={Photometric redshift uncertainties in weak gravitational lensing shear analysis: models and marginalization},
   volume={518},
   ISSN={1365-2966},
   url={http://dx.doi.org/10.1093/mnras/stac3090},
   DOI={10.1093/mnras/stac3090},
   number={1},
   journal={Monthly Notices of the Royal Astronomical Society},
   publisher={Oxford University Press (OUP)},
   author={Zhang, Tianqing and Rau, Markus Michael and Mandelbaum, Rachel and Li, Xiangchong and Moews, Ben},
   year={2022},
   month=oct, pages={709–723} 
}

@article{Rana2025HSCY3CosmicShearRatios,
      title={Hyper Suprime-Cam Y3 results: photo-$z$ bias calibration with lensing shear ratios and cosmological constraints from cosmic shea}, 
      author={Divya Rana and Surhud More and Hironao Miyatake and Sunao Sugiyama and Tianqing Zhang and Masato Shirasaki},
      year={2025},
      journal={Physical Review D},
      publisher={American Physical Society (APS)},
      eprint={2508.21681},
      archivePrefix={arXiv},
      primaryClass={astro-ph.CO},
      url={https://arxiv.org/abs/2508.21681}, 
}

@ARTICLE{Zhang2025RedshiftGCPointMass,
       author = {{Zhang}, Tianqing and {Li}, Xiangchong and {Sugiyama}, Sunao and {Mandelbaum}, Rachel and {More}, Surhud and {Dalal}, Roohi and {Kannawadi}, Arun and {Miyatake}, Hironao and {Nishizawa}, Atsushi J. and {Nishimichi}, Takahiro and {Oguri}, Masamune and {Osato}, Ken and {Rau}, Markus M. and {Shirasaki}, Masato and {Sunayama}, Tomomi and {Takada}, Masahiro},
        title = "{Cosmology and Source Redshift Constraints from Galaxy Clustering and Tomographic Weak Lensing with HSC Y3 and SDSS using the Point-Mass Correction Model}",
      journal = {arXiv e-prints},
         year = 2025,
        month = jul,
          eid = {arXiv:2507.01386},
        pages = {arXiv:2507.01386},
          doi = {10.48550/arXiv.2507.01386},
archivePrefix = {arXiv},
       eprint = {2507.01386},
 primaryClass = {astro-ph.CO},
       adsurl = {https://ui.adsabs.harvard.edu/abs/2025arXiv250701386Z}
}

@article{HSCOverviewAihara2017,
   title={The Hyper Suprime-Cam SSP Survey: Overview and survey design},
   volume={70},
   ISSN={2053-051X},
   url={http://dx.doi.org/10.1093/pasj/psx066},
   DOI={10.1093/pasj/psx066},
   number={SP1},
   journal={Publications of the Astronomical Society of Japan},
   publisher={Oxford University Press (OUP)},
   author={Aihara, Hiroaki and Arimoto, Nobuo and Armstrong, Robert and Arnouts, Stéphane and Bahcall, Neta A and Bickerton, Steven and Bosch, James and Bundy, Kevin and Capak, Peter L and Chan, James H H and Chiba, Masashi and Coupon, Jean and Egami, Eiichi and Enoki, Motohiro and Finet, Francois and Fujimori, Hiroki and Fujimoto, Seiji and Furusawa, Hisanori and Furusawa, Junko and Goto, Tomotsugu and Goulding, Andy and Greco, Johnny P and Greene, Jenny E and Gunn, James E and Hamana, Takashi and Harikane, Yuichi and Hashimoto, Yasuhiro and Hattori, Takashi and Hayashi, Masao and Hayashi, Yusuke and Hełminiak, Krzysztof G and Higuchi, Ryo and Hikage, Chiaki and Ho, Paul T P and Hsieh, Bau-Ching and Huang, Kuiyun and Huang, Song and Ikeda, Hiroyuki and Imanishi, Masatoshi and Inoue, Akio K and Iwasawa, Kazushi and Iwata, Ikuru and Jaelani, Anton T and Jian, Hung-Yu and Kamata, Yukiko and Karoji, Hiroshi and Kashikawa, Nobunari and Katayama, Nobuhiko and Kawanomoto, Satoshi and Kayo, Issha and Koda, Jin and Koike, Michitaro and Kojima, Takashi and Komiyama, Yutaka and Konno, Akira and Koshida, Shintaro and Koyama, Yusei and Kusakabe, Haruka and Leauthaud, Alexie and Lee, Chien-Hsiu and Lin, Lihwai and Lin, Yen-Ting and Lupton, Robert H and Mandelbaum, Rachel and Matsuoka, Yoshiki and Medezinski, Elinor and Mineo, Sogo and Miyama, Shoken and Miyatake, Hironao and Miyazaki, Satoshi and Momose, Rieko and More, Anupreeta and More, Surhud and Moritani, Yuki and Moriya, Takashi J and Morokuma, Tomoki and Mukae, Shiro and Murata, Ryoma and Murayama, Hitoshi and Nagao, Tohru and Nakata, Fumiaki and Niida, Mana and Niikura, Hiroko and Nishizawa, Atsushi J and Obuchi, Yoshiyuki and Oguri, Masamune and Oishi, Yukie and Okabe, Nobuhiro and Okamoto, Sakurako and Okura, Yuki and Ono, Yoshiaki and Onodera, Masato and Onoue, Masafusa and Osato, Ken and Ouchi, Masami and Price, Paul A and Pyo, Tae-Soo and Sako, Masao and Sawicki, Marcin and Shibuya, Takatoshi and Shimasaku, Kazuhiro and Shimono, Atsushi and Shirasaki, Masato and Silverman, John D and Simet, Melanie and Speagle, Joshua and Spergel, David N and Strauss, Michael A and Sugahara, Yuma and Sugiyama, Naoshi and Suto, Yasushi and Suyu, Sherry H and Suzuki, Nao and Tait, Philip J and Takada, Masahiro and Takata, Tadafumi and Tamura, Naoyuki and Tanaka, Manobu M and Tanaka, Masaomi and Tanaka, Masayuki and Tanaka, Yoko and Terai, Tsuyoshi and Terashima, Yuichi and Toba, Yoshiki and Tominaga, Nozomu and Toshikawa, Jun and Turner, Edwin L and Uchida, Tomohisa and Uchiyama, Hisakazu and Umetsu, Keiichi and Uraguchi, Fumihiro and Urata, Yuji and Usuda, Tomonori and Utsumi, Yousuke and Wang, Shiang-Yu and Wang, Wei-Hao and Wong, Kenneth C and Yabe, Kiyoto and Yamada, Yoshihiko and Yamanoi, Hitomi and Yasuda, Naoki and Yeh, Sherry and Yonehara, Atsunori and Yuma, Suraphong},
   year={2017},
   month=sep }

@ARTICLE{2010PhRvD..81f3531B,
       author = {{Baldauf}, Tobias and {Smith}, Robert E. and {Seljak}, Uro{\v{s}} and {Mandelbaum}, Rachel},
        title = "{Algorithm for the direct reconstruction of the dark matter correlation function from weak lensing and galaxy clustering}",
      journal = {\prd},
         year = 2010,
        month = mar,
       volume = {81},
       number = {6},
          eid = {063531},
        pages = {063531},
          doi = {10.1103/PhysRevD.81.063531},
archivePrefix = {arXiv},
       eprint = {0911.4973},
 primaryClass = {astro-ph.CO},
       adsurl = {https://ui.adsabs.harvard.edu/abs/2010PhRvD..81f3531B}
}

@ARTICLE{DavisPeebles1983,
       author = {{Davis}, M. and {Peebles}, P.~J.~E.},
        title = "{A survey of galaxy redshifts. V. The two-point position and velocity correlations.}",
      journal = {\apj},
         year = 1983,
        month = apr,
       volume = {267},
        pages = {465-482},
          doi = {10.1086/160884},
       adsurl = {https://ui.adsabs.harvard.edu/abs/1983ApJ...267..465D}
}

@ARTICLE{LandySzalay1993,
       author = {{Landy}, Stephen D. and {Szalay}, Alexander S.},
        title = "{Bias and Variance of Angular Correlation Functions}",
      journal = {\apj},
         year = 1993,
        month = jul,
       volume = {412},
        pages = {64},
          doi = {10.1086/172900},
       adsurl = {https://ui.adsabs.harvard.edu/abs/1993ApJ...412...64L}
}

@ARTICLE{FKPWeights1994,
       author = {{Feldman}, Hume A. and {Kaiser}, Nick and {Peacock}, John A.},
        title = "{Power-Spectrum Analysis of Three-dimensional Redshift Surveys}",
      journal = {\apj},
         year = 1994,
        month = may,
       volume = {426},
        pages = {23},
          doi = {10.1086/174036},
archivePrefix = {arXiv},
       eprint = {astro-ph/9304022},
 primaryClass = {astro-ph},
       adsurl = {https://ui.adsabs.harvard.edu/abs/1994ApJ...426...23F}
}

@article{ChoiCFHTLensAngularCross2016,
    author = {Choi, A. and Heymans, C. and Blake, C. and Hildebrandt, H. and Duncan, C. A. J. and Erben, T. and Nakajima, R. and Van Waerbeke, L. and Viola, M.},
    title = {CFHTLenS and RCSLenS: testing photometric redshift distributions using angular cross-correlations with spectroscopic galaxy surveys},
    journal = {Monthly Notices of the Royal Astronomical Society},
    volume = {463},
    number = {4},
    pages = {3737-3754},
    year = {2016},
    month = {09},
    issn = {0035-8711},
    doi = {10.1093/mnras/stw2241},
    url = {https://doi.org/10.1093/mnras/stw2241},
    eprint = {https://academic.oup.com/mnras/article-pdf/463/4/3737/18513459/stw2241.pdf},
}

@ARTICLE{Limber1953Approximation,
       author = {{Limber}, D. Nelson},
        title = "{The Analysis of Counts of the Extragalactic Nebulae in Terms of a Fluctuating Density Field.}",
      journal = {\apj},
         year = 1953,
        month = jan,
       volume = {117},
        pages = {134},
          doi = {10.1086/145672},
       adsurl = {https://ui.adsabs.harvard.edu/abs/1953ApJ...117..134L}
}

@article{Mohammad2022jackknife,
   title={Creating jackknife and bootstrap estimates of the covariance matrix for the two-point correlation function},
   volume={514},
   ISSN={1365-2966},
   url={http://dx.doi.org/10.1093/mnras/stac1458},
   DOI={10.1093/mnras/stac1458},
   number={1},
   journal={Monthly Notices of the Royal Astronomical Society},
   publisher={Oxford University Press (OUP)},
   author={Mohammad, Faizan G and Percival, Will J},
   year={2022},
   month=may, pages={1289–1301} 
}

@article{MagnificationBiasComplexSelections,
    author = {von Wietersheim-Kramsta, Maximilian and Joachimi, Benjamin and van den Busch, Jan Luca and Heymans, Catherine and Hildebrandt, Hendrik and Asgari, Marika and Tr’oster, Tilman and Unruh, Sandra and Wright, Angus H},
    title = {Magnification bias in galaxy surveys with complex sample selection functions},
    journal = {Monthly Notices of the Royal Astronomical Society},
    volume = {504},
    number = {1},
    pages = {1452-1465},
    year = {2021},
    month = {04},
    issn = {0035-8711},
    doi = {10.1093/mnras/stab1000},
    url = {https://doi.org/10.1093/mnras/stab1000},
    eprint = {https://academic.oup.com/mnras/article-pdf/504/1/1452/39571307/stab1000.pdf},
}

@ARTICLE{LePhare1,
       author = {{Ilbert}, O. and {Arnouts}, S. and {McCracken}, H.~J. and {Bolzonella}, M. and {Bertin}, E. and {Le F{\`e}vre}, O. and {Mellier}, Y. and {Zamorani}, G. and {Pell{\`o}}, R. and {Iovino}, A. and {Tresse}, L. and {Le Brun}, V. and {Bottini}, D. and {Garilli}, B. and {Maccagni}, D. and {Picat}, J.~P. and {Scaramella}, R. and {Scodeggio}, M. and {Vettolani}, G. and {Zanichelli}, A. and {Adami}, C. and {Bardelli}, S. and {Cappi}, A. and {Charlot}, S. and {Ciliegi}, P. and {Contini}, T. and {Cucciati}, O. and {Foucaud}, S. and {Franzetti}, P. and {Gavignaud}, I. and {Guzzo}, L. and {Marano}, B. and {Marinoni}, C. and {Mazure}, A. and {Meneux}, B. and {Merighi}, R. and {Paltani}, S. and {Pollo}, A. and {Pozzetti}, L. and {Radovich}, M. and {Zucca}, E. and {Bondi}, M. and {Bongiorno}, A. and {Busarello}, G. and {de La Torre}, S. and {Gregorini}, L. and {Lamareille}, F. and {Mathez}, G. and {Merluzzi}, P. and {Ripepi}, V. and {Rizzo}, D. and {Vergani}, D.},
        title = "{Accurate photometric redshifts for the CFHT legacy survey calibrated using the VIMOS VLT deep survey}",
      journal = {\aap},
         year = 2006,
        month = oct,
       volume = {457},
       number = {3},
        pages = {841-856},
          doi = {10.1051/0004-6361:20065138},
archivePrefix = {arXiv},
}

@ARTICLE{LePhare2,
       author = {{Arnouts}, S. and {Cristiani}, S. and {Moscardini}, L. and {Matarrese}, S. and {Lucchin}, F. and {Fontana}, A. and {Giallongo}, E.},
        title = "{Measuring and modelling the redshift evolution of clustering: the Hubble Deep Field North}",
      journal = {\mnras},
         year = 1999,
        month = dec,
       volume = {310},
       number = {2},
        pages = {540-556},
          doi = {10.1046/j.1365-8711.1999.02978.x},
archivePrefix = {arXiv},
       eprint = {astro-ph/9902290},
 primaryClass = {astro-ph},
       adsurl = {https://ui.adsabs.harvard.edu/abs/1999MNRAS.310..540A}
}

@article{CarrascoKind2014PhotozMLz,
   title={SOMz: photometric redshift PDFs with self-organizing maps and random atlas},
   volume={438},
   ISSN={0035-8711},
   url={http://dx.doi.org/10.1093/mnras/stt2456},
   DOI={10.1093/mnras/stt2456},
   number={4},
   journal={Monthly Notices of the Royal Astronomical Society},
   publisher={Oxford University Press (OUP)},
   author={Carrasco Kind, Matias and Brunner, Robert J.},
   year={2014},
   month=jan, pages={3409–3421} }

@article{Hsieh2014DEMp,
   title={ESTIMATING LUMINOSITIES AND STELLAR MASSES OF GALAXIES PHOTOMETRICALLY WITHOUT DETERMINING REDSHIFTS},
   volume={792},
   ISSN={1538-4357},
   url={http://dx.doi.org/10.1088/0004-637X/792/2/102},
   DOI={10.1088/0004-637x/792/2/102},
   number={2},
   journal={The Astrophysical Journal},
   publisher={American Astronomical Society},
   author={Hsieh, B. C. and Yee, H. K. C.},
   year={2014},
   month=aug, pages={102} }

@article{Brammer2008EAZY,
   title={EAZY: A Fast, Public Photometric Redshift Code},
   volume={686},
   ISSN={1538-4357},
   url={http://dx.doi.org/10.1086/591786},
   DOI={10.1086/591786},
   number={2},
   journal={The Astrophysical Journal},
   publisher={American Astronomical Society},
   author={Brammer, Gabriel B. and van Dokkum, Pieter G. and Coppi, Paolo},
   year={2008},
   month=oct, pages={1503–1513} }

@article{Collister2004ANNz,
   title={ANNz: Estimating Photometric Redshifts Using Artificial Neural Networks},
   volume={116},
   ISSN={1538-3873},
   url={http://dx.doi.org/10.1086/383254},
   DOI={10.1086/383254},
   number={818},
   journal={Publications of the Astronomical Society of the Pacific},
   publisher={IOP Publishing},
   author={Collister, Adrian A. and Lahav, Ofer},
   year={2004},
   month=apr, pages={345–351} }

@article{Laurent2017QSOBias,
   title={Clustering of quasars in SDSS-IV eBOSS: study of potential systematics and bias determination},
   volume={2017},
   ISSN={1475-7516},
   url={http://dx.doi.org/10.1088/1475-7516/2017/07/017},
   DOI={10.1088/1475-7516/2017/07/017},
   number={07},
   journal={Journal of Cosmology and Astroparticle Physics},
   publisher={IOP Publishing},
   author={Laurent, Pierre and Eftekharzadeh, Sarah and Goff, Jean-Marc Le and Myers, Adam and Burtin, Etienne and White, Martin and Ross, Ashley J. and Tinker, Jeremy and Tojeiro, Rita and Bautista, Julian and Brinkmann, Jonathan and Comparat, Johan and Dawson, Kyle and Bourboux, Hélion du Mas des and Kneib, Jean-Paul and McGreer, Ian D. and Palanque-Delabrouille, Nathalie and Percival, Will J. and Prada, Francisco and Rossi, Graziano and Schneider, Donald P. and Weinberg, David and Yèche, Christophe and Zarrouk, Pauline and Zhao, Gong-Bo},
   year={2017},
   month=jul, pages={017–017} }

@article{WrightSOMz,
   title={Photometric redshift calibration with self-organising maps},
   volume={637},
   ISSN={1432-0746},
   url={http://dx.doi.org/10.1051/0004-6361/201936782},
   DOI={10.1051/0004-6361/201936782},
   journal={\aap},
   publisher={EDP Sciences},
   author={Wright, Angus H. and Hildebrandt, Hendrik and van den Busch, Jan Luca and Heymans, Catherine},
   year={2020},
   month=may, pages={A100} }

@article{SOMPZDES2024,
      title={Enhancing weak lensing redshift distribution characterization by optimizing the Dark Energy Survey Self-Organizing Map Photo-z method}, 
      author={A. Campos and B. Yin and S. Dodelson and A. Amon and A. Alarcon and C. Sánchez and G. M. Bernstein and G. Giannini and J. Myles and S. Samuroff and O. Alves and F. Andrade-Oliveira and K. Bechtol and M. R. Becker and J. Blazek and H. Camacho and A. Carnero Rosell and M. Carrasco Kind and R. Cawthon and C. Chang and R. Chen and A. Choi and J. Cordero and C. Davis and J. DeRose and H. T. Diehl and C. Doux and A. Drlica-Wagner and K. Eckert and T. F. Eifler and J. Elvin-Poole and S. Everett and X. Fang and A. Ferté and O. Friedrich and M. Gatti and D. Gruen and R. A. Gruendl and I. Harrison and W. G. Hartley and K. Herner and H. Huang and E. M. Huff and M. Jarvis and E. Krause and N. Kuropatkin and P. -F. Leget and N. MacCrann and J. McCullough and A. Navarro-Alsina and S. Pandey and J. Prat and M. Raveri and R. P. Rollins and A. Roodman and R. Rosenfeld and A. J. Ross and E. S. Rykoff and J. Sanchez and L. F. Secco and I. Sevilla-Noarbe and E. Sheldon and T. Shin and M. A. Troxel and I. Tutusaus and T. N. Varga and R. H. Wechsler and B. Yanny and Y. Zhang and J. Zuntz and M. Aguena and J. Annis and D. Bacon and S. Bocquet and D. Brooks and D. L. Burke and J. Carretero and F. J. Castander and M. Costanzi and L. N. da Costa and J. De Vicente and P. Doel and I. Ferrero and B. Flaugher and J. Frieman and J. García-Bellido and E. Gaztanaga and G. Gutierrez and S. R. Hinton and D. L. Hollowood and K. Honscheid and D. J. James and K. Kuehn and M. Lima and H. Lin and J. L. Marshall and J. Mena-Fernández and F. Menanteau and R. Miquel and R. L. C. Ogando and M. Paterno and M. E. S. Pereira and A. Pieres and A. A. Plazas Malagón and A. Porredon and E. Sanchez and D. Sanchez Cid and M. Smith and E. Suchyta and M. E. C. Swanson and G. Tarle and C. To and V. Vikram and N. Weaverdyck},
      year={2024},
      eprint={2408.00922},
      archivePrefix={arXiv},
      primaryClass={astro-ph.CO},
      url={https://arxiv.org/abs/2408.00922}, 
}

@article{Hildebrandt2021ClusteringRedshiftsSOM,
   title={KiDS-1000 catalogue: Redshift distributions and their calibration},
   volume={647},
   ISSN={1432-0746},
   url={http://dx.doi.org/10.1051/0004-6361/202039018},
   DOI={10.1051/0004-6361/202039018},
   journal={\aap},
   publisher={EDP Sciences},
   author={Hildebrandt, H. and van den Busch, J. L. and Wright, A. H. and Blake, C. and Joachimi, B. and Kuijken, K. and Tröster, T. and Asgari, M. and Bilicki, M. and de Jong, J. T. A. and Dvornik, A. and Erben, T. and Getman, F. and Giblin, B. and Heymans, C. and Kannawadi, A. and Lin, C.-A. and Shan, H.-Y.},
}

@ARTICLE{LimaDIRMethod2008,
       author = {{Lima}, Marcos and {Cunha}, Carlos E. and {Oyaizu}, Hiroaki and {Frieman}, Joshua and {Lin}, Huan and {Sheldon}, Erin S.},
        title = "{Estimating the redshift distribution of photometric galaxy samples}",
      journal = {\mnras},
         year = 2008,
        month = oct,
       volume = {390},
       number = {1},
        pages = {118-130},
          doi = {10.1111/j.1365-2966.2008.13510.x},
archivePrefix = {arXiv},
       eprint = {0801.3822},
 primaryClass = {astro-ph},
       adsurl = {https://ui.adsabs.harvard.edu/abs/2008MNRAS.390..118L}
}

@article{BianchiPIP2017,
    author = {Bianchi, Davide and Percival, Will J.},
    title = {Unbiased clustering estimation in the presence of missing observations},
    journal = {Monthly Notices of the Royal Astronomical Society},
    volume = {472},
    number = {1},
    pages = {1106-1118},
    year = {2017},
    month = {08},
    issn = {0035-8711},
    doi = {10.1093/mnras/stx2053},
    url = {https://doi.org/10.1093/mnras/stx2053},
    eprint = {https://academic.oup.com/mnras/article-pdf/472/1/1106/19720617/stx2053.pdf},
}

@article{Buchs2019SOMRedshifts,
   title={Phenotypic redshifts with self-organizing maps: A novel method to characterize redshift distributions of source galaxies for weak lensing},
   volume={489},
   ISSN={1365-2966},
   url={http://dx.doi.org/10.1093/mnras/stz2162},
   DOI={10.1093/mnras/stz2162},
   number={1},
   journal={Monthly Notices of the Royal Astronomical Society},
   publisher={Oxford University Press (OUP)},
   author={Buchs, R and Davis, C and Gruen, D and DeRose, J and Alarcon, A and Bernstein, G M and Sánchez, C and Myles, J and Roodman, A and Allen, S and Amon, A and Choi, A and Masters, D C and Miquel, R and Troxel, M A and Wechsler, R H and Abbott, T M C and Annis, J and Avila, S and Bechtol, K and Bridle, S L and Brooks, D and Buckley-Geer, E and Burke, D L and Carnero Rosell, A and Carrasco Kind, M and Carretero, J and Castander, F J and Cawthon, R and D’Andrea, C B and da Costa, L N and De Vicente, J and Desai, S and Diehl, H T and Doel, P and Drlica-Wagner, A and Eifler, T F and Evrard, A E and Flaugher, B and Fosalba, P and Frieman, J and García-Bellido, J and Gaztanaga, E and Gruendl, R A and Gschwend, J and Gutierrez, G and Hartley, W G and Hollowood, D L and Honscheid, K and James, D J and Kuehn, K and Kuropatkin, N and Lima, M and Lin, H and Maia, M A G and March, M and Marshall, J L and Melchior, P and Menanteau, F and Ogando, R L C and Plazas, A A and Rykoff, E S and Sanchez, E and Scarpine, V and Serrano, S and Sevilla-Noarbe, I and Smith, M and Soares-Santos, M and Sobreira, F and Suchyta, E and Swanson, M E C and Tarle, G and Thomas, D and Vikram, V},
   year={2019},
   month=aug, pages={820–841} }

@article{Masters2017C3R2SOM,
   title={The Complete Calibration of the Color–Redshift Relation (C3R2) Survey: Survey Overview and Data Release 1},
   volume={841},
   ISSN={1538-4357},
   url={http://dx.doi.org/10.3847/1538-4357/aa6f08},
   DOI={10.3847/1538-4357/aa6f08},
   number={2},
   journal={The Astrophysical Journal},
   publisher={American Astronomical Society},
   author={Masters, Daniel C. and Stern, Daniel K. and Cohen, Judith G. and Capak, Peter L. and Rhodes, Jason D. and Castander, Francisco J. and Paltani, Stéphane},
   year={2017},
   month=may, pages={111} }

@article{KaiserRSD,
    author = {Kaiser, Nick},
    title = {Clustering in real space and in redshift space},
    journal = {Monthly Notices of the Royal Astronomical Society},
    volume = {227},
    number = {1},
    pages = {1-21},
    year = {1987},
    month = {07},
    issn = {0035-8711},
    doi = {10.1093/mnras/227.1.1},
    url = {https://doi.org/10.1093/mnras/227.1.1},
    eprint = {https://academic.oup.com/mnras/article-pdf/227/1/1/18522208/mnras227-0001.pdf},
}

@article{KiDS2025datarelease5,
    author = "Wright, Angus H. and others",
    title = "{The fifth data release of the Kilo Degree Survey: Multi-epoch optical/NIR imaging covering wide and legacy-calibration fields}",
    eprint = "2503.19439",
    archivePrefix = "arXiv",
    primaryClass = "astro-ph.GA",
    doi = "10.1051/0004-6361/202346730",
    journal = "Astron. Astrophys.",
    volume = "686",
    pages = "A170",
    year = "2024"
}

@ARTICLE{DES2018datarelease1,
       author = {{Abbott}, T.~M.~C. and {Abdalla}, F.~B. and {Allam}, S. and {Amara}, A. and {Annis}, J. and {Asorey}, J. and {Avila}, S. and {Ballester}, O. and {Banerji}, M. and {Barkhouse}, W. and {Baruah}, L. and {Baumer}, M. and {Bechtol}, K. and {Becker}, M.~R. and {Benoit-L{\'e}vy}, A. and {Bernstein}, G.~M. and {Bertin}, E. and {Blazek}, J. and {Bocquet}, S. and {Brooks}, D. and {Brout}, D. and {Buckley-Geer}, E. and {Burke}, D.~L. and {Busti}, V. and {Campisano}, R. and {Cardiel-Sas}, L. and {Carnero Rosell}, A. and {Carrasco Kind}, M. and {Carretero}, J. and {Castander}, F.~J. and {Cawthon}, R. and {Chang}, C. and {Chen}, X. and {Conselice}, C. and {Costa}, G. and {Crocce}, M. and {Cunha}, C.~E. and {D'Andrea}, C.~B. and {da Costa}, L.~N. and {Das}, R. and {Daues}, G. and {Davis}, T.~M. and {Davis}, C. and {De Vicente}, J. and {DePoy}, D.~L. and {DeRose}, J. and {Desai}, S. and {Diehl}, H.~T. and {Dietrich}, J.~P. and {Dodelson}, S. and {Doel}, P. and {Drlica-Wagner}, A. and {Eifler}, T.~F. and {Elliott}, A.~E. and {Evrard}, A.~E. and {Farahi}, A. and {Fausti Neto}, A. and {Fernandez}, E. and {Finley}, D.~A. and {Flaugher}, B. and {Foley}, R.~J. and {Fosalba}, P. and {Friedel}, D.~N. and {Frieman}, J. and {Garc{\'\i}a-Bellido}, J. and {Gaztanaga}, E. and {Gerdes}, D.~W. and {Giannantonio}, T. and {Gill}, M.~S.~S. and {Glazebrook}, K. and {Goldstein}, D.~A. and {Gower}, M. and {Gruen}, D. and {Gruendl}, R.~A. and {Gschwend}, J. and {Gupta}, R.~R. and {Gutierrez}, G. and {Hamilton}, S. and {Hartley}, W.~G. and {Hinton}, S.~R. and {Hislop}, J.~M. and {Hollowood}, D. and {Honscheid}, K. and {Hoyle}, B. and {Huterer}, D. and {Jain}, B. and {James}, D.~J. and {Jeltema}, T. and {Johnson}, M.~W.~G. and {Johnson}, M.~D. and {Kacprzak}, T. and {Kent}, S. and {Khullar}, G. and {Klein}, M. and {Kovacs}, A. and {Koziol}, A.~M.~G. and {Krause}, E. and {Kremin}, A. and {Kron}, R. and {Kuehn}, K. and {Kuhlmann}, S. and {Kuropatkin}, N. and {Lahav}, O. and {Lasker}, J. and {Li}, T.~S. and {Li}, R.~T. and {Liddle}, A.~R. and {Lima}, M. and {Lin}, H. and {L{\'o}pez-Reyes}, P. and {MacCrann}, N. and {Maia}, M.~A.~G. and {Maloney}, J.~D. and {Manera}, M. and {March}, M. and {Marriner}, J. and {Marshall}, J.~L. and {Martini}, P. and {McClintock}, T. and {McKay}, T. and {McMahon}, R.~G. and {Melchior}, P. and {Menanteau}, F. and {Miller}, C.~J. and {Miquel}, R. and {Mohr}, J.~J. and {Morganson}, E. and {Mould}, J. and {Neilsen}, E. and {Nichol}, R.~C. and {Nogueira}, F. and {Nord}, B. and {Nugent}, P. and {Nunes}, L. and {Ogando}, R.~L.~C. and {Old}, L. and {Pace}, A.~B. and {Palmese}, A. and {Paz-Chinch{\'o}n}, F. and {Peiris}, H.~V. and {Percival}, W.~J. and {Petravick}, D. and {Plazas}, A.~A. and {Poh}, J. and {Pond}, C. and {Porredon}, A. and {Pujol}, A. and {Refregier}, A. and {Reil}, K. and {Ricker}, P.~M. and {Rollins}, R.~P. and {Romer}, A.~K. and {Roodman}, A. and {Rooney}, P. and {Ross}, A.~J. and {Rykoff}, E.~S. and {Sako}, M. and {Sanchez}, M.~L. and {Sanchez}, E. and {Santiago}, B. and {Saro}, A. and {Scarpine}, V. and {Scolnic}, D. and {Serrano}, S. and {Sevilla-Noarbe}, I. and {Sheldon}, E. and {Shipp}, N. and {Silveira}, M.~L. and {Smith}, M. and {Smith}, R.~C. and {Smith}, J.~A. and {Soares-Santos}, M. and {Sobreira}, F. and {Song}, J. and {Stebbins}, A. and {Suchyta}, E. and {Sullivan}, M. and {Swanson}, M.~E.~C. and {Tarle}, G. and {Thaler}, J. and {Thomas}, D. and {Thomas}, R.~C. and {Troxel}, M.~A. and {Tucker}, D.~L. and {Vikram}, V. and {Vivas}, A.~K. and {Walker}, A.~R. and {Wechsler}, R.~H. and {Weller}, J. and {Wester}, W. and {Wolf}, R.~C. and {Wu}, H. and {Yanny}, B. and {Zenteno}, A. and {Zhang}, Y. and {Zuntz}, J. and {DES Collaboration} and {Juneau}, S. and {Fitzpatrick}, M. and {Nikutta}, R.},
        title = "{The Dark Energy Survey: Data Release 1}",
      journal = {\apjs},
         year = 2018,
        month = dec,
       volume = {239},
       number = {2},
          eid = {18},
        pages = {18},
          doi = {10.3847/1538-4365/aae9f0},
archivePrefix = {arXiv},
       eprint = {1801.03181},
 primaryClass = {astro-ph.IM},
       adsurl = {https://ui.adsabs.harvard.edu/abs/2018ApJS..239...18A}
}

@ARTICLE{LSST2019survey,
       author = {{Ivezi{\'c}}, {\v{Z}}eljko and {Kahn}, Steven M. and {Tyson}, J. Anthony and {Abel}, Bob and {Acosta}, Emily and {Allsman}, Robyn and {Alonso}, David and {AlSayyad}, Yusra and {Anderson}, Scott F. and {Andrew}, John and {Angel}, James Roger P. and {Angeli}, George Z. and {Ansari}, Reza and {Antilogus}, Pierre and {Araujo}, Constanza and {Armstrong}, Robert and {Arndt}, Kirk T. and {Astier}, Pierre and {Aubourg}, {\'E}ric and {Auza}, Nicole and {Axelrod}, Tim S. and {Bard}, Deborah J. and {Barr}, Jeff D. and {Barrau}, Aurelian and {Bartlett}, James G. and {Bauer}, Amanda E. and {Bauman}, Brian J. and {Baumont}, Sylvain and {Bechtol}, Ellen and {Bechtol}, Keith and {Becker}, Andrew C. and {Becla}, Jacek and {Beldica}, Cristina and {Bellavia}, Steve and {Bianco}, Federica B. and {Biswas}, Rahul and {Blanc}, Guillaume and {Blazek}, Jonathan and {Blandford}, Roger D. and {Bloom}, Josh S. and {Bogart}, Joanne and {Bond}, Tim W. and {Booth}, Michael T. and {Borgland}, Anders W. and {Borne}, Kirk and {Bosch}, James F. and {Boutigny}, Dominique and {Brackett}, Craig A. and {Bradshaw}, Andrew and {Brandt}, William Nielsen and {Brown}, Michael E. and {Bullock}, James S. and {Burchat}, Patricia and {Burke}, David L. and {Cagnoli}, Gianpietro and {Calabrese}, Daniel and {Callahan}, Shawn and {Callen}, Alice L. and {Carlin}, Jeffrey L. and {Carlson}, Erin L. and {Chandrasekharan}, Srinivasan and {Charles-Emerson}, Glenaver and {Chesley}, Steve and {Cheu}, Elliott C. and {Chiang}, Hsin-Fang and {Chiang}, James and {Chirino}, Carol and {Chow}, Derek and {Ciardi}, David R. and {Claver}, Charles F. and {Cohen-Tanugi}, Johann and {Cockrum}, Joseph J. and {Coles}, Rebecca and {Connolly}, Andrew J. and {Cook}, Kem H. and {Cooray}, Asantha and {Covey}, Kevin R. and {Cribbs}, Chris and {Cui}, Wei and {Cutri}, Roc and {Daly}, Philip N. and {Daniel}, Scott F. and {Daruich}, Felipe and {Daubard}, Guillaume and {Daues}, Greg and {Dawson}, William and {Delgado}, Francisco and {Dellapenna}, Alfred and {de Peyster}, Robert and {de Val-Borro}, Miguel and {Digel}, Seth W. and {Doherty}, Peter and {Dubois}, Richard and {Dubois-Felsmann}, Gregory P. and {Durech}, Josef and {Economou}, Frossie and {Eifler}, Tim and {Eracleous}, Michael and {Emmons}, Benjamin L. and {Fausti Neto}, Angelo and {Ferguson}, Henry and {Figueroa}, Enrique and {Fisher-Levine}, Merlin and {Focke}, Warren and {Foss}, Michael D. and {Frank}, James and {Freemon}, Michael D. and {Gangler}, Emmanuel and {Gawiser}, Eric and {Geary}, John C. and {Gee}, Perry and {Geha}, Marla and {Gessner}, Charles J.~B. and {Gibson}, Robert R. and {Gilmore}, D. Kirk and {Glanzman}, Thomas and {Glick}, William and {Goldina}, Tatiana and {Goldstein}, Daniel A. and {Goodenow}, Iain and {Graham}, Melissa L. and {Gressler}, William J. and {Gris}, Philippe and {Guy}, Leanne P. and {Guyonnet}, Augustin and {Haller}, Gunther and {Harris}, Ron and {Hascall}, Patrick A. and {Haupt}, Justine and {Hernandez}, Fabio and {Herrmann}, Sven and {Hileman}, Edward and {Hoblitt}, Joshua and {Hodgson}, John A. and {Hogan}, Craig and {Howard}, James D. and {Huang}, Dajun and {Huffer}, Michael E. and {Ingraham}, Patrick and {Innes}, Walter R. and {Jacoby}, Suzanne H. and {Jain}, Bhuvnesh and {Jammes}, Fabrice and {Jee}, M. James and {Jenness}, Tim and {Jernigan}, Garrett and {Jevremovi{\'c}}, Darko and {Johns}, Kenneth and {Johnson}, Anthony S. and {Johnson}, Margaret W.~G. and {Jones}, R. Lynne and {Juramy-Gilles}, Claire and {Juri{\'c}}, Mario and {Kalirai}, Jason S. and {Kallivayalil}, Nitya J. and {Kalmbach}, Bryce and {Kantor}, Jeffrey P. and {Karst}, Pierre and {Kasliwal}, Mansi M. and {Kelly}, Heather and {Kessler}, Richard and {Kinnison}, Veronica and {Kirkby}, David and {Knox}, Lloyd and {Kotov}, Ivan V. and {Krabbendam}, Victor L. and {Krughoff}, K. Simon and {Kub{\'a}nek}, Petr and {Kuczewski}, John and {Kulkarni}, Shri and {Ku}, John and {Kurita}, Nadine R. and {Lage}, Craig S. and {Lambert}, Ron and {Lange}, Travis and {Langton}, J. Brian and {Le Guillou}, Laurent and {Levine}, Deborah and {Liang}, Ming and {Lim}, Kian-Tat and {Lintott}, Chris J. and {Long}, Kevin E. and {Lopez}, Margaux and {Lotz}, Paul J. and {Lupton}, Robert H. and {Lust}, Nate B. and {MacArthur}, Lauren A. and {Mahabal}, Ashish and {Mandelbaum}, Rachel and {Markiewicz}, Thomas W. and {Marsh}, Darren S. and {Marshall}, Philip J. and {Marshall}, Stuart and {May}, Morgan and {McKercher}, Robert and {McQueen}, Michelle and {Meyers}, Joshua and {Migliore}, Myriam and {Miller}, Michelle and {Mills}, David J.},
        title = "{LSST: From Science Drivers to Reference Design and Anticipated Data Products}",
      journal = {\apj},
         year = 2019,
        month = mar,
       volume = {873},
       number = {2},
          eid = {111},
        pages = {111},
          doi = {10.3847/1538-4357/ab042c},
archivePrefix = {arXiv},
       eprint = {0805.2366},
 primaryClass = {astro-ph},
       adsurl = {https://ui.adsabs.harvard.edu/abs/2019ApJ...873..111I}
}

@article{dassignies2025_clusteringDESY6,
      title={Dark Energy Survey Year 6 Results: Clustering-redshifts and importance sampling of Self-Organised-Maps $n(z)$ realizations for $3\times2$pt samples}, 
      author={W. d'Assignies and G. M. Bernstein and B. Yin and G. Giannini and A. Alarcon and M. Manera and C. To and M. Yamamoto and N. Weaverdyck and R. Cawthon and M. Gatti and A. Amon and D. Anbajagane and S. Avila and M. R. Becker and K. Bechtol and C. Chang and M. Crocce and J. De Vicente and S. Dodelson and J. Fang and A. Ferté and D. Gruen and E. Legnani and A. Porredon and J. Prat and M. Rodriguez-Monroy and C. Sánchez and T. Schutt and I. Sevilla-Noarbe and D. Sanchez Cid and M. A. Troxel and T. M. C. Abbott and M. Aguena and O. Alves and D. Bacon and S. Bocquet and D. Brooks and R. Camilleri and A. Carnero Rosell and M. Carrasco Kind and J. Carretero and F. J. Castander and L. N. da Costa and M. E. da Silva Pereira and T. M. Davis and S. Desai and P. Doel and C. Doux and A. Drlica-Wagner and T. Eifler and J. Elvin-Poole and S. Everett and B. Flaugher and P. Fosalba and J. Frieman and J. Garcia-Bellido and E. Gaztanaga and P. Giles and G. Gutierrez and S. R. Hinton and D. L. Hollowood and K. Honscheid and D. Huterer and B. Jain and D. J. James and K. Kuehn and O. Lahav and S. Lee and J. L. Marshall and J. Mena-Fernandez and F. Menanteau and R. Miquel and J. Muir and J. Myles and R. L. C. Ogando and A. Palmese and M. Paterno and P. Petravick and A. A. Plazas Malagon and M. Raveri and A. Roodman and S. Samuroff and E. Sanchez and E. Sheldon and T. Shin and M. Smith and E. Suchyta and M. E. C. Swanson and G. Tarle and D. Thomas and V. Vikram and A. R. Walker},
      year={2025},
      eprint={2510.23565},
      archivePrefix={arXiv},
      primaryClass={astro-ph.CO},
      url={https://arxiv.org/abs/2510.23565}, 
}

@article{roman2019wfirst,
      title={WFIRST: The Essential Cosmology Space Observatory for the Coming Decade}, 
      author={O. Doré and C. Hirata and Y. Wang and D. Weinberg and T. Eifler and R. J. Foley and C. He Heinrich and E. Krause and S. Perlmutter and A. Pisani and D. Scolnic and D. N. Spergel and N. Suntzeff and G. Aldering and C. Baltay and P. Capak and A. Choi and S. Deustua and C. Dvorkin and S. M. Fall and X. Fang and A. Fruchter and L. Galbany and S. Ho and R. Hounsell and A. Izard and B. Jain and A. M. Koekemoer and J. Kruk and A. Leauthaud and S. Malhotra and R. Mandelbaum and E. Massara and D. Masters and H. Miyatake and A. Plazas and J. Rhoads and J. Rhodes and B. Rose and D. Rubin and M. Sako and L. Samushia and M. Shirasaki and M. Simet and M. Takada and M. A. Troxel and H. Wu and N. Yoshida and Z. Zhai},
      year={2019},
      eprint={1904.01174},
      archivePrefix={arXiv},
      primaryClass={astro-ph.CO},
      url={https://arxiv.org/abs/1904.01174}, 
}

@ARTICLE{DES2021Krause,
       author = {{Krause}, E. and {Fang}, X. and {Pandey}, S. and {Secco}, L.~F. and {Alves}, O. and {Huang}, H. and {Blazek}, J. and {Prat}, J. and {Zuntz}, J. and {Eifler}, T.~F. and {MacCrann}, N. and {DeRose}, J. and {Crocce}, M. and {Porredon}, A. and {Jain}, B. and {Troxel}, M.~A. and {Dodelson}, S. and {Huterer}, D. and {Liddle}, A.~R. and {Leonard}, C.~D. and {Amon}, A. and {Chen}, A. and {Elvin-Poole}, J. and {Fert{\'e}}, A. and {Muir}, J. and {Park}, Y. and {Samuroff}, S. and {Brandao-Souza}, A. and {Weaverdyck}, N. and {Zacharegkas}, G. and {Rosenfeld}, R. and {Campos}, A. and {Chintalapati}, P. and {Choi}, A. and {Di Valentino}, E. and {Doux}, C. and {Herner}, K. and {Lemos}, P. and {Mena-Fern{\'a}ndez}, J. and {Omori}, Y. and {Paterno}, M. and {Rodriguez-Monroy}, M. and {Rogozenski}, P. and {Rollins}, R.~P. and {Troja}, A. and {Tutusaus}, I. and {Wechsler}, R.~H. and {Abbott}, T.~M.~C. and {Aguena}, M. and {Allam}, S. and {Andrade-Oliveira}, F. and {Annis}, J. and {Bacon}, D. and {Baxter}, E. and {Bechtol}, K. and {Bernstein}, G.~M. and {Brooks}, D. and {Buckley-Geer}, E. and {Burke}, D.~L. and {Carnero Rosell}, A. and {Carrasco Kind}, M. and {Carretero}, J. and {Castander}, F.~J. and {Cawthon}, R. and {Chang}, C. and {Costanzi}, M. and {da Costa}, L.~N. and {Pereira}, M.~E.~S. and {De Vicente}, J. and {Desai}, S. and {Diehl}, H.~T. and {Doel}, P. and {Everett}, S. and {Evrard}, A.~E. and {Ferrero}, I. and {Flaugher}, B. and {Fosalba}, P. and {Frieman}, J. and {Garc{\'\i}a-Bellido}, J. and {Gaztanaga}, E. and {Gerdes}, D.~W. and {Giannantonio}, T. and {Gruen}, D. and {Gruendl}, R.~A. and {Gschwend}, J. and {Gutierrez}, G. and {Hartley}, W.~G. and {Hinton}, S.~R. and {Hollowood}, D.~L. and {Honscheid}, K. and {Hoyle}, B. and {Huff}, E.~M. and {James}, D.~J. and {Kuehn}, K. and {Kuropatkin}, N. and {Lahav}, O. and {Lima}, M. and {Maia}, M.~A.~G. and {Marshall}, J.~L. and {Martini}, P. and {Melchior}, P. and {Menanteau}, F. and {Miquel}, R. and {Mohr}, J.~J. and {Morgan}, R. and {Myles}, J. and {Palmese}, A. and {Paz-Chinch{\'o}n}, F. and {Petravick}, D. and {Pieres}, A. and {Plazas Malag{\'o}n}, A.~A. and {Sanchez}, E. and {Scarpine}, V. and {Schubnell}, M. and {Serrano}, S. and {Sevilla-Noarbe}, I. and {Smith}, M. and {Soares-Santos}, M. and {Suchyta}, E. and {Tarle}, G. and {Thomas}, D. and {To}, C. and {Varga}, T.~N. and {Weller}, J.},
        title = "{Dark Energy Survey Year 3 Results: Multi-Probe Modeling Strategy and Validation}",
      journal = {arXiv e-prints},
         year = 2021,
        month = may,
          eid = {arXiv:2105.13548},
        pages = {arXiv:2105.13548},
          doi = {10.48550/arXiv.2105.13548},
archivePrefix = {arXiv},
       eprint = {2105.13548},
 primaryClass = {astro-ph.CO},
       adsurl = {https://ui.adsabs.harvard.edu/abs/2021arXiv210513548K}
}

@article{Gatti2021ClusteringDESWLsrcDistrib,
   title={Dark Energy Survey Year 3 Results: clustering redshifts – calibration of the weak lensing source redshift distributions with redMaGiC and BOSS/eBOSS},
   volume={510},
   ISSN={1365-2966},
   url={http://dx.doi.org/10.1093/mnras/stab3311},
   DOI={10.1093/mnras/stab3311},
   number={1},
   journal={Monthly Notices of the Royal Astronomical Society},
   publisher={Oxford University Press (OUP)},
   author={Gatti, M and Giannini, G and Bernstein, G M and Alarcon, A and Myles, J and Amon, A and Cawthon, R and Troxel, M and DeRose, J and Everett, S and Ross, A J and Rykoff, E S and Elvin-Poole, J and Cordero, J and Harrison, I and Sanchez, C and Prat, J and Gruen, D and Lin, H and Crocce, M and Rozo, E and Abbott, T M C and Aguena, M and Allam, S and Annis, J and Avila, S and Bacon, D and Bertin, E and Brooks, D and Burke, D L and Rosell, A Carnero and Kind, M Carrasco and Carretero, J and Castander, F J and Choi, A and Conselice, C and Costanzi, M and Crocce, M and da Costa, L N and Pereira, M E S and Dawson, K and Desai, S and Diehl, H T and Eckert, K and Eifler, T F and Evrard, A E and Ferrero, I and Flaugher, B and Fosalba, P and Frieman, J and García-Bellido, J and Gaztanaga, E and Giannantonio, T and Gruendl, R A and Gschwend, J and Hinton, S R and Hollowood, D L and Honscheid, K and Hoyle, B and Huterer, D and James, D J and Kuehn, K and Kuropatkin, N and Lahav, O and Lima, M and MacCrann, N and Maia, M A G and March, M and Marshall, J L and Melchior, P and Menanteau, F and Miquel, R and Mohr, J J and Morgan, R and Ogando, R L C and Palmese, A and Paz-Chinchón, F and Percival, W J and Plazas, A A and Rodriguez-Monroy, M and Roodman, A and Rossi, G and Samuroff, S and Sanchez, E and Scarpine, V and Secco, L F and Serrano, S and Sevilla-Noarbe, I and Smith, M and Soares-Santos, M and Suchyta, E and Swanson, M E C and Tarle, G and Thomas, D and To, C and Varga, T N and Weller, J and Wilkinson, R D},
   year={2021},
   month=nov, pages={1223–1247} }

@article{Gatti2018DESY1ClusteringRedshifts,
    author = {Gatti, M and Vielzeuf, P and Davis, C and Cawthon, R and Rau, M M and DeRose, J and De Vicente, J and Alarcon, A and Rozo, E and Gaztanaga, E and Hoyle, B and Miquel, R and Bernstein, G M and Bonnett, C and Carnero Rosell, A and Castander, F J and Chang, C and da Costa, L N and Gruen, D and Gschwend, J and Hartley, W G and Lin, H and MacCrann, N and Maia, M A G and Ogando, R L C and Roodman, A and Sevilla-Noarbe, I and Troxel, M A and Wechsler, R H and Asorey, J and Davis, T M and Glazebrook, K and Hinton, S R and Lewis, G and Lidman, C and Macaulay, E and Möller, A and O'Neill, C R and Sommer, N E and Uddin, S A and Yuan, F and Zhang, B and Abbott, T M C and Allam, S and Annis, J and Bechtol, K and Brooks, D and Burke, D L and Carollo, D and Carrasco Kind, M and Carretero, J and Cunha, C E and D'Andrea, C B and DePoy, D L and Desai, S and Eifler, T F and Evrard, A E and Flaugher, B and Fosalba, P and Frieman, J and García-Bellido, J and Gerdes, D W and Goldstein, D A and Gruendl, R A and Gutierrez, G and Honscheid, K and Hoormann, J K and Jain, B and James, D J and Jarvis, M and Jeltema, T and Johnson, M W G and Johnson, M D and Krause, E and Kuehn, K and Kuhlmann, S and Kuropatkin, N and Li, T S and Lima, M and Marshall, J L and Melchior, P and Menanteau, F and Nichol, R C and Nord, B and Plazas, A A and Reil, K and Rykoff, E S and Sako, M and Sanchez, E and Scarpine, V and Schubnell, M and Sheldon, E and Smith, M and Smith, R C and Soares-Santos, M and Sobreira, F and Suchyta, E and Swanson, M E C and Tarle, G and Thomas, D and Tucker, B E and Tucker, D L and Vikram, V and Walker, A R and Weller, J and Wester, W and Wolf, R C},
    title = {Dark Energy Survey Year 1 results: cross-correlation redshifts – methods and systematics characterization},
    journal = {Monthly Notices of the Royal Astronomical Society},
    volume = {477},
    number = {2},
    pages = {1664-1682},
    year = {2018},
    month = {02},
    issn = {0035-8711},
    doi = {10.1093/mnras/sty466},
    url = {https://doi.org/10.1093/mnras/sty466},
    eprint = {https://academic.oup.com/mnras/article-pdf/477/2/1664/25009657/sty466.pdf},
}

@article{Davis2017DESyear1ClusterZWLsrcdistrib,
      title={Dark Energy Survey Year 1 Results: Cross-Correlation Redshifts in the DES -- Calibration of the Weak Lensing Source Redshift Distributions}, 
      author={C. Davis and M. Gatti and P. Vielzeuf and R. Cawthon and E. Rozo and A. Alarcon and G. M. Bernstein and C. Bonnett and A. Carnero Rosell and F. J. Castander and C. Chang and L. N. da Costa and T. M. Davis and J. De Vicente and J. DeRose and A. Drlica-Wagner and J. Elvin-Poole and E. Gaztanaga and D. Gruen and J. Gschwend and W. G. Hartley and B. Hoyle and H. Lin and M. A. G. Maia and R. Miquel and R. L. C. Ogando and M. M. Rau and A. Roodman and E. S. Rykoff and I. Sevilla-Noarbe and M. A. Troxel and R. H. Wechsler and T. M. C. Abbott and F. B. Abdalla and S. Allam and J. Annis and K. Bechtol and A. Benoit-Lévy and D. Brooks and E. Buckley-Geer and D. L. Burke and M. Carrasco Kind and J. Carretero and M. Crocce and C. E. Cunha and S. Desai and H. T. Diehl and P. Doel and T. F. Eifler and B. Flaugher and J. Frieman and J. García-Bellido and D. W. Gerdes and R. A. Gruendl and G. Gutierrez and K. Honscheid and D. J. James and T. Jeltema and E. Krause and R. Kron and K. Kuehn and N. Kuropatkin and O. Lahav and M. Lima and M. March and J. L. Marshall and F. Menanteau and R. C. Nichol and B. Nord and A. A. Plazas and E. Sanchez and V. Scarpine and R. Schindler and M. Smith and M. Soares-Santos and F. Sobreira and E. Suchyta and M. E. C. Swanson and G. Tarle and D. Thomas and D. L. Tucker and V. Vikram and A. R. Walker and J. Zuntz},
      year={2017},
      eprint={1710.02517},
      archivePrefix={arXiv},
      primaryClass={astro-ph.CO},
      url={https://arxiv.org/abs/1710.02517}, 
}

@article{Secco2022CosmoShearDES,
   title={Dark Energy Survey Year 3 results: Cosmology from cosmic shear and robustness to modeling uncertainty},
   volume={105},
   ISSN={2470-0029},
   url={http://dx.doi.org/10.1103/PhysRevD.105.023515},
   DOI={10.1103/physrevd.105.023515},
   number={2},
   journal={Physical Review D},
   publisher={American Physical Society (APS)},
   author={Secco, L. F. and Samuroff, S. and Krause, E. and Jain, B. and Blazek, J. and Raveri, M. and Campos, A. and Amon, A. and Chen, A. and Doux, C. and Choi, A. and Gruen, D. and Bernstein, G. M. and Chang, C. and DeRose, J. and Myles, J. and Ferté, A. and Lemos, P. and Huterer, D. and Prat, J. and Troxel, M. A. and MacCrann, N. and Liddle, A. R. and Kacprzak, T. and Fang, X. and Sánchez, C. and Pandey, S. and Dodelson, S. and Chintalapati, P. and Hoffmann, K. and Alarcon, A. and Alves, O. and Andrade-Oliveira, F. and Baxter, E. J. and Bechtol, K. and Becker, M. R. and Brandao-Souza, A. and Camacho, H. and Carnero Rosell, A. and Carrasco Kind, M. and Cawthon, R. and Cordero, J. P. and Crocce, M. and Davis, C. and Di Valentino, E. and Drlica-Wagner, A. and Eckert, K. and Eifler, T. F. and Elidaiana, M. and Elsner, F. and Elvin-Poole, J. and Everett, S. and Fosalba, P. and Friedrich, O. and Gatti, M. and Giannini, G. and Gruendl, R. A. and Harrison, I. and Hartley, W. G. and Herner, K. and Huang, H. and Huff, E. M. and Jarvis, M. and Jeffrey, N. and Kuropatkin, N. and Leget, P.-F. and Muir, J. and Mccullough, J. and Navarro Alsina, A. and Omori, Y. and Park, Y. and Porredon, A. and Rollins, R. and Roodman, A. and Rosenfeld, R. and Ross, A. J. and Rykoff, E. S. and Sanchez, J. and Sevilla-Noarbe, I. and Sheldon, E. S. and Shin, T. and Troja, A. and Tutusaus, I. and Varga, T. N. and Weaverdyck, N. and Wechsler, R. H. and Yanny, B. and Yin, B. and Zhang, Y. and Zuntz, J. and Abbott, T. M. C. and Aguena, M. and Allam, S. and Annis, J. and Bacon, D. and Bertin, E. and Bhargava, S. and Bridle, S. L. and Brooks, D. and Buckley-Geer, E. and Burke, D. L. and Carretero, J. and Costanzi, M. and da Costa, L. N. and De Vicente, J. and Diehl, H. T. and Dietrich, J. P. and Doel, P. and Ferrero, I. and Flaugher, B. and Frieman, J. and García-Bellido, J. and Gaztanaga, E. and Gerdes, D. W. and Giannantonio, T. and Gschwend, J. and Gutierrez, G. and Hinton, S. R. and Hollowood, D. L. and Honscheid, K. and Hoyle, B. and James, D. J. and Jeltema, T. and Kuehn, K. and Lahav, O. and Lima, M. and Lin, H. and Maia, M. A. G. and Marshall, J. L. and Martini, P. and Melchior, P. and Menanteau, F. and Miquel, R. and Mohr, J. J. and Morgan, R. and Ogando, R. L. C. and Palmese, A. and Paz-Chinchón, F. and Petravick, D. and Pieres, A. and Plazas Malagón, A. A. and Rodriguez-Monroy, M. and Romer, A. K. and Sanchez, E. and Scarpine, V. and Schubnell, M. and Scolnic, D. and Serrano, S. and Smith, M. and Soares-Santos, M. and Suchyta, E. and Swanson, M. E. C. and Tarle, G. and Thomas, D. and To, C.},
   year={2022},
   month=jan }

\appendix
\section{Tomographic bin widths in the photometric galaxy bias approximation} \label{sec:appendix_pz}
In this appendix, the effects of the width and photo-$z$ spread for the galaxy bias measurement described in section \ref{sec:modeling:photoz_gal_bias} are investigated. 
In the aforementioned section, the choice is to use $\dz=0.1$ tomographic bins in order to recover auto-correlation evolution. In the context of this appendix we will compare deviations in galaxy bias $b_p(z)$ inferred from the approximated $\bo^\mathrm{meas}_{pp}$ with the the galaxy bias found using the underlying true $\bo^\mathrm{true}_{pp}$, and thus evaluate efficiency of the method with deviations from the chosen bias model. Therefore, to compare deviations from bias expressions instead of $\bo_{pp}$, we will use the linear bias approximation in this appendix. We note that the linear bias assumption is not involved in the computation of $\bo^\mathrm{meas}_{pp}$ and thus not included in the $n(z)$ measurement.
The purpose of this appendix is to assess if binning by $\dz_{p_k}=0.1$ is valid and verify if the fairly wide profiles of the distributions presented in Figure \ref{fig:photoz_gal_bias} do not significantly affect the recovered $\bo^\mathrm{meas}_{pp}$ expression, compared to a correction where one would use much narrower tomographic bins such as in \cite{Cawthon2022DESY3Boss} ($\dz=0.02$). 
In fact, using very narrow tomographic bins to probe galaxy bias leads to its own set of difficulties, as one must take into account new systematics such as border effects with adjacent bins, validity of the Limber equation, and use of spectroscopic bins with minimal data.
Effects of using different tomographic bin widths for this correction were explored in \cite{EuclidClusterRedshifts2025}.

For this model, we assume that the underlying photometric galaxy bias $b_p^{\rm true}$ follows three different forms (case a, b and c): 
\begin{equation}\label{eq:app:bias_models}
b_p^{\rm true}(z)=
    \begin{cases}
        1\quad\quad\quad\;\;\text{(a, constant evolution)}\\ 1/D(z)\quad\text{(b, passive evolution)}\\ 1+z^2\quad\;\,\text{(c, strong evolution)}
    \end{cases}
\end{equation}

\noindent where $D(z)$ is the factor of growth of the universe. 
The measured distributions for tomographic bins are assumed to follow $n^\mathrm{meas}_{p_k}(z)$ distributions where $p_k$ are the successive tomographic bins. 
In this context, the $p$ subscript is used to describe a generic tomographic bin, different from the fiducial HSC tomographic bins presented in the main corpus of this work.

\begin{figure}[h]
    \centering
    \includegraphics[width=0.92\textwidth]{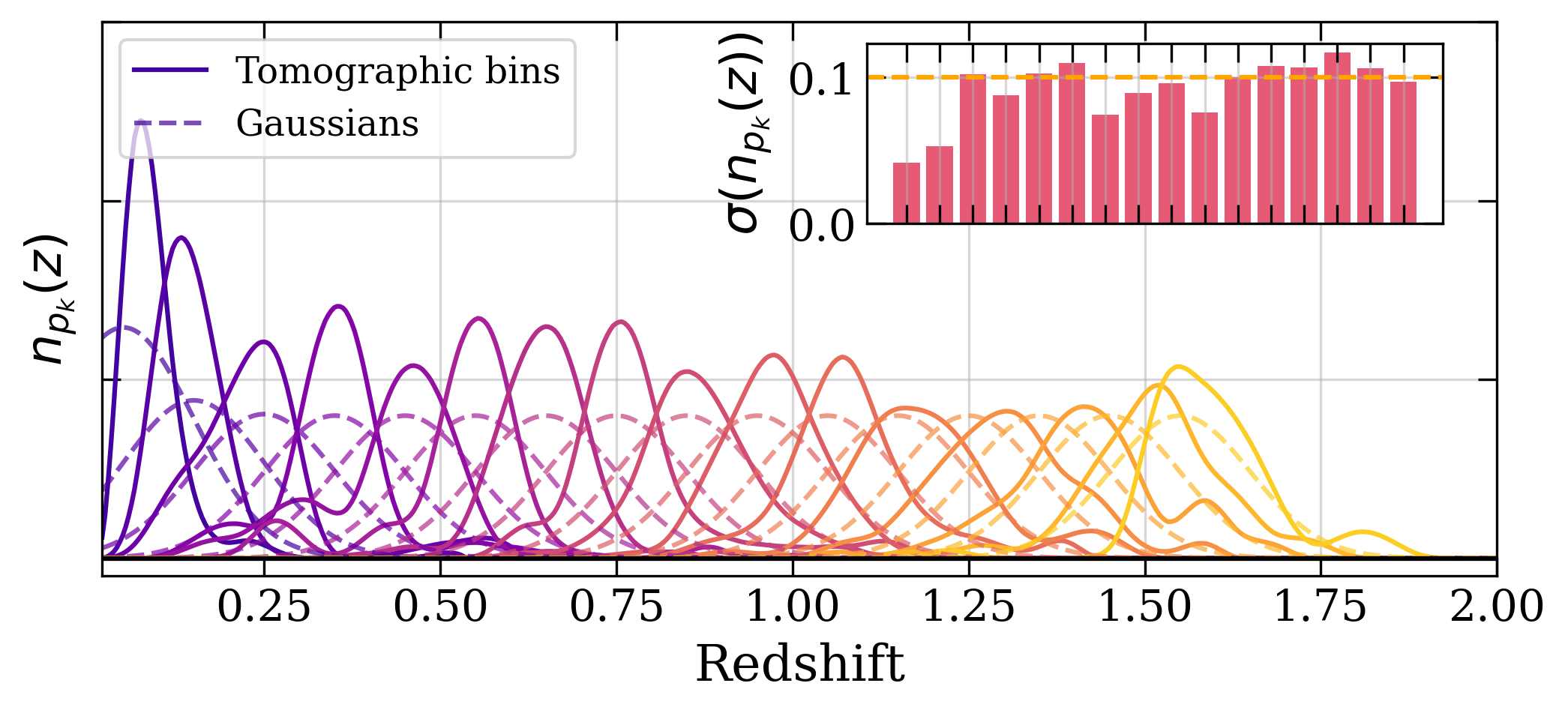}
    \caption{Photo-z $n_{pk}^{\mathrm{meas}}$ tomographic bins used in this model for (i) Tomographic bins derived from section \ref{sec:modeling:photoz_gal_bias} and (ii) gaussian tomographic bins $\sim\mathcal{N}(z_k,\,0.1)$ centered on the midpoint of the photometric redshift bins $z_k$. The inset plot of the figure showcases the standard deviation $\sigma(n_{p_k}(z))$ for the successive bins, compared to the chosen $0.1$ level for the Gaussian bins scenario.}
    \label{fig:appendix_nz_models}
\end{figure}

Case 1 will assume the $n_{p_k}$ distributions measured in section \ref{sec:modeling:photoz_gal_bias} are the $n^\mathrm{meas}_{p_k}$ distributions for the tomographic bins.
Case 2 will be treated with gaussian tomographic bins $n^\mathrm{meas}_{p_k}\sim\mathcal{N}(z_k,\,0.1)$ where $z_k$ is the center redshift of the tomographic bin $p_k$. We choose $0.1$ in order to compare with the standard deviations observed on the tomographic bins, as displayed in figure \ref{fig:appendix_nz_models}. Case 1 is shown in full lines, while case 2 is shown in dashed lines. 

We further add a inset figure characterizing the standard deviation of the distributions for each bin in case 1 (crimson histogram); compared to the chosen $0.1$ standard deviation value assumed for the gaussian scenario (gold line).
In section \ref{sec:modeling:photoz_gal_bias}, $b_p(z)\sqrt{\bo_{mm}(z)}$ is assumed to be constant over the tomographic bin. Other works assume $b_p(z)$ is constant over the tomographic bin: we note the quantity we consider constant in our approximation to be $\gamma(z)$, where $\gamma(z)=b(z)\sqrt{\bo_{mm}(z)}$ (method 1, $\tilde{n}_1(z)\propto\bo_{sp}(z)/\sqrt{\bo_{ss}(z)}$, used in \cite{Cawthon2022DESY3Boss} and in section \ref{sec:modeling:photoz_gal_bias}) or $\gamma(z)=b(z)$ (method 2, $\tilde{n}_2(z)\propto\bo_{sp}(z)/\sqrt{\bo_{ss}(z)\bo_{mm}(z)}$, used in \cite{EuclidClusterRedshifts2025}). 
As such, the theoretical true distributions $n_{p_k}^{\rm true}$ with respect to the measurements follow:
\begin{equation}
    n_{p_k}^{\mathrm{true}}(z)=\frac{n^{\mathrm{meas}}_{p_k}(z)\frac{1}{\gamma(z)}}{\int_0^{+\infty} n^{\mathrm{meas}}_{p_k}(z')\frac{1}{\gamma(z')}\mathrm{d}z'}
\end{equation}

\noindent Given $b_p^\mathrm{true}(z)$ usually increases with redshift, $\sqrt{\bo_{mm}(z)}$ and $b_p(z)$ are competing effects. 
Therefore, $\tilde{n}_1(z)$ better approximates the underlying true $n(z)$ than $\tilde{n}_2(z)$ for biases $b_p^\mathrm{true}(z)$ that evolve sufficiently fast. 

\begin{figure}[h]
    \centering
    \includegraphics[width=0.92\textwidth]{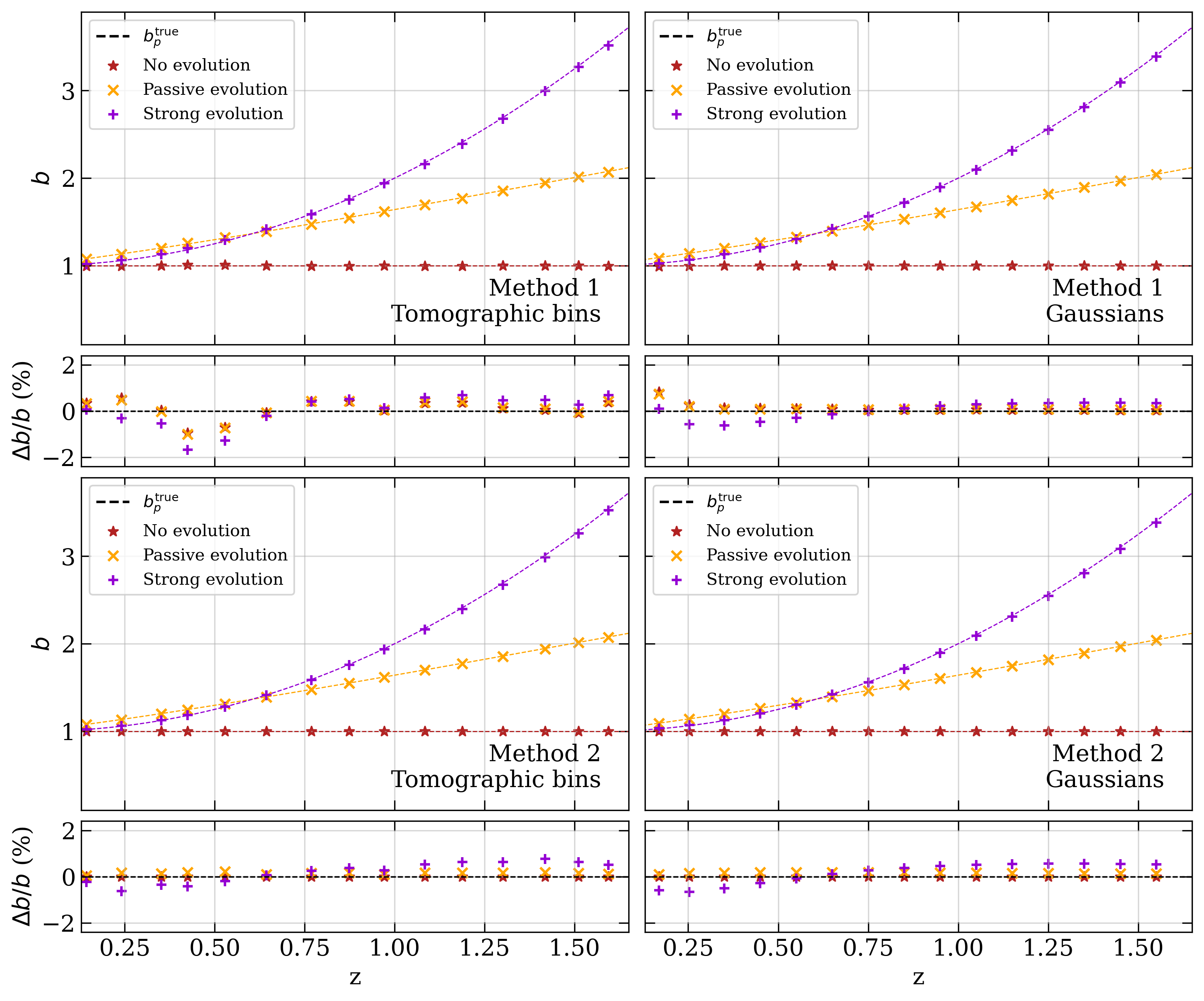}
    \caption{Relative error measurements on the bias for each scenario. We find that the bias deviations are contained to the $\%$ level.}
    \label{fig:appendix_gal_bias}
\end{figure}

Here, sufficiently fast implies that the optimal case to recover the exact $n(z)$ with $\tilde{n}_1$ is obtained for $b_p(z)\propto1/\sqrt{\bo_{mm}(z)}$, while the optimal case for $\tilde{n}_2$ is $b_p$ constant (case a in equation \ref{eq:app:bias_models}): one can compare the local amplitudes of $\left|\frac{\mathrm{d}\gamma(z)}{\mathrm{dz}}\frac{1}{\gamma(z)}\right|$.

The theoretical measured photometric sample cross-correlation $\bo_{pp}(z)$ is given by using equation \ref{eq:cross} in the case of auto-correlations:
\begin{equation}
    \bo_{p_kp_k}(z_k)=\int_0^{+\infty}b^{\mathrm{true}}_p(z)^2n^{\mathrm{true}}_{p_k}(z)^2\bo_{mm}(z)\mathrm{d}z
\end{equation}

\noindent Using the expression derived in section \ref{sec:modeling:photoz_gal_bias}, the theoretical measured galaxy bias (assuming linear galaxy bias) $b_p^{\rm meas}$ with the tomographic bins can be obtained:
\begin{equation}
    b_p^{\rm meas}(z_k)=\sqrt{\frac{\bo_{p_kp_k}(z_k)}{\int_0^{+\infty}n_{p_k}^{\rm meas}(z)^2\bo_{mm}(z)\mathrm{d}z}}
\end{equation}

\noindent In figure \ref{fig:appendix_gal_bias}, we display respectively the relative error $\Delta b_p/b_p(z)=(b_p^{\rm meas}(z)-b_p^{\rm true}(z))/b_p^{\rm true}(z)$ in percentage for each bin, as well as the reconstructed bias with tomographic bins. The relative error, just like the power law fit of section \ref{sec:modeling:photoz_gal_bias}, is computed by evaluating at the expectation value of the tomographic bin. Each subplot corresponds to a different method ($\tilde{n}_1$ or $\tilde{n}_2$) or a different tomographic bin case (1 or 2) with different $b^\mathrm{true}_p(z)$ expressions.
We conclude with this appendix that the tomographic bin approximation at $\Delta z_p=0.1$ presents no significant issues when recovering the general bias trend, since we observe local deviations from the true galaxy bias model of at most $\sim1.7\%$. This result shows that the method is highly effective at recovering the global trend for galaxy bias evolution.

\section{Towards probing high-redshift source distributions with quasars}\label{sec:appendix_qso}

This appendix provides some insights on the sensitivity of calibration measurements for high redshift sources with the QSO sample on the HSC dataset. Future and ongoing Stage-IV optical surveys such as Euclid \cite{EuclidI2024overview}, the Roman telescope \cite{roman2019wfirst}, the Vera C. Rubin telescope (LSST) \cite{LSST2019survey} and in the context of this work HSC \cite{HSCOverviewAihara2017} will probe large weak lensing samples with unprecedented depths.

\begin{figure}[h]
    \centering
    \includegraphics[width=0.92\textwidth]{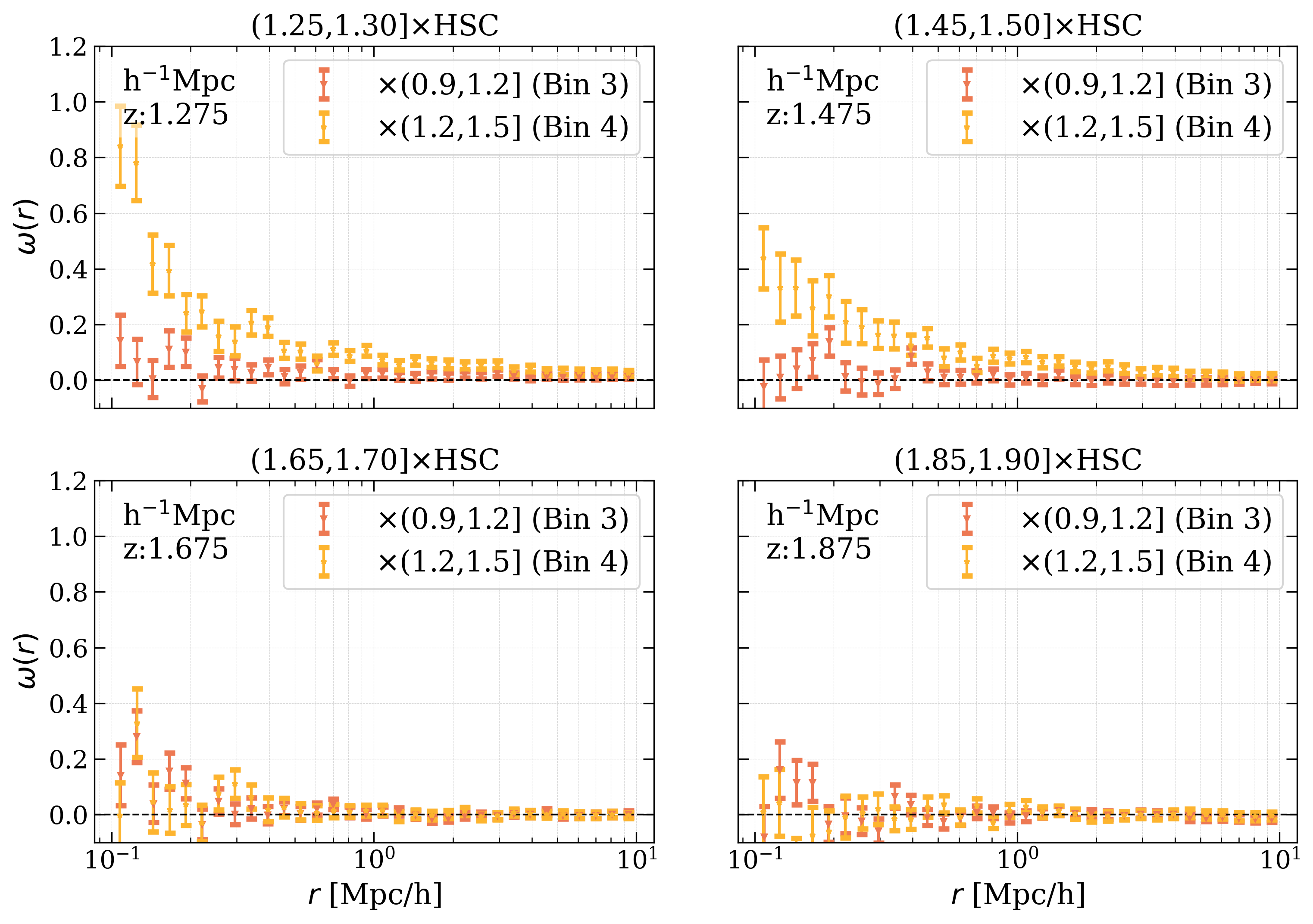}
    \caption{Cross-correlation functions between the DESI DR1 QSOs at 4 different fine spectroscopic bins and the two last tomographic bins of the HSC catalog.}
    \label{fig:appendix_qso_cross}
\end{figure}

\noindent This raises the question of calibrations of high-redshift source populations, beyond $z\sim1.6$ (where ELGs are not available). Moreover, high-redshift calibration can be useful to break degeneracies in redshift populations and identifying outlier populations confused by the chosen photometric redshift algorithm: for example, the calibration cut mentioned in section \ref{sec:data:hsc} removes a potentially degenerate high-redshift population with the first and second bin. High-redshift calibration could probe such populations and avoid conservative cuts on the redshift populations. 

In Figure \ref{fig:appendix_qso_cross} show the cross-correlations between the DR1 QSO sample and the two last redshift bins for HSC at different redshift ranges in order to get a sense of the $\omega(r_p)$ sensitivity. Note that the measurement error is comparable to the DR2 sensitivity due to the already high coverage of DR1 in HSC, as showcased by the density difference in Figure \ref{fig:density_z}. In DR1, density of QSOs on the HSC footprint is $\sim0.047\;\mathrm{arcmin}^{-2}$, and increases to $\sim0.057\;\mathrm{arcmin}^{-2}$ for DR2.

\begin{figure}[h]
    \centering
    \includegraphics[width=0.9\textwidth]{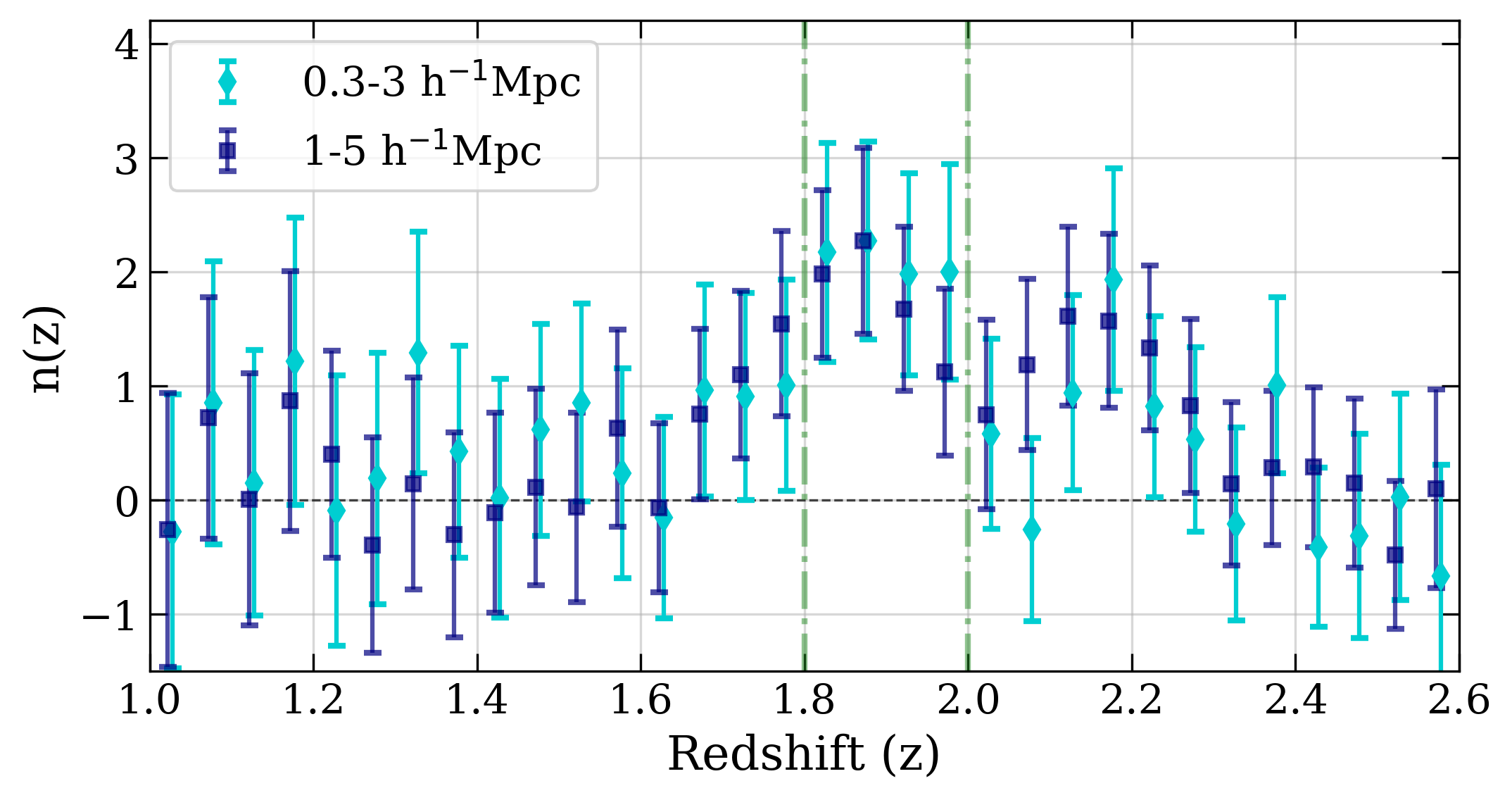}
    \caption{Measuring a $0.1$ arcmin$^{-2}$ population of galaxies with quasars over $z_\mathrm{phot}\in]1.8,\,2.0)$. Both scale cuts are displayed. In green, we display the boundaries of Bin 5 for reference. Measurements are slightly offset for readability.}
    \label{fig:appendix_tomobin_highz}
\end{figure}

To further assess the sensitivity to galaxy populations $z\gtrsim1.6$, we create a fifth tomographic bin for redshifts $z_\mathrm{phot}\in]1.8,\,2.0)$. This fifth tomographic bin is further randomly down-sampled to $25\%$ of its original size, in order to obtain an average raw angular density for HSC sources of $\sim0.1\;\mathrm{arcmin}^{-2}$ in the bin. The measurements use DESI DR2 QSOs alone and are presented in Figure \ref{fig:appendix_tomobin_highz}.

\noindent For this measurement, we assume the photometric sample galaxy bias reported in section \ref{sec:modeling:photoz_gal_bias} remains valid at high redshifts. We display measurements at both scale cuts. One can notice that at these redshifts, the statistical power of QSOs is broadly similar for both scale cuts. This can be explained the sparsity of the QSO population \cite{DESI.Selection.QSO.Chaussidon2023} and lack of satellites, which leads to low signal at small scales, especially below the halo virial radius. 
Still, Figure \ref{fig:appendix_tomobin_highz} reports a significant detection of the "Bin 5" population for both scale cuts, thus showing QSO sensitivity for high redshift calibration. For example, the QSOs reveal a photometric redshift scatter towards higher redshifts in this bin ($z\sim2.0-2.3$).
While quasar clustering redshift calibration is relatively novel, recent DES Y6 works \cite{dassignies2025_clusteringDESY6} use eBOSS \cite{Dawson2016_eBOSS} quasars for high-redshift calibration of the tomographic bins and finds comparable sensitivities. Therefore, quasars offer promising avenue for high redshift source calibration as well as breaking degeneracies for outlier populations in photometric redshift catalogs. 

\end{document}